\documentclass[12pt]{article}
\usepackage{jheppub}

\pdfoutput=1

\usepackage{amsmath,bbm,array,amsfonts,graphicx,wrapfig,lscape,float,mathtools,multirow,longtable}
\usepackage[dvipsnames]{xcolor}

\newcommand{\be}{\begin{equation}}
\newcommand{\ee}{\end{equation}}
\newcommand{\beq}{\begin{equation}}
\newcommand{\beql}[1]{\begin{equation}\label{#1}}
\newcommand{\eeq}{\end{equation}}
\newcommand{\ba}{\begin{array}}
\newcommand{\ea}{\end{array}}
\newcommand{\bea}{\begin{eqnarray}}
\newcommand{\beal}[1]{\begin{eqnarray}\label{#1}}
\newcommand{\eea}{\end{eqnarray}}
\newcommand{\ben}{\begin{enumerate}}
\newcommand{\een}{\end{enumerate}}
\newcommand{\bean}{\begin{eqnarray*}}
\newcommand{\eean}{\end{eqnarray*}}
\newcommand{\eref}[1]{(\ref{#1})}
\newcommand{\sref}[1]{\S\ref{#1}}
\newcommand{\tref}[1]{Table~\ref{#1}}
\newcommand{\nn}{\nonumber}

\newcommand{\fref}[1]{Figure \ref{#1}}
\newcommand{\btab}[1]{\begin{tabular}{#1}}
\newcommand{\etab}{\end{tabular}}

\def \Black {\pdfsetcolor{0 0 0 1}}

\newcommand{\comment}[1]{}

\newcommand{\ud}{\mathrm{d}}

\newcommand{\qed}{\nobreak \ifvmode \relax \else
      \ifdim\lastskip<1.5em \hskip-\lastskip
      \hskip1.5em plus0em minus0.5em \fi \nobreak
      \vrule height0.75em width0.5em depth0.25em\fi}

\definecolor{darkspringgreen}{rgb}{0.09, 0.45, 0.27}
\definecolor{forestgreen}{rgb}{0.13, 0.55, 0.13}

\usepackage{array}
\newcolumntype{C}[1]{>{\centering\let\newline\\\arraybackslash\hspace{0pt}}m{#1}}

\usepackage{breqn}

\title{Brane Brick Models for the Sasaki-Einstein 7-Manifolds $Y^{p,k}(\mathbb{CP}^{1}\times \mathbb{CP}^{1})$ and $Y^{p,k}(\mathbb{CP}^{2})$} 

\author[a,b,c]{Sebasti\'an Franco,} 
\author[d]{Dongwook Ghim,}
\author[e,f]{and Rak-Kyeong Seong}

\affiliation[a]{Physics Department, The City College of the CUNY\\
	160 Convent Avenue, New York, NY 10031, USA}
	
\affiliation[b]{Physics Program and \textsuperscript{$c$}Initiative for the Theoretical Sciences\\
	The Graduate School and University Center, The City University of New York\\
	365 Fifth Avenue, New York NY 10016, USA}

\affiliation[d]{
Yukawa Institute for Theoretical Physics, Kyoto University, \\
Sakyo-ku, Kyoto 606-8502, Japan
}
	
\affiliation[e]{
Department of Mathematical Sciences, and 
\textsuperscript{$f$}Department of Physics,\\ 
Ulsan National Institute of Science and Technology,\\
50 UNIST-gil, Ulsan 44919, South Korea
}

\emailAdd{sfranco@ccny.cuny.edu}
\emailAdd{dghim@yukawa.kyoto-u.ac.jp}
\emailAdd{seong@unist.ac.kr}

\preprint{
\begin{flushright}
UNIST-MTH-22-RS-02 \\
YITP-22-139\\
\end{flushright}
}

\abstract{
The $2d$ $(0,2)$ supersymmetric gauge theories corresponding to the classes of $Y^{p,k}(\mathbb{CP}^1\times\mathbb{CP}^1)$ and $Y^{p,k}(\mathbb{CP}^2)$ manifolds are identified. 
The complex cones over these Sasaki-Einstein 7-manifolds are non-compact toric Calabi-Yau 4-folds.
These infinite families of geometries are the largest ones for Sasaki-Einstein 7-manifolds whose metrics, toric diagrams, and volume functions are known explicitly.
This work therefore presents the largest classification of $2d$ $(0,2)$ supersymmetric gauge theories corresponding to Calabi-Yau 4-folds with known metrics.
}

\begin{document}

\maketitle

\section{Introduction}

The realization of gauge theories in various dimensions in terms of D-branes probing Calabi-Yau (CY) singularities is a powerful approach with multiple applications, including the gauge gravity correspondence, a geometric perspective on gauge theory dynamics, and model building, to name a few (see e.g. \cite{Klebanov:1998hh,Morrison:1998cs,Klebanov:2000hb,Aldazabal:2000sa,Verlinde:2005jr}). 
When the CYs are toric, the gauge theories are additionally endowed with beautiful combinatorial structures. For example, the $4d$ gauge theories on D3-branes probing toric CY 3-folds are captured by brane tilings \cite{Hanany:2005ve,Franco:2005rj}, also known as dimer models.
Brane tilings and their generalizations led to trailblazing developments in connection with the AdS/CFT correspondence \cite{Maldacena:1997re,Gubser:1998bc,Witten:1998qj,Franco:2005sm}, cluster algebras \cite{Baur:2013hwa, Franco:2014nca}, integrable systems \cite{Goncharov:2011hp,Franco:2011sz,Eager:2011dp,Franco:2015rnr}, scattering amplitudes \cite{Arkani-Hamed:2012zlh,Franco:2012mm} and many other research areas in mathematics and theoretical physics.

Brane tilings provide a general solution to the correspondence between $4d$ gauge theories and toric CY$_3$’s. They are also powerful enough to concretely deal with infinite families of CY$_3$’s, among which those with explicitly known metrics are particularly interesting. A prominent example is given by the $Y^{p,q}$ family of Sasaki-Einstein manifolds, where $p\geq 0$ and $0\leq q \leq p$, whose metrics were determined in \cite{Gauntlett:2004yd}.\footnote{For brevity, throughout the paper, we will often use the same name to refer to the Sasaki-Einstein base and the corresponding cone. The distinction should be clear from the context.} Before the discovery of the metrics for the $Y^{p,q}$ manifolds, the only Sasaki-Einstein 5-manifolds with known metrics were $S^5$ and $T^{1,1}$ \cite{Romans:1984an} (and there orbifolds), whose corresponding cones are $\mathbb{C}^3$ and the conifold, respectively. Explicit metrics were later introduced for the $L^{a,b,c}$ family, which contain and generalize the $Y^{p,q}$’s, and the corresponding brane tilings were found in \cite{Franco:2005sm,Butti:2005sw}. The matching in these theories between non-trivial gauge theory data, such as central charges and R-charges, and the corresponding geometric computations of volumes represented striking confirmations of the AdS/CFT correspondence with $\mathcal{N}=1$ SUSY. Remarkably, subsequent developments made it possible to determine many of these geometric quantities directly from the toric data, without requiring knowledge of the metric \cite{Martelli:2005tp,Butti:2005vn,Martelli:2006yb,Feng:2007ur}. Brane tilings also allow for an elegant treatment of infinite families of orbifolds of various geometries \cite{Hanany:2010cx,Davey:2010px}.

More recently, a similar program has focused on understanding the $2d$ $\mathcal{N}=(0,2)$ gauge theories on D1-branes probing toric CY 4-folds \cite{Franco:2015tna}. This program culminated with the introduction of {\it brane brick models}, a new class of type IIA brane configurations that are connected to the D1-branes at the singular CY 4-folds by T-duality \cite{Franco:2015tya}. Brane brick models fully encode the corresponding $2d$ (0,2) theories. Namely, they specify both the quivers and the $J$- and $E$-terms. For this reason, throughout this paper, we will consider determining the brane brick model and finding the $2d$ gauge theory for a given CY$_4$ as synonyms.\footnote{More precisely, the correspondence between toric CY$_4$'s and brane brick models is not one-to-one but one-to-many, due to the existence of a low energy equivalence between $2d$ (0,2) gauge theories denoted triality \cite{Gadde:2013lxa,Franco:2016nwv}.} Brane brick models significantly simplify the map between geometry and the corresponding gauge theories. We refer the interested reader to \cite{Franco:2016nwv,Franco:2016qxh,Franco:2017cjj} for further developments. Interestingly, in some cases, the brane brick models for certain CY 4-folds can be derived from brane tilings associated to appropriate CY 3-folds. Algorithms for achieving this include dimensional reduction \cite{Franco:2015tna}, orbifold reduction \cite{Franco:2016fxm} and $3d$ printing \cite{Franco:2018qsc}.

Brane brick models have been explicitly constructed for several toric CY 4-folds (see e.g. \cite{Franco:2015tna,Franco:2015tya,Franco:2016nwv,Franco:2016qxh} for various examples).
More recently, brane brick models were found for all the 18 smooth Fano 3-folds, whose toric diagrams are given by regular reflexive lattice polytopes in 3 dimensions \cite{0025-5726-19-1-A02, Franco:2022gvl}.
In this work, we concentrate on a particular class of toric CY 4-folds that consists of cones over two families of Sasaki-Einstein 7-manifolds known as $Y^{p,k}(\mathbb{CP}^{1}\times\mathbb{CP}^{1})$ and $Y^{p,k}(\mathbb{CP}^{2})$ \cite{Martelli:2008rt},   
which can be regarded as natural generalizations of the $Y^{p,q}$ 5-manifolds. 

\begin{figure}[H]
\begin{center}
\resizebox{0.7\hsize}{!}{
  \includegraphics[trim=0mm 0mm 0mm 0mm, width=8in]{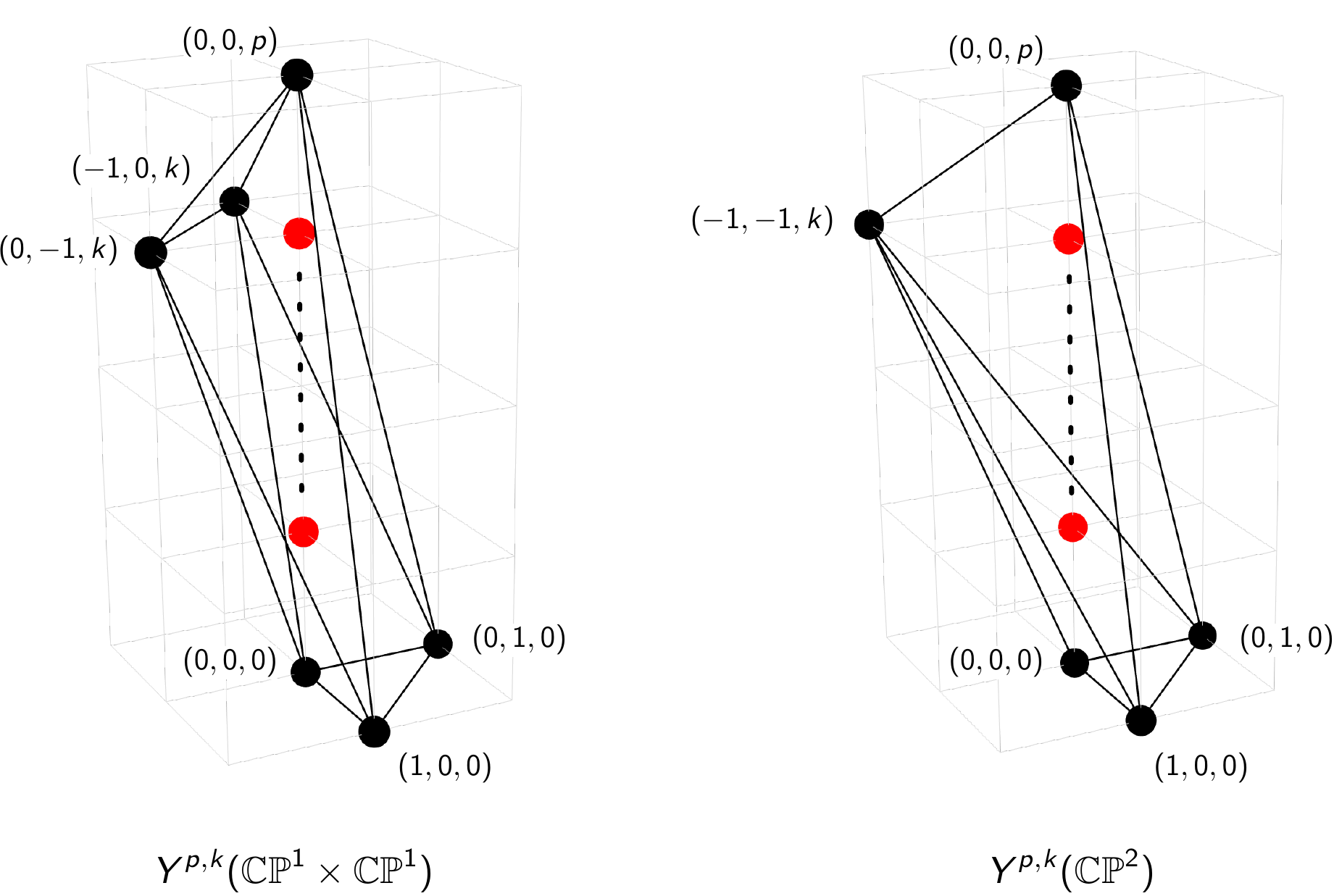}
}
\caption{
General toric diagrams corresponding to the $Y^{p,k}(\mathbb{CP}^2)$ and $Y^{p,k}(\mathbb{CP}^1\times \mathbb{CP}^1)$ families of toric CY 4-folds. 
The ranges for the parameters $p$ and $k$ are summarized in \eref{es05a04}.
\label{f_toric_ypk}
}
\end{center}
\end{figure}

In \cite{Gauntlett:2004hh}, it was shown that for any positive curvature K\"ahler-Einstein manifold $B_{2n}$, there is an infinite family of corresponding Sasaki-Einstein manifolds of the form $Y_{2n+3}(B_{2n})$.
When $B_4$ is either $\mathbb{CP}^{1}\times \mathbb{CP}^{1}$ or $\mathbb{CP}^2$, we have the two families of 7-dimensional Sasaki-Einstein manifolds known as $Y^{p,k}(\mathbb{CP}^{1}\times\mathbb{CP}^{1})$ and $Y^{p,k}(\mathbb{CP}^{2})$, respectively.
Our primary focus in this paper will be on these two families of Sasaki-Einstein 7-manifolds and the corresponding toric CY 4-folds.
As previously mentioned, from here onwards we will use the names $Y^{p,k}(\mathbb{CP}^{1}\times\mathbb{CP}^{1})$ and $Y^{p,k}(\mathbb{CP}^{2})$ interchangeably for both the Sasaki-Einstein 7-manifolds and for the corresponding CY 4-folds. The two families $Y^{p,k}(\mathbb{CP}^{1}\times\mathbb{CP}^{1})$ and $Y^{p,k}(\mathbb{CP}^{2})$ stand out because their Sasaki-Einstein metrics are known explicitly \cite{Gauntlett:2004hh,Martelli:2008rt}. 
The general toric diagrams for the corresponding CY 4-folds are show in \fref{f_toric_ypk}.

In this work, we take these two infinite families of toric CY 4-folds and construct the $2d$ $(0,2)$ gauge theories on D1-branes probing them.\footnote{More precisely, we construct one gauge theory for each CY$_4$ in the family. Generically, we expect multiple dual theories related by triality for each geometry. We focus on phases described by brane brick models, which are commonly referred to as toric phases \cite{Franco:2016nwv}.} The mesonic moduli space for each of these theories agrees with the desired CY$_4$, confirming that the proposed gauge theories are indeed correct.

The following work is structured as follows. In section \sref{sec:background}, a brief overview of brane brick models and the families of Sasaki-Einstein 7-manifolds $Y^{p,k}(\mathbb{CP}^{1}\times\mathbb{CP}^{1})$ and $Y^{p,k}(\mathbb{CP}^{2})$ is given.
Then, in sections \sref{sec:cp1cp1} and \sref{sec:cp2}, we present the general form of the field content and $J$- and $E$-terms for the $2d$ $(0,2)$ supersymmetric gauge theories for $Y^{p,k}(\mathbb{CP}^{1}\times\mathbb{CP}^{1})$ and $Y^{p,k}(\mathbb{CP}^{2})$, respectively. 
We illustrate the constructions with explicit examples for cases $p=2,1$ where $p\leq k$. 
By calculating the mesonic moduli space, we verify that the proposed $2d$ $(0,2)$ gauge theories and the associated brane brick models correspond to the desired geometries. Moreover, we identify the global symmetries of these theories, which match the isometry groups for the corresponding Sasaki-Einstein 7-manifolds identified in \cite{Gauntlett:2004hh,Martelli:2008rt}.
\\

\section{Background \label{sec:background}}

The following sections give a brief introductory overview of the two infinite families of Sasaki-Einstein 7-manifolds $Y^{p,k}(\mathbb{CP}^{1}\times\mathbb{CP}^{1})$ and $Y^{p,k}(\mathbb{CP}^{2})$ and the realization of $2d$ $(0,2)$ supersymmetric gauge theories through brane brick models.

\subsection{$Y^{p,k}$ Sasaki-Einstein 7-manifolds \label{sec:se7}}

In \cite{Gauntlett:2004hh}, it was found that for every $2n$-dimensional K\"ahler-Einstein manifold $B_{2n}$ there is an infinite family of compact Sasaki-Einstein manifolds $Y_{2n+3}$ of dimension $2n+3$ \cite{Martelli:2008rt}. 
For $n=2$, we have 4-dimensional K\"ahler-Einstein bases $B_4$, either $\mathbb{CP}^{1}\times \mathbb{CP}^{1}$ or $\mathbb{CP}^{2}$, and two associated infinite families of Sasaki-Einstein 7-manifolds of the form $Y^{p,k}(\mathbb{CP}^{1}\times\mathbb{CP}^{1})$ and $Y^{p,k}(\mathbb{CP}^{2})$, respectively. 
These infinite families of Sasaki-Einstein manifolds can be thought of as Lens space bundles $S^{3}/\mathbb{Z}_p$ over the base K\"ahler-Einstein manifolds $\mathbb{CP}^{1}\times \mathbb{CP}^{1}$ and $\mathbb{CP}^{2}$ \cite{Martelli:2008rt}.

The isometry of the Sasaki-Einstein 7-manifolds takes the general form $H \times U(1)^2$, where $H$ is the isometry of the base $B_4$. 
For $Y^{p,k}(\mathbb{CP}^{1}\times\mathbb{CP}^{1})$ and $Y^{p,k}(\mathbb{CP}^{2})$, the isometries are $SU(2)\times SU(2) \times U(1)^2$ and $SU(3)\times U(1)^2$, respectively.  These isometries translate into the global symmetries of the corresponding brane brick models. 

\begin{figure}[ht!!]
\begin{center}
\resizebox{0.92\hsize}{!}{
  \includegraphics[trim=0mm 0mm 0mm 0mm, width=8in]{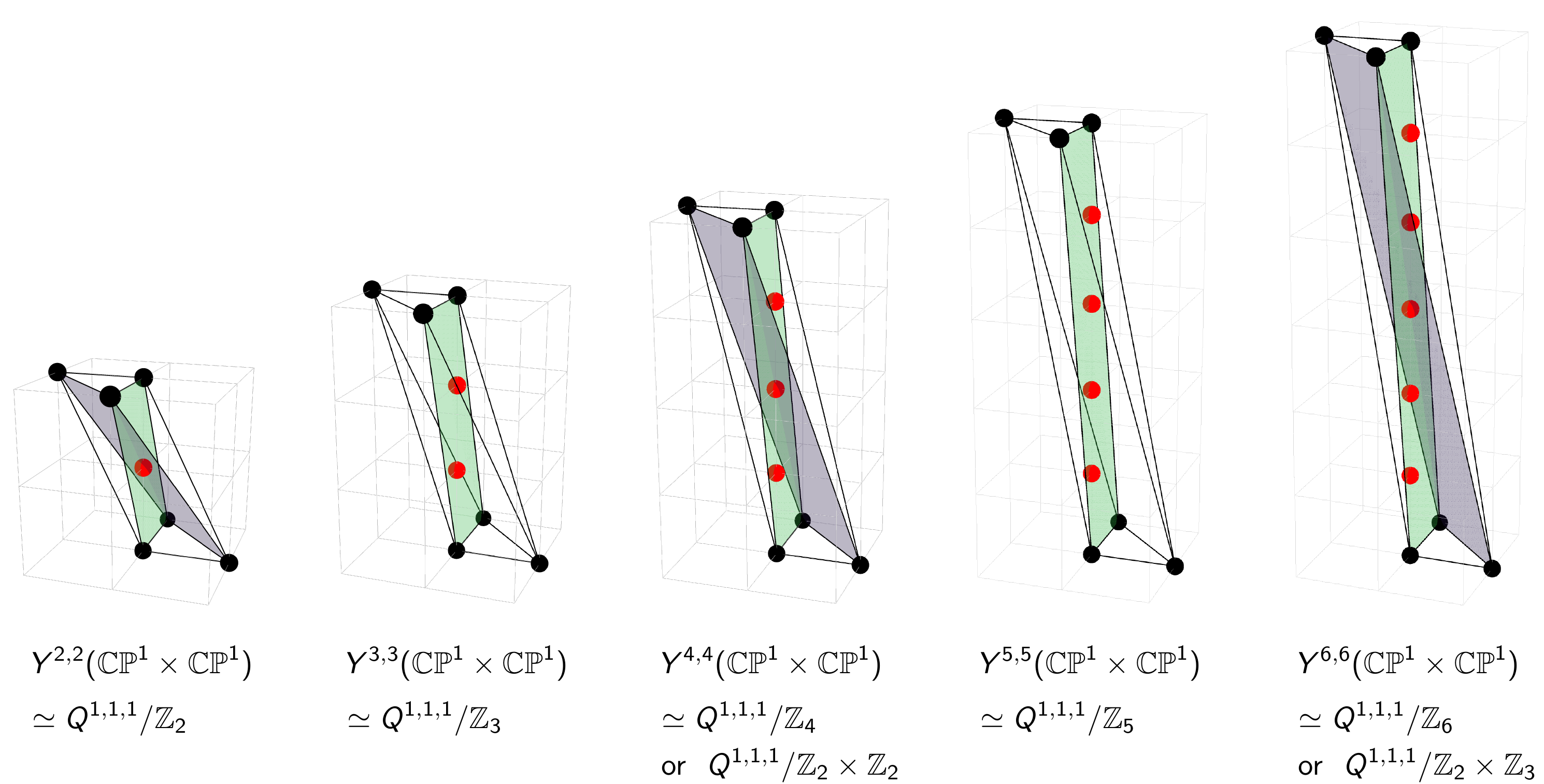}
}
\caption{
$Y^{p,k}(\mathbb{CP}^{1}\times\mathbb{CP}^{1})$ with $p=k$
is equivalent to the abelian orbifolds of the form $Q^{1,1,1}/\mathbb{Z}_p$. 
In addition, if $p=k=2m$ where $m\in \mathbb{Z}^{+}$, $Y^{p,k}(\mathbb{CP}^{1}\times\mathbb{CP}^{1})$ is also equivalent to another abelian orbifold of the form $Q^{1,1,1}/\mathbb{Z}_2 \times \mathbb{Z}_p$.
The corresponding $2d$ $(0,2)$ theories can be obtained from orbifold reduction \cite{Franco:2016fxm} of the $4d$ $\mathcal{N}=1$ theory corresponding to $F_0$ or an abelian orbifold of $F_0$.
The planes shown in gray correspond to the toric diagram for $F_0$ and the planes shown in green correspond to the toric diagrams for abelian orbifolds of the form $\mathcal{C}/\mathbb{Z}_m$. 
\label{fyppcp1cp1equivalences}
}
\end{center}
\end{figure}

The toric CY 4-folds $X_{4}$ have the following metric
\beal{es05a01}
\ud s^2 (X_4) = \ud r^2 + r^2 \ud s^2 (Y_7) ~,~
\eea
where the local metric $ \ud s^2 (Y_7)$ of the Sasaki-Einstein 7-manifold $Y_7$ is presented explicitly in \cite{Martelli:2008rt}. 
Here, rather than looking at the precise form of the metric, we note that a coordinate $\alpha$ of the metric can be periodically identified where the periods are related to $p,k\in \mathbb{Z}$.
The periodic identification of $\alpha$ is associated to a principle $U(1)$ bundle over a space $M_{6}$, which is the total space of an $S^2$ bundle over the base $B_{4}$.
The resulting range of $k$ is fixed to be of the following form
\beal{es05a02}
\frac{hp}{2} \leq k \leq h p ~,~
\eea
where $h\in \mathbb{Z}$ is related to Chern numbers of the principle $U(1)$ bundles according to \cite{Martelli:2008rt}.
For our two infinite families, the range of $k$ takes therefore the following forms
\beal{es05a03}
Y^{p,k}(\mathbb{CP}^{1}\times\mathbb{CP}^{1}) &~:~ 
p \leq k \leq 2 p 
~,~
\nn\\
Y^{p,k}(\mathbb{CP}^{2}) &~:~ 
\frac{3p}{2} \leq k \leq 3 p 
~.~
\eea
Under these ranges for $k$, the toric diagrams for $Y^{p,k}(\mathbb{CP}^{1}\times\mathbb{CP}^{1})$ and $Y^{p,k}(\mathbb{CP}^{2})$ were identified in \cite{Martelli:2008rt} as the convex hull of the following extremal points
\beal{es05a5}
Y^{p,k}(\mathbb{CP}^{1}\times\mathbb{CP}^{1}) &~:~ 
\left(
\begin{array}{cccccc}
w_1 & w_2 & w_3 & w_4 & w_5 & w _6 \\
\hline
0 & 0 & -1 & 1 & 0 & 0 \\
0 & 0 & 0 & 0 & -1 & 1 \\
0 & p & 0 & k & 0 & k \\
\end{array}
\right)
~,~
\nn\\
Y^{p,k}(\mathbb{CP}^{2}) &~:~ 
\left(
\begin{array}{ccccc}
w_1 & w_2 & w_3 & w_4 & w_5 \\
\hline
0 & 0 & 1 & 0 & -1 \\
0 & 0 & 0 & 1 & -1 \\
0 & p & 0 & 0 & k \\
\end{array}
\right)
~,~
\eea
where all points lie on a $4$-dimensional plane at distance $1$ from the origin.

\begin{figure}[ht!!]
\begin{center}
\resizebox{0.55\hsize}{!}{
  \includegraphics[trim=0mm 0mm 0mm 0mm, width=8in]{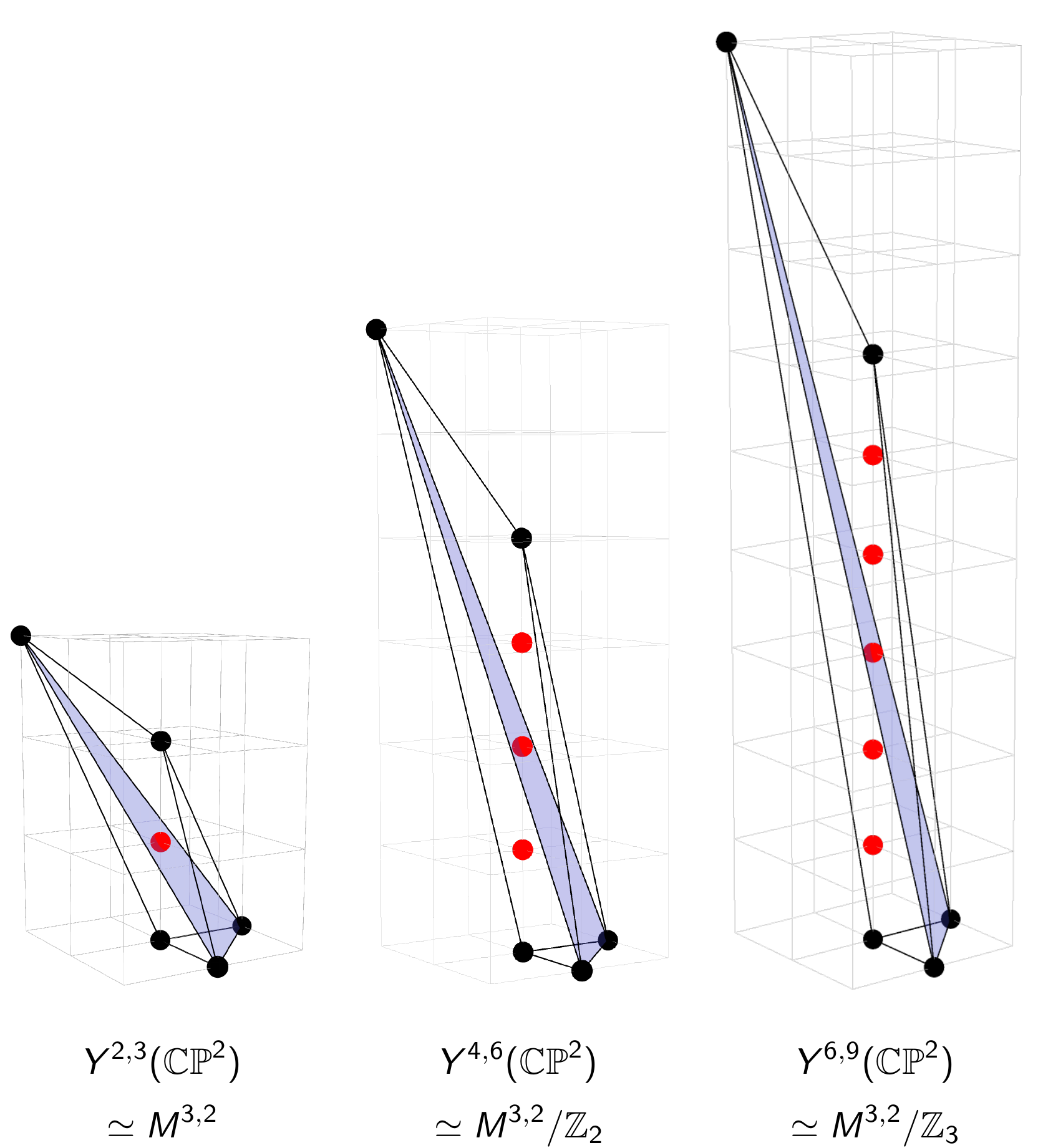}
}
\caption{
$Y^{p,k}(\mathbb{CP}^{2})$ with $p=\frac{2}{3} k = 2m$ and $m\in\mathbb{Z}^{+}$
is equivalent to the abelian orbifolds of the form $M^{3,2}/\mathbb{Z}_m$. 
The corresponding $2d$ $(0,2)$ theories can be obtained via orbifold reduction \cite{Franco:2016fxm} of the $4d$ $\mathcal{N}=1$ theory corresponding to $dP_0$.
The highlighted plane indicates the toric diagram for $dP_0$. 
\label{fyppcp2equivalences}
}
\end{center}
\end{figure}

The toric diagrams of toric CY 4-folds are equivalent up to $GL(3,\mathbb{Z})$ transformations. 
As such, we can find a suitable transformation to redefine the coordinates of the toric diagrams for $Y^{p,k}(\mathbb{CP}^{1}\times\mathbb{CP}^{1})$ and $Y^{p,k}(\mathbb{CP}^{2})$ in \eref{es05a5} as follows
\beal{es05a6}
Y^{p,k}(\mathbb{CP}^{1}\times\mathbb{CP}^{1}) &~:~ 
w_{a}
\cdot
\left(
\begin{array}{ccc}
1 & 0 & 0\\
0 & 1 & 0\\
p & p & -1
\ea
\right) 
\rightarrow
\left(
\begin{array}{cccccc}
w_1 & w_2 & w_3 & w_4 & w_5 & w _6 \\
\hline
 0 & 0 & -1 & 1 & 0 & 0 \\
 0 & 0 & 0 & 0 & -1 & 1 \\
 0 & -p & -p & p-k & -p & p-k \\
\end{array}
\right)
~,~
\nn\\
Y^{p,k}(\mathbb{CP}^{2}) &~:~ 
w_{a}
\cdot
\left(
\begin{array}{ccc}
1 & 0 & 0\\
0 & 1 & 0\\
-p & -p & -1
\ea
\right) 
\rightarrow
\left(
\begin{array}{ccccc}
w_1 & w_2 & w_3 & w_4 & w_5 \\
\hline
 0 & 0 & 1 & 0 & -1 \\
 0 & 0 & 0 & 1 & -1 \\
 0 & -p & -p & -p & 2 p-k \\
\end{array}
\right)
~,~
\eea
where for each point the $x^{(3)}$-coordinate can be shifted in the positive direction by $p$ to give respectively,
\beal{es05a7}
Y^{p,k}(\mathbb{CP}^{1}\times\mathbb{CP}^{1}) &~:~ 
\left(
\begin{array}{cccccc}
w_1 & w_2 & w_3 & w_4 & w_5 & w _6 \\
\hline
 0 & 0 & -1 & 1 & 0 & 0 \\
 0 & 0 & 0 & 0 & -1 & 1 \\
 p & 0 & 0 & 2p-k & 0 & 2p-k \\
\end{array}
\right)
~,~
\nn\\
Y^{p,k}(\mathbb{CP}^{2}) &~:~ 
\left(
\begin{array}{ccccc}
w_1 & w_2 & w_3 & w_4 & w_5 \\
\hline
 0 & 0 & 1 & 0 & -1 \\
 0 & 0 & 0 & 1 & -1 \\
 p & 0 & 0 & 0 & 3 p-k \\
\end{array}
\right)
~.~
\eea

By swapping the roles of $w_1$ and $w_2$, and by redefining the parameter $k$ as $k^\prime = 2p - k$ and $k^\prime = 3p - k$ for 
$Y^{p,k}(\mathbb{CP}^{1}\times\mathbb{CP}^{1})$ and $Y^{p,k}(\mathbb{CP}^{2})$, respectively, we can redefine the bounds on $k$ in \eref{es05a03} into a more convenient form as follows
\beal{es05a04}
Y^{p,k}(\mathbb{CP}^{1}\times\mathbb{CP}^{1}) &~:~ 
0 \leq k \leq p 
~,~
\nn\\
Y^{p,k}(\mathbb{CP}^{2}) &~:~ 
0 \leq k \leq \frac{3}{2} p 
~.~
\eea
In this work, for both $Y^{p,k}(\mathbb{CP}^{1}\times\mathbb{CP}^{1})$ and $Y^{p,k}(\mathbb{CP}^{2})$ families, we will include the upper and lower limits for $k$, since they correspond to additional distinct toric CY 4-folds. 
Using these new bounds on the redefined parameter $k$, we note that for $Y^{p,k}(\mathbb{CP}^{1}\times\mathbb{CP}^{1})$ when $k=0$, the toric CY 4-fold is the abelian orbifold of the form $\mathcal{C} \times \mathbb{C}/\mathbb{Z}_p$. In the other limit when $k=p$, $Y^{p,k}(\mathbb{CP}^{1}\times\mathbb{CP}^{1})$ becomes the abelian orbifold of the form $Q^{1,1,1}/\mathbb{Z}_p$ as illustrated in \fref{fyppcp1cp1equivalences}.
We further note that when $k=p=2m$ with $m\in \mathbb{Z}^{+}$, $Y^{p,k}(\mathbb{CP}^{1}\times\mathbb{CP}^{1})$ corresponds to another abelian orbifold of the form $Q^{1,1,1}/\mathbb{Z}_2 \times \mathbb{Z}_m$ as illustrated in \fref{fyppcp1cp1equivalences}.
Similarly, we note that for $Y^{p,k}(\mathbb{CP}^{2})$, when $k=0$, we have the abelian orbifold of the form $\mathbb{C}^4/\mathbb{Z}_3 \times \mathbb{Z}_p$, and in the limit when $p=\frac{2}{3} k = 2m$ with $m\in\mathbb{Z}^{+}$, the toric CY 4-fold is the abelian orbifold of the form $M^{3,2}/\mathbb{Z}_m$ as illustrated in \fref{fyppcp2equivalences}.

\subsection{$2d$ $(0,2)$ theories and brane brick models \label{sec:bbm}}

Brane brick models are Type IIA brane configurations connected by T-duality to D1-branes probing toric CY 4-folds. They fully encode the $2d$ $(0,2)$ gauge theories on the worldvolume of the D1-branes and significantly simplify the connection between the probed geometry and the corresponding field theory \cite{Franco:2015tya}. Brane brick models consist of D4-branes wrapping a 3-torus $T^3$, suspended from an NS5-brane that wraps a holomorphic surface $\Sigma$.
Here, the holomorphic surface $\Sigma$ is the zero locus of the Newton polynomial of the toric diagram of the corresponding toric CY 4-fold. 
\tref{tbbconfig} summarizes the main ingredients in these configurations. 

\begin{table}[ht!!]
\centering
\begin{tabular}{c|cccccccccc}
\; &  0 & 1 & 2 & 3 & 4 & 5 & 6 & 7 & 8 & 9\\
\hline
\text{D4} & 
$\times$  & $\times$ & $\times$ & $\cdot$ & $\times$ & $\cdot$ & $\times$ & $\cdot$ & $\cdot$ & $\cdot$ \\
\text{NS5} & 
$\times$  & $\times$ & \multicolumn{6}{c}{--------- \ $\Sigma$ \ ---------} & $\cdot$ & $\cdot$\\
\end{tabular}
\caption{
The Type IIA brane configuration for brane brick models. 
}
\label{tbbconfig}
\end{table}

The gauge symmetry and matter content of the type of $2d$ $(0,2)$ gauge theories considered in this paper can be encoded in terms of generalized quiver diagrams. $U(N_i)$ gauge groups are represented by nodes, whereas chiral $X_{ij}$ and Fermi $\Lambda_{ij}$ fields are represented by oriented black and unoriented red lines, respectively, as illustrated in \fref{f_quiver}. 

\begin{figure}[ht!!]
\begin{center}
\resizebox{0.6\hsize}{!}{
\includegraphics[height=6cm]{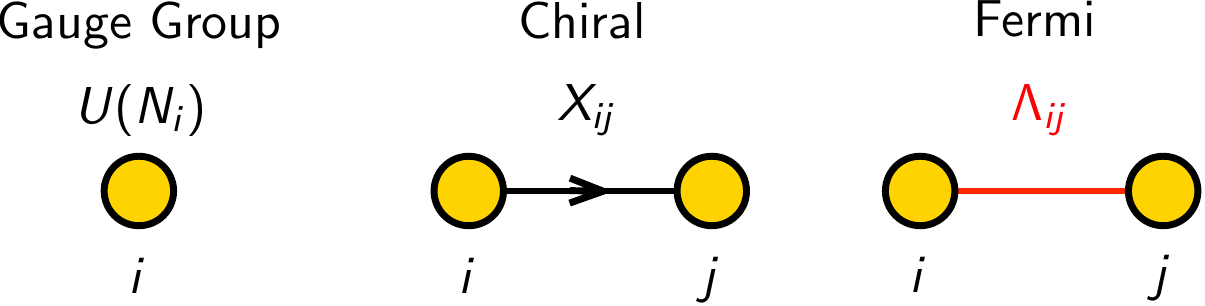} 
}
\caption{
Basic building blocks of $2d$ $(0,2)$ quiver diagrams. 
\label{f_quiver}}
 \end{center}
 \end{figure}

Every Fermi field $\Lambda_{ij}$ is associated to a pair of holomorphic functions of chiral fields $J[\Lambda_{ij}]$ and $E[\Lambda_{ij}]$ \cite{Garcia-Compean:1998sla,Gadde:2013lxa,Franco:2015tna}. In order to form gauge invariant terms in the Lagrangian, $E[\Lambda_{ij}]$ has the same gauge quantum numbers as $\Lambda_{ij}$ while $J[\Lambda_{ij}]$ has the conjugate gauge quantum numbers. The structure of $J$- and $E$-terms in theories associated to brane brick models are further constrained to have the following binomial form
\beal{es06a02}
J[\Lambda_{ij}] = J_{+}[\Lambda_{ij}] - J_{-}[\Lambda_{ij}] ~,~
E[\Lambda_{ij}] = E_{+}[\Lambda_{ij}] - E_{-}[\Lambda_{ij}] ~,~
\eea
with $J_{\pm}[\Lambda_{ij}]$ and $E_{\pm}[\Lambda_{ij}]$ holomorphic monomials in chiral fields. This constraint is commonly known as the \textit{toric condition} \cite{Franco:2015tna} and ensures that the associated CY 4-folds are toric. 

The $2d$ $(0,2)$ gauge theories for toric CY 4-folds can be equivalently encoded by periodic quivers living on $T^3$, which are connected to brane brick models by graph dualization. In this periodic quiver, $J$- and $E$-terms are mapped to minimal gauge invariant closed loops, which we denote as plaquettes, which consist of an oriented path of chiral fields and a single Fermi field, the corresponding $\Lambda_{ij}$. The Fermi field in a plaquette appears without conjugation for $J$-terms and conjugated for $E$-terms. Therefore, as a result of the toric condition, a Fermi field $\Lambda_{ij}$ is associated to four plaquettes in the periodic quiver, where pairs of plaquettes are identified with the binomials in \eref{es06a02}. This structure is illustrated in \fref{f_plaquettes}.

\begin{figure}[htt!]
\begin{center}
\resizebox{0.9\hsize}{!}{
\includegraphics[height=6cm]{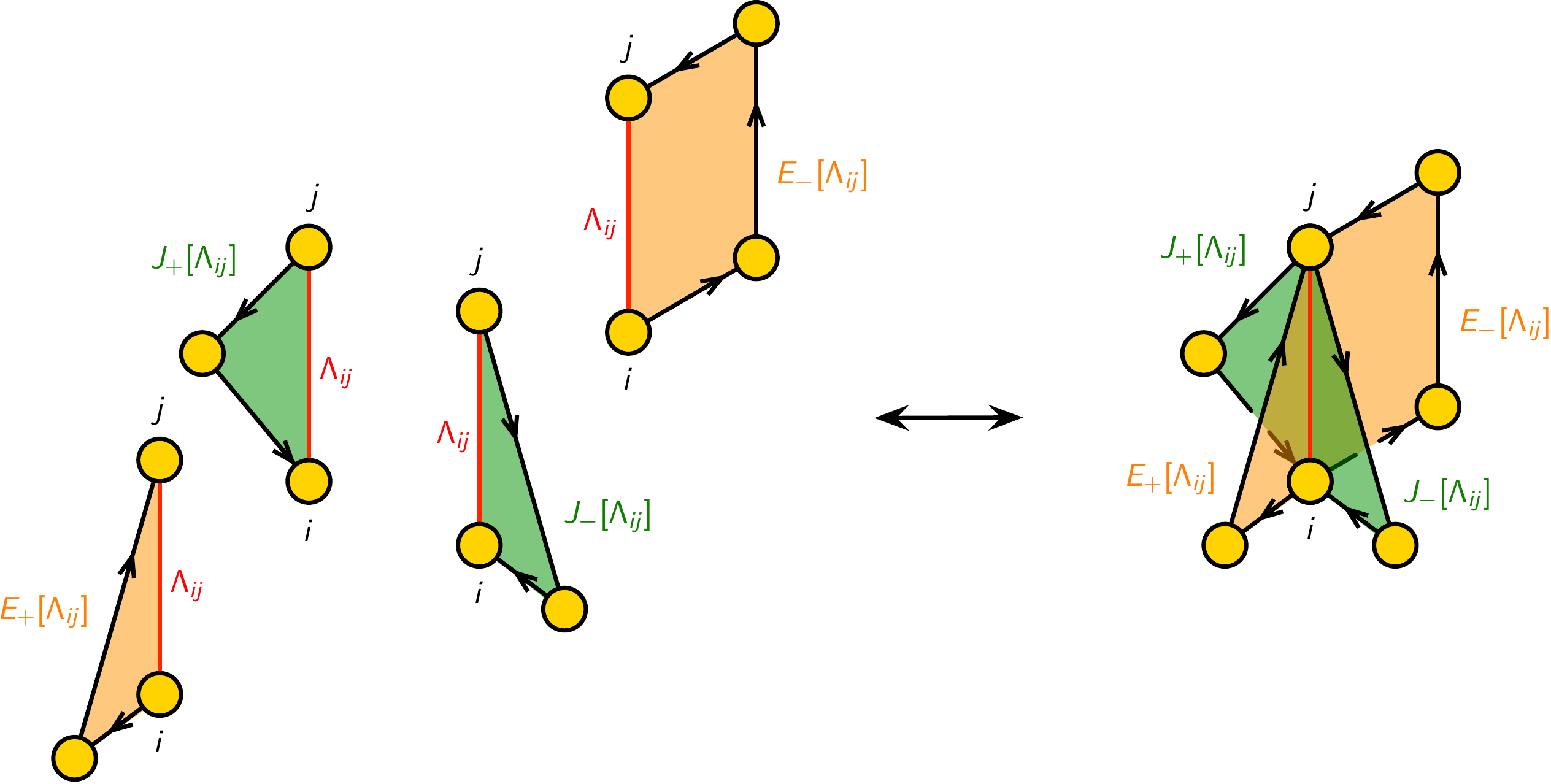} 
}
\caption{
The four plaquettes associated to $J_{\pm}[\Lambda_{ij}]$ and $E_{\pm}[\Lambda_{ij}]$ for a Fermi field $\Lambda_{ij}$ in the periodic quiver dual to a brane brick model.
\label{f_plaquettes}}
 \end{center}
 \end{figure}

The $J$- and $E$-terms of any $2d$ $(0,2)$ gauge theory must satisfy the {\it trace condition} \cite{Franco:2015tna}
\beal{es06a05}
\sum_{a} \text{Tr}(E[\Lambda_a] J[\Lambda_a]) = 0 ~,~
\eea
where the sum is over the Fermi fields $\Lambda_a$. 
$2d$ $(0,2)$ theories are invariant under the individual exchanges $\Lambda_a \leftrightarrow \overline{\Lambda}_a$, which simultaneously exchanges the roles of $J_a \leftrightarrow E_a$. A discussion on how brane brick models implement the trace condition can be found in \cite{Franco:2021elb}.

We focus on the case when the ranks of all gauge groups are equal, such that $N_i=N$. In this case, the non-abelian $SU(N_i)^2$ anomaly cancellation condition at each node $i$ \cite{Franco:2015tna} reduces to
\beal{es06a01}
n_i^{\chi} - n_i^F = 2 ~,~
\eea
where $n_i^\chi$ is the total number of incoming plus outgoing chiral bifundamental fields at node $i$, and $n_i^F$ is the total number of bifundamental Fermi fields connected to node $i$. Since adjoint fields correspond to lines in the quiver that start and end on the same node, each of them contributes 2 to these numbers. 
\\

\section{$Y^{p,k}(\mathbb{CP}^1 \times \mathbb{CP}^1)$ models \label{sec:cp1cp1}}

\subsection{Construction of the $2d$ $(0,2)$ theories \label{sec:cp1cp1intro}}

The $2d$ $(0,2)$ gauge theories associated to the $Y^{p,k}(\mathbb{CP}^{1}\times\mathbb{CP}^{1})$ family are parameterized in  terms of $p,k \in \mathbb{Z}$, where we can set the range for $k$ to be $0\leq k \leq p$ following the discussion in section \sref{sec:se7}.

The total number of gauge groups is $4p$.
For the gauge group labels on the bifundamental chiral and Fermi fields we use the following notation,
\beal{es08a4}
[m] \equiv  ( (m-1) ~\text{mod}~ 4p ) + 1 ~.~
\eea
The chiral and Fermi fields form a periodic layered structure with a total number of $p$ layers. 
Each layer contains 4 gauge groups, which is the same number of gauge groups as in the $4d$ $\mathcal{N}=1$ theories corresponding to $F_0$  \cite{Feng:2001xr}, and the periodicity is manifested by the modularity in the notation in \eref{es08a4}.
This periodic layered structure in the quiver originates from the fact that the $2d$ $(0,2)$ theories were obtained using orbifold reduction \cite{Franco:2016fxm}, $3d$ printing \cite{Franco:2018qsc} and partial resolution via higgsing \cite{Franco:2015tna}, starting with the $4d$ $\mathcal{N}=1$ gauge theory corresponding to the first phase of $F_0$ \cite{Feng:2001xr,Feng:2002zw}.

The gauge group indices are shifted by additional indices between each layer of the quiver,
\beal{es08a3}
a = 0,\dots,p-1 ~,~
b = k, \dots, p-1 ~,~
c = 0,\dots,k-1 ~.~
\eea
Finally, we also have indices for the two non-abelian factors of the global symmetry $SU(2)^2 \times U(1)^2$,
\beal{es08a5}
i,j,k,l = 1,2 ~.~
\eea

Under these labels, the Fermi fields take the form
\beal{es08a20}
&
\Lambda_{[4a+1][4a+2]}^{i}~,~
\Lambda_{[4b+6][4b+3]}^{2i}~,~
\Lambda_{[4b+7][4b+4]}^{2i}~,~
\Lambda_{[4b+8][4b+1]}^{2i}~,~
&
\nn\\
&
\Lambda_{[4c+4][4c+7]}^{1i}~,~
\Lambda_{[4c+8][4c+3]}^{2i}~,~
&
\eea
and the chiral fields take the form
\beal{es08a21}
&
X_{[4a+2][4a+3]}^{i}~,~
X_{[4a+3][4a+4]}^{i}~,~
X_{[4a+4][4a+1]}^{i}~,~
X_{[4b+1][4b+6]}^{i}~,~
&
\nn\\
&
Q_{[4c+5][4c+1]}~,~
Q_{[4c+6][4c+2]}~,~
Q_{[4c+7][4c+3]}~,~
Q_{[4c+8][4c+4]}~,~
&
\nn\\
&
Q_{[4c+5][4c+3]}~,~
Q_{[4c+8][4c+2]}~,~
P_{[4c+1][4c+7]}~,~
P_{[4c+4][4c+6]}~.~
&
\eea

The $J$-terms take the following general form
\beal{es08a0}
J[\Lambda_{[4a+1][4a+2]}^{i}] 
&=&
\epsilon_{ij} \epsilon_{kl} 
X_{[4a+2][4a+3]}^{k} X_{[4a+3][4a+4]}^{j} X_{[4a+4][4a+1]}^{l}
~,~
\nn\\
J[\Lambda_{[4b+6][4b+3]}^{2i}] 
&=&
\epsilon_{ij} \epsilon_{kl} 
X_{[4b+3][4b+4]}^{k} X_{[4b+4][4b+1]}^{j} X_{[4b+1][4b+6]}^{l}
~,~
\nn\\
J[\Lambda_{[4b+7][4b+4]}^{2i}] 
&=& 
\epsilon_{ij} \epsilon_{kl} 
X_{[4b+4][4b+1]}^{k} X_{[4b+1][4b+6]}^{j} X_{[4b+6][4b+7]}^{l}
~,~
\nn\\
J[\Lambda_{[4b+8][4b+1]}^{2i}] 
&=& 
\epsilon_{ij} \epsilon_{kl} 
X_{[4b+1][4b+6]}^{k} X_{[4b+6][4b+7]}^{j} X_{[4b+7][4b+8]}^{l}
~,~
\nn\\
J[\Lambda_{[4c+4][4c+7]}^{1i}] 
&=& 
\epsilon_{ij} \epsilon_{kl} 
X_{[4c+7][4c+8]}^{k} X_{[4c+8][4c+5]}^{j} Q_{[4c+5][4c+3]} X_{[4c+3][4c+4]}^{l}
~,~
\nn\\
J[\Lambda_{[4c+8][4c+3]}^{2i}] 
&=& 
\epsilon_{ij} \epsilon_{kl} 
X_{[4c+3][4c+4]}^{k} X_{[4c+4][4c+1]}^{j} P_{[4c+1][4c+7]} X_{[4c+7][4c+8]}^{l}
~,~
\eea
with the corresponding $E$-terms taking the form
\beal{es08a1}
E[\Lambda_{[4b+6][4b+3]}^{2i}] 
&=&
X_{[4b+6][4b+7]}^{i} Q_{[4b+7][4b+3]}
-
Q_{[4b+6][4b+2]} X_{[4b+2][4b+3]}^{i}
~,~
\nn\\
E[\Lambda_{[4b+7][4b+4]}^{2i}] 
&=&
X_{[4b+7][4b+8]}^{i} Q_{[4b+8][4b+4]}
-
Q_{[4b+7][4b+3]} X_{[4b+3][4b+4]}^{i}
~,~
\nn\\
E[\Lambda_{[4b+8][4b+1]}^{2i}] 
&=&
X_{[4b+8][4b+5]}^{i} Q_{[4b+5][4b+1]}
-
Q_{[4b+8][4b+4]} X_{[4b+4][4b+1]}^{i}
~,~
\nn\\
E[\Lambda_{[4c+4][4c+7]}^{1i}] 
&=&
X_{[4c+4][4c+1]}^{i} P_{[4c+1][4c+7]}
-
P_{[4c+4][4c+6]} X_{[4c+6][4c+7]}^{i}
~,~
\nn\\
E[\Lambda_{[4c+8][4c+3]}^{2i}] 
&=&
X_{[4c+8][4c+5]}^{i} Q_{[4c+5][4c+3]}
-
Q_{[4c+8][4c+2]} X_{[4c+2][4c+3]}^{i}
~.~
\eea
Additional $E$-terms plaquettes remain as follows
\beal{es08a2}
E_{+}[\Lambda_{[4b+1][4b+2]}^{i}] 
&=& 
X_{[4b+1][4b+6]}^{i} Q_{[b+5][b+1]}
~,~
\nn\\
E_{-}[\Lambda_{[4b+5][4b+6]}^{i}] 
&=&
- Q_{[4b+5][4b+1]} X_{[4b+1][4b+6]}^{i}
~,~
\nn\\
E_{+}[\Lambda_{[4c+1][4c+2]}^{i}] 
&=& 
P_{[4c+1][4c+7]} X_{[4c+7][4c+8]}^{i} Q_{[4c+8][4c+2]}
~,~
\nn\\
E_{-}[\Lambda_{[4c+5][4c+6]}^{i}] 
&=& 
- Q_{[4c+5][4c+3]} X_{[4c+3][4c+4]}^{i} P_{[4c+4][4c+6]}
~,~
\eea
which combine to proper $E$-terms of the form $E=E_+ - E_-$.
The $E$-terms plaquettes are shown here separately because quadratic and cubic plaquettes are combined depending on the value of $k$ in order to form a single $E$-term.
\\

\subsection{Examples \label{sec:cp20}}

In order to illustrate the construction of the $2d$ $(0,2)$ theories corresponding to $Y^{p,k}(\mathbb{CP}^1 \times \mathbb{CP}^1)$, in this section we go through all the cases where $p=2$. 
We note that geometries with $p=2$ and $k>0$ correspond to smooth Fano $3$-folds that have been studied extensively in \cite{Franco:2022gvl}.
For completeness, appendix \sref{sec:sap1} summarizes also the $J$- and $E$-terms for the $2d$ $(0,2)$ theories when $p=1$.
\\

\subsubsection{$Y^{2,0}(\mathbb{CP}^1 \times \mathbb{CP}^1)$ \label{sec:cp20}}

\begin{figure}[ht!!]
\begin{center}
\resizebox{0.35\hsize}{!}{
\includegraphics[height=6cm]{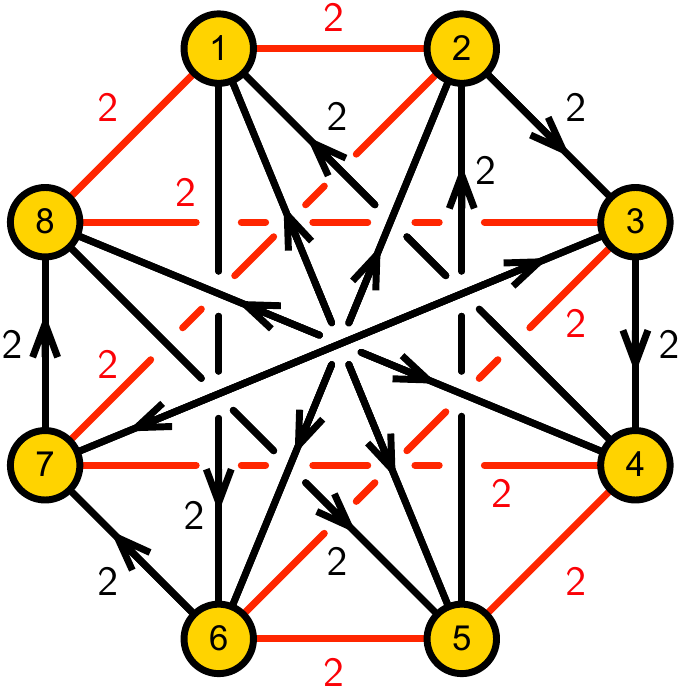} 
}
\caption{
Quiver for $Y^{2,0}(\mathbb{CP}^1 \times \mathbb{CP}^1)$.
\label{f_quiver_cp1_cp1_20}}
 \end{center}
 \end{figure}

We begin with the $2d$ $(0,2)$ theory for $Y^{2,0}(\mathbb{CP}^1 \times \mathbb{CP}^1)$, whose quiver diagram is shown in \fref{f_quiver_cp1_cp1_20}.
\noindent
The $J$- and $E$-terms for this theory take the following form
  \beq
\begin{array}{rccrcl}
 &   J &   & & E &  \\
\Lambda_{12}^{i}  : & \epsilon_{ij} \epsilon_{kl} X^{k}_{23}  X^{j}_{34}  X^{l}_{41}   &~&  X^{i}_{16}  Q_{62}  &-&   Q_{15}  X^{i}_{52} \\ 
\Lambda_{27}^{2i}  : &  \epsilon_{ij} \epsilon_{kl} X^{k}_{78}  X^{j}_{85}  X^{l}_{52}  &~&  X^{i}_{23}  Q_{37}  &-&   Q_{26}  X^{i}_{67} \\ 
\Lambda_{38}^{2i}  : &  \epsilon_{ij} \epsilon_{kl} X^{k}_{85}  X^{j}_{52}  X^{l}_{23}  &~&  X^{i}_{34}  Q_{48}  &-&   Q_{37}  X^{i}_{78} \\ 
\Lambda_{45}^{2i}  : &   \epsilon_{ij} \epsilon_{kl} X^{k}_{52}  X^{j}_{23}  X^{l}_{34}  &~&  X^{i}_{41}  Q_{15}  &-&   Q_{48}  X^{i}_{85} \\ 
\Lambda_{56}^{i}  : &  \epsilon_{ij} \epsilon_{kl} X^{k}_{67}  X^{j}_{78}  X^{l}_{85}   &~&  X^{i}_{52}  Q_{26}  &-&   Q_{51}  X^{i}_{16} \\ 
\Lambda_{63}^{2i}  : &  \epsilon_{ij} \epsilon_{kl} X^{k}_{34}  X^{j}_{41}  X^{l}_{16}  &~&  X^{i}_{67}  Q_{73}  &-&   Q_{62}  X^{i}_{23} \\ 
\Lambda_{74}^{2i}  : &  \epsilon_{ij} \epsilon_{kl} X^{k}_{41}  X^{j}_{16}  X^{l}_{67}  &~&  X^{i}_{78}  Q_{84}  &-&   Q_{73}  X^{i}_{34} \\ 
\Lambda_{81}^{2i}  : &  \epsilon_{ij} \epsilon_{kl} X^{k}_{16}  X^{j}_{67}  X^{l}_{78}  &~&  X^{i}_{85}  Q_{51}  &-&   Q_{84}  X^{i}_{41} \\ 
 \end{array} 
~,~
\label{es0100}
 \eeq
where the global symmetry indices are $i,j,k,l=1,2$.

The $P$-matrix takes the form
 \beal{es0101}
{\footnotesize
 P=
\resizebox{0.725\textwidth}{!}{$
\left(
\begin{array}{r|ccccc|cccc|cccccccc|cccccccc}
\; & p_1 & p_2 & p_3 & p_4 & p_5 & q_1 & q_2 & q_3 & q_4 & s_1 & s_2 & s_3 & s_4 & s_5 & s_6 & s_7 & s_8 & o_1 & o_2 & o_3 & o_4 & o_5 & o_6 & o_7 & o_8\\
\hline
X_{16}^{1} & 1 & 0 & 0 & 0 & 0 & 1 & 0 & 0 & 0 & 1 & 0 & 0 & 0 & 0 & 0 & 0 & 0 & 0 & 0 & 0 & 0 & 0 & 1 & 1 & 1 \\
X_{23}^{1} & 0 & 1 & 0 & 0 & 0 & 0 & 1 & 0 & 0 & 0 & 1 & 0 & 0 & 0 & 0 & 0 & 0 & 0 & 0 & 0 & 0 & 1 & 1 & 1 & 0 \\
X_{34}^{1} & 1 & 0 & 0 & 0 & 0 & 0 & 0 & 1 & 0 & 0 & 0 & 1 & 0 & 0 & 0 & 0 & 0 & 0 & 1 & 1 & 0 & 0 & 0 & 0 & 1 \\
X_{41}^{1} & 0 & 1 & 0 & 0 & 0 & 0 & 0 & 0 & 1 & 0 & 0 & 0 & 1 & 0 & 0 & 0 & 0 & 1 & 0 & 0 & 1 & 1 & 0 & 0 & 0 \\
X_{52}^{1} & 1 & 0 & 0 & 0 & 0 & 1 & 0 & 0 & 0 & 0 & 0 & 0 & 0 & 1 & 0 & 0 & 0 & 1 & 0 & 1 & 1 & 0 & 0 & 0 & 0 \\
X_{67}^{1} & 0 & 1 & 0 & 0 & 0 & 0 & 1 & 0 & 0 & 0 & 0 & 0 & 0 & 0 & 1 & 0 & 0 & 0 & 1 & 1 & 1 & 0 & 0 & 0 & 0 \\
X_{78}^{1} & 1 & 0 & 0 & 0 & 0 & 0 & 0 & 1 & 0 & 0 & 0 & 0 & 0 & 0 & 0 & 1 & 0 & 1 & 0 & 0 & 0 & 1 & 0 & 1 & 0 \\
X_{85}^{1} & 0 & 1 & 0 & 0 & 0 & 0 & 0 & 0 & 1 & 0 & 0 & 0 & 0 & 0 & 0 & 0 & 1 & 0 & 1 & 0 & 0 & 0 & 1 & 0 & 1 \\
\hline
X_{16}^{2} & 0 & 0 & 1 & 0 & 0 & 1 & 0 & 0 & 0 & 1 & 0 & 0 & 0 & 0 & 0 & 0 & 0 & 0 & 0 & 0 & 0 & 0 & 1 & 1 & 1 \\
X_{23}^{2} & 0 & 0 & 0 & 1 & 0 & 0 & 1 & 0 & 0 & 0 & 1 & 0 & 0 & 0 & 0 & 0 & 0 & 0 & 0 & 0 & 0 & 1 & 1 & 1 & 0 \\
X_{34}^{2} & 0 & 0 & 1 & 0 & 0 & 0 & 0 & 1 & 0 & 0 & 0 & 1 & 0 & 0 & 0 & 0 & 0 & 0 & 1 & 1 & 0 & 0 & 0 & 0 & 1 \\
X_{41}^{2} & 0 & 0 & 0 & 1 & 0 & 0 & 0 & 0 & 1 & 0 & 0 & 0 & 1 & 0 & 0 & 0 & 0 & 1 & 0 & 0 & 1 & 1 & 0 & 0 & 0 \\
X_{52}^{2} & 0 & 0 & 1 & 0 & 0 & 1 & 0 & 0 & 0 & 0 & 0 & 0 & 0 & 1 & 0 & 0 & 0 & 1 & 0 & 1 & 1 & 0 & 0 & 0 & 0 \\
X_{67}^{2} & 0 & 0 & 0 & 1 & 0 & 0 & 1 & 0 & 0 & 0 & 0 & 0 & 0 & 0 & 1 & 0 & 0 & 0 & 1 & 1 & 1 & 0 & 0 & 0 & 0 \\
X_{78}^{2} & 0 & 0 & 1 & 0 & 0 & 0 & 0 & 1 & 0 & 0 & 0 & 0 & 0 & 0 & 0 & 1 & 0 & 1 & 0 & 0 & 0 & 1 & 0 & 1 & 0 \\
X_{85}^{2} & 0 & 0 & 0 & 1 & 0 & 0 & 0 & 0 & 1 & 0 & 0 & 0 & 0 & 0 & 0 & 0 & 1 & 0 & 1 & 0 & 0 & 0 & 1 & 0 & 1 \\
\hline
Q_{15} & 0 & 0 & 0 & 0 & 1 & 0 & 0 & 0 & 0 & 1 & 0 & 0 & 0 & 0 & 1 & 1 & 1 & 0 & 1 & 0 & 0 & 0 & 1 & 1 & 1 \\
Q_{26} & 0 & 0 & 0 & 0 & 1 & 0 & 0 & 0 & 0 & 1 & 1 & 1 & 1 & 0 & 0 & 0 & 0 & 0 & 0 & 0 & 0 & 1 & 1 & 1 & 1 \\
Q_{37} & 0 & 0 & 0 & 0 & 1 & 0 & 0 & 0 & 0 & 1 & 0 & 1 & 1 & 0 & 1 & 0 & 0 & 0 & 1 & 1 & 1 & 0 & 0 & 0 & 1 \\
Q_{48} & 0 & 0 & 0 & 0 & 1 & 0 & 0 & 0 & 0 & 1 & 0 & 0 & 1 & 0 & 1 & 1 & 0 & 1 & 0 & 0 & 1 & 1 & 0 & 1 & 0 \\
Q_{51} & 0 & 0 & 0 & 0 & 1 & 0 & 0 & 0 & 0 & 0 & 1 & 1 & 1 & 1 & 0 & 0 & 0 & 1 & 0 & 1 & 1 & 1 & 0 & 0 & 0 \\
Q_{62} & 0 & 0 & 0 & 0 & 1 & 0 & 0 & 0 & 0 & 0 & 0 & 0 & 0 & 1 & 1 & 1 & 1 & 1 & 1 & 1 & 1 & 0 & 0 & 0 & 0 \\
Q_{73} & 0 & 0 & 0 & 0 & 1 & 0 & 0 & 0 & 0 & 0 & 1 & 0 & 0 & 1 & 0 & 1 & 1 & 1 & 0 & 0 & 0 & 1 & 1 & 1 & 0 \\
Q_{84} & 0 & 0 & 0 & 0 & 1 & 0 & 0 & 0 & 0 & 0 & 1 & 1 & 0 & 1 & 0 & 0 & 1 & 0 & 1 & 1 & 0 & 0 & 1 & 0 & 1 \\
\end{array}
\right)
$}
}~.~
\eea

\noindent
The associated $J$- and $E$-term $U(1)$-charge matrix takes the following form
 \beal{es0102}
{\footnotesize
 Q_{JE}=
 \resizebox{0.725\textwidth}{!}{$
\left(
\begin{array}{ccccc|cccc|cccccccc|cccccccc}
p_1 & p_2 & p_3 & p_4 & p_5 & q_1 & q_2 & q_3 & q_4 & s_1 & s_2 & s_3 & s_4 & s_5 & s_6 & s_7 & s_8 & o_1 & o_2 & o_3 & o_4 & o_5 & o_6 & o_7 & o_8\\
\hline
 1 & 0 & 1 & 0 & 0 & -1 & 0 & -1 & 0 & 0 & 0 & 0 & 0 & 0 & 0 & 0 & 0 & 0 & 0 & 0 & 0 & 0 & 0 & 0 & 0 \\
 0 & 1 & 0 & 1 & 0 & 0 & -1 & 0 & -1 & 0 & 0 & 0 & 0 & 0 & 0 & 0 & 0 & 0 & 0 & 0 & 0 & 0 & 0 & 0 & 0 \\
 0 & 0 & 0 & 0 & 1 & 0 & 1 & 1 & 1 & 0 & -1 & 0 & 0 & 2 & 0 & 0 & -1 & -1 & 0 & -1 & 0 & 0 & 0 & 0 & 0 \\
 0 & 0 & 0 & 0 & 0 & -1 & 1 & 1 & 1 & 1 & -1 & 0 & 0 & 3 & 0 & 0 & -1 & -1 & 0 & -1 & 0 & 0 & 0 & 0 & 0 \\
 0 & 0 & 0 & 0 & 0 & -1 & 1 & 0 & 0 & 0 & -1 & 0 & 0 & 2 & 0 & 0 & -1 & 0 & 0 & -1 & 0 & 0 & 0 & 0 & 1 \\
 0 & 0 & 0 & 0 & 0 & -1 & 0 & 0 & 1 & 0 & -1 & 0 & 0 & 2 & 0 & 0 & -1 & -1 & 0 & 0 & 0 & 0 & 0 & 1 & 0 \\
 0 & 0 & 0 & 0 & 0 & -1 & 0 & 0 & 0 & 0 & -1 & 0 & 0 & 1 & 0 & 0 & -1 & 0 & 0 & 0 & 0 & 0 & 1 & 0 & 0 \\
 0 & 0 & 0 & 0 & 0 & 0 & 0 & 1 & 1 & 0 & 0 & 0 & 0 & 2 & 1 & 0 & -1 & -1 & 0 & -1 & 0 & 0 & 0 & 0 & 0 \\
 0 & 0 & 0 & 0 & 0 & 0 & 0 & 0 & 0 & 0 & 0 & 0 & 0 & 1 & 0 & 0 & -1 & 0 & 1 & -1 & 0 & 0 & 0 & 0 & 0 \\
 0 & 0 & 0 & 0 & 0 & 0 & 0 & 0 & 1 & 0 & 0 & 0 & 0 & 1 & 0 & 1 & -1 & -1 & 0 & 0 & 0 & 0 & 0 & 0 & 0 \\
 0 & 0 & 0 & 0 & 0 & 0 & 1 & 1 & 0 & 0 & -1 & 0 & 1 & 2 & 0 & 0 & 0 & -1 & 0 & -1 & 0 & 0 & 0 & 0 & 0 \\
 0 & 0 & 0 & 0 & 0 & 0 & 1 & 0 & 0 & 0 & -1 & 1 & 0 & 1 & 0 & 0 & 0 & 0 & 0 & -1 & 0 & 0 & 0 & 0 & 0 \\
 0 & 0 & 0 & 0 & 0 & 0 & 0 & 0 & 0 & 0 & -1 & 0 & 0 & 1 & 0 & 0 & 0 & -1 & 0 & 0 & 0 & 1 & 0 & 0 & 0 \\
 0 & 0 & 0 & 0 & 0 & 0 & 0 & 1 & 0 & 0 & 0 & 0 & 0 & 1 & 0 & 0 & 0 & -1 & 0 & -1 & 1 & 0 & 0 & 0 & 0 \\
\end{array}
\right)
$}
}~,~
\eea
and the corresponding $D$-term $U(1)$-charge matrix takes the form
 \beal{es0103}
{\footnotesize
 Q_{D}=
 \resizebox{0.725\textwidth}{!}{$
\left(
\begin{array}{ccccc|cccc|cccccccc|cccccccc}
p_1 & p_2 & p_3 & p_4 & p_5 & q_1 & q_2 & q_3 & q_4 & s_1 & s_2 & s_3 & s_4 & s_5 & s_6 & s_7 & s_8 & o_1 & o_2 & o_3 & o_4 & o_5 & o_6 & o_7 & o_8\\
\hline
 0 & 0 & 0 & 0 & 0 & 0 & 0 & 0 & 0 & 0 & 1 & 0 & 0 & -1 & 0 & 0 & 0 & 0 & 0 & 0 & 0 & 0 & 0 & 0 & 0 \\
 0 & 0 & 0 & 0 & 0 & 0 & -1 & 0 & 0 & 0 & 0 & 0 & 0 & -1 & 0 & 0 & 0 & 0 & 0 & 1 & 0 & 0 & 0 & 0 & 0 \\
 0 & 0 & 0 & 0 & 0 & 0 & 0 & -1 & 0 & 0 & 0 & 0 & 0 & -1 & 0 & 0 & 0 & 1 & 0 & 0 & 0 & 0 & 0 & 0 & 0 \\
 0 & 0 & 0 & 0 & 0 & 0 & 0 & 0 & 0 & 0 & 0 & 0 & 0 & 1 & 0 & 0 & -1 & 0 & 0 & 0 & 0 & 0 & 0 & 0 & 0 \\
 0 & 0 & 0 & 0 & 0 & -1 & 1 & 0 & 0 & 0 & -1 & 0 & 0 & 1 & 0 & 0 & 0 & 0 & 0 & 0 & 0 & 0 & 0 & 0 & 0 \\
 0 & 0 & 0 & 0 & 0 & 0 & 0 & 1 & 0 & 0 & 0 & 0 & 0 & 1 & 0 & 0 & 0 & 0 & 0 & -1 & 0 & 0 & 0 & 0 & 0 \\
 0 & 0 & 0 & 0 & 0 & 0 & 0 & 0 & 1 & 0 & 0 & 0 & 0 & 1 & 0 & 0 & 0 & -1 & 0 & 0 & 0 & 0 & 0 & 0 & 0 \\
\end{array}
\right)
$}
}~.~
\eea
We observe that the charge matrices are invariant under
\beal{es104}
p_1 \leftrightarrow p_3 ~,~
p_2 \leftrightarrow p_4 ~.~
\eea
Each of the above $Z_2$-symmetries correspond to an $SU(2)$ factor in the global symmetry of the $2d$ $(0,2)$ gauge theory.

\begin{figure}[H]
\begin{center}
\resizebox{0.35\hsize}{!}{
\includegraphics[height=6cm]{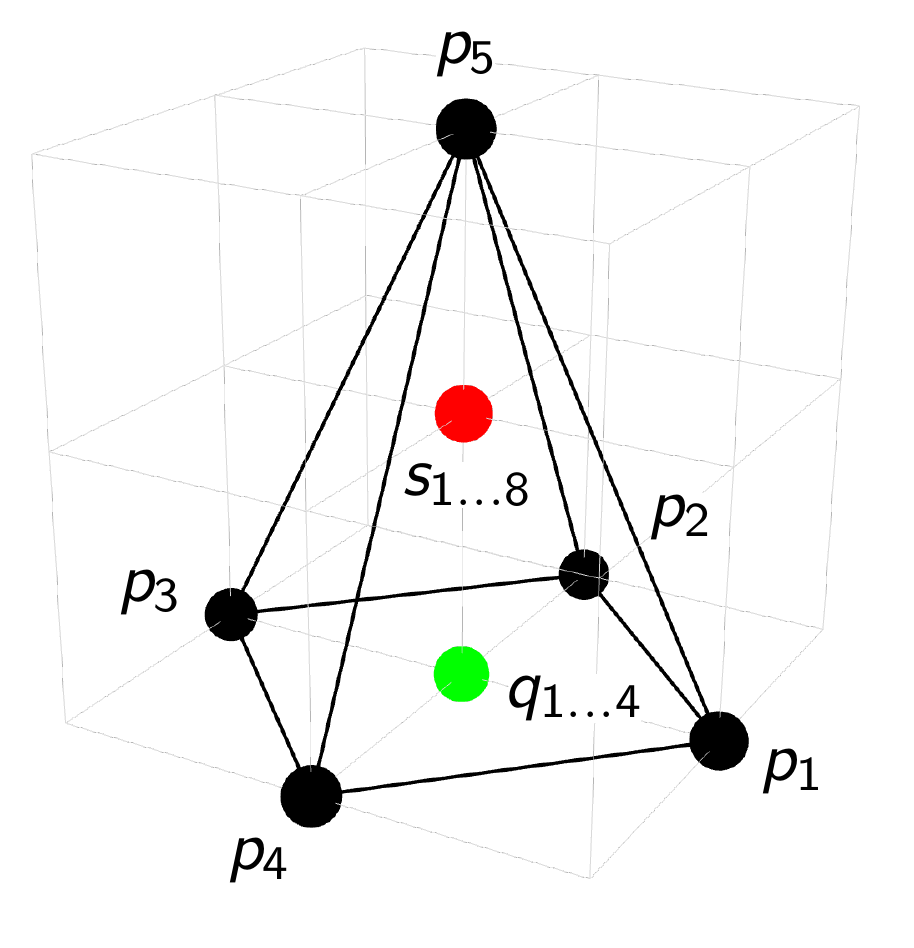} 
}
\caption{
Toric diagram of $Y^{2,0}(\mathbb{CP}^{1}\times\mathbb{CP}^{1})$.
\label{f_toric_cp1_cp1_20}}
 \end{center}
 \end{figure}

The corresponding toric data is given by
 \beal{es0105}
{\footnotesize
 G_{t}=
 \resizebox{0.725\textwidth}{!}{$
\left(
\begin{array}{ccccc|cccc|cccccccc|cccccccc}
p_1 & p_2 & p_3 & p_4 & p_5 & q_1 & q_2 & q_3 & q_4 & s_1 & s_2 & s_3 & s_4 & s_5 & s_6 & s_7 & s_8 & o_1 & o_2 & o_3 & o_4 & o_5 & o_6 & o_7 & o_8\\
\hline
 1 & 1 & 1 & 1 & 1 & 1 & 1 & 1 & 1 & 1 & 1 & 1 & 1 & 1 & 1 & 1 & 1 & 2 & 2 & 2 & 2 & 2 & 2 & 2 & 2 \\
 1 & 0 & -1 & 0 & 0 & 0 & 0 & 0 & 0 & 0 & 0 & 0 & 0 & 0 & 0 & 0 & 0 & 0 & 0 & 0 & 0 & 0 & 0 & 0 & 0 \\
 0 & 1 & 0 & -1 & 0 & 0 & 0 & 0 & 0 & 0 & 0 & 0 & 0 & 0 & 0 & 0 & 0 & 0 & 0 & 0 & 0 & 0 & 0 & 0 & 0 \\
 0 & 0 & 0 & 0 & 2 & 0 & 0 & 0 & 0 & 1 & 1 & 1 & 1 & 1 & 1 & 1 & 1 & 1 & 1 & 1 & 1 & 1 & 1 & 1 & 1 \\
\end{array}
\right)
$}
}~,~
\eea
where the GLSM fields $o_k$ correspond to extra GLSM fields in the nomenclature of \cite{Franco:2015tna}.
The toric diagram is shown in \fref{f_toric_cp1_cp1_20} and it agrees with the expectation.
We note that the $2d$ $(0,2)$ theory for $Y^{2,0}(\mathbb{CP}^1 \times \mathbb{CP}^1)$ can be constructed using orbifold reduction  \cite{Franco:2016fxm}, starting from the $4d$ $\mathcal{N}=1$ theory for the first phase of $F_0$. 
\\

\subsubsection{$Y^{2,1}(\mathbb{CP}^1 \times \mathbb{CP}^1)$ \label{sec:cp21}}

\begin{figure}[H]
\begin{center}
\resizebox{0.35\hsize}{!}{
\includegraphics[height=6cm]{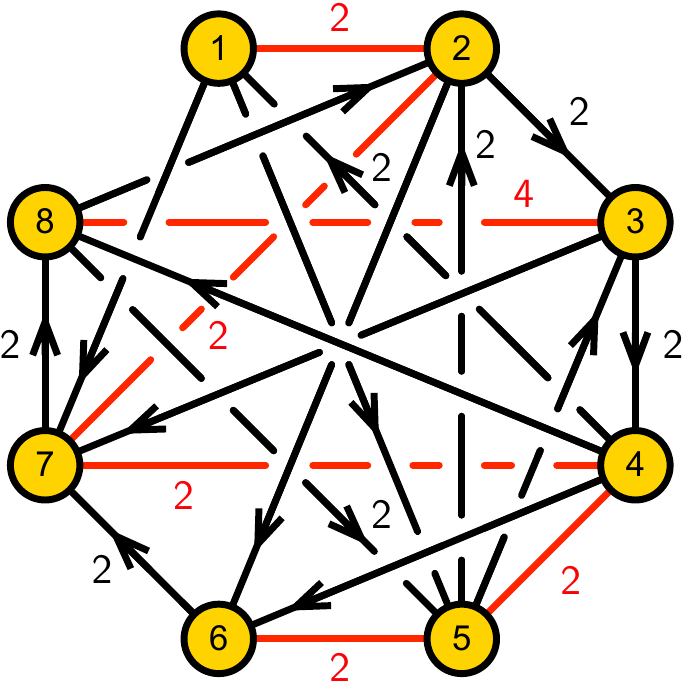} 
}
\caption{
Quiver for $Y^{2,1}(\mathbb{CP}^1 \times \mathbb{CP}^1)$.
\label{f_quiver_cp1_cp1_21}}
 \end{center}
 \end{figure}

The $J$- and $E$-terms of the $2d$ $(0,2)$ theory corresponding to $Y^{2,1}(\mathbb{CP}^1 \times \mathbb{CP}^1)$ are
  \beq
\begin{array}{rccrcl}
 &   J &   & & E &  \\
\Lambda_{12}^{i}  : & \epsilon_{ij} \epsilon_{kl} X^{k}_{23}  X^{j}_{34}  X^{l}_{41} &~&   P_{17}  X^{i}_{78}  Q_{82}   &-& Q_{15}  X^{i}_{52} \\   
\Lambda_{27}^{2i}  : &  \epsilon_{ij} \epsilon_{kl} X^{k}_{78}  X^{j}_{85}  X^{l}_{52}   &~&   X^{i}_{23}  Q_{37}   &-& Q_{26}  X^{i}_{67} \\   
\Lambda_{38}^{2i}  : & \epsilon_{ij} \epsilon_{kl}  X^{k}_{85}  X^{j}_{52}  X^{l}_{23}   &~&   X^{i}_{34}  Q_{48}   &-& Q_{37}  X^{i}_{78} \\   
\Lambda_{45}^{2i}  : & \epsilon_{ij} \epsilon_{kl} X^{k}_{52}  X^{j}_{23}  X^{l}_{34}   &~&   X^{i}_{41}  Q_{15}   &-& Q_{48}  X^{i}_{85} \\   
\Lambda_{47}^{1i}  : & \epsilon_{ij} \epsilon_{kl}  X^{k}_{78}  X^{j}_{85}  Q_{53}  X^{l}_{34}  &~&   X^{i}_{41}  P_{17}   &-& P_{46}  X^{i}_{67} \\   
\Lambda_{56}^{i}  : & \epsilon_{ij} \epsilon_{kl}  X^{k}_{67}  X^{j}_{78}  X^{l}_{85}   &~&   X^{i}_{52}  Q_{26}   &-& Q_{53}  X^{i}_{34}  P_{46} \\   
\Lambda_{83}^{2i}  : & \epsilon_{ij} \epsilon_{kl}  X^{k}_{34}  X^{j}_{41}  P_{17}  X^{l}_{78}   &~&   X^{i}_{85}  Q_{53}   &-& Q_{82}  X^{i}_{23} \\   
 \end{array} 
~,~
\label{es0200}
 \eeq
 where the corresponding quiver diagram is shown in \fref{f_quiver_cp1_cp1_21}.
The global symmetry indices are $i,j,k,l=1,2$.
 
The corresponding $P$-matrix takes the form
 \beal{es0201}
{\footnotesize
 P=
\resizebox{0.725\textwidth}{!}{$
\left(
\begin{array}{r|cccccc|ccccccccc|ccccccc|ccc}
\; & p_1 & p_2 & p_3 & p_4 & p_5 & p_6 & s_1 & s_2 & s_3 & s_4 & s_5 & s_6 & s_7 & s_8 & s_9 
& o_1 & o_2 & o_3 & o_4 & o_5 & o_6 & o_7
& o_8 & o_9 & o_{10}
\\
\hline
X_{23}^{1} & 1 & 0 & 0 & 0 & 0 & 0 & 0 & 0 & 1 & 0 & 0 & 0 & 0 & 0 & 0 & 0 & 0 & 1 & 1 & 0 & 1 & 1 & 1 & 1 & 0 \\
X_{34}^{1} & 0 & 1 & 0 & 0 & 0 & 0 & 0 & 0 & 0 & 1 & 0 & 0 & 0 & 0 & 0 & 0 & 0 & 0 & 0 & 1 & 0 & 0 & 1 & 0 & 1 \\
X_{41}^{1} & 1 & 0 & 0 & 0 & 0 & 0 & 0 & 1 & 0 & 0 & 0 & 0 & 1 & 0 & 0 & 1 & 1 & 1 & 0 & 0 & 0 & 0 & 0 & 1 & 1 \\
X_{52}^{1} & 0 & 1 & 0 & 0 & 0 & 1 & 1 & 0 & 0 & 0 & 0 & 0 & 0 & 0 & 0 & 1 & 1 & 0 & 1 & 1 & 0 & 0 & 0 & 1 & 1 \\
X_{67}^{1} & 1 & 0 & 0 & 0 & 0 & 0 & 0 & 1 & 0 & 0 & 0 & 0 & 0 & 1 & 0 & 0 & 0 & 0 & 1 & 1 & 1 & 0 & 0 & 1 & 1 \\
X_{78}^{1} & 0 & 1 & 0 & 0 & 0 & 0 & 0 & 0 & 0 & 0 & 0 & 1 & 0 & 0 & 0 & 0 & 1 & 0 & 0 & 0 & 0 & 0 & 1 & 1 & 0 \\
X_{85}^{1} & 1 & 0 & 0 & 0 & 0 & 0 & 0 & 0 & 0 & 0 & 1 & 0 & 0 & 0 & 0 & 1 & 0 & 1 & 0 & 0 & 1 & 1 & 1 & 0 & 1 \\
\hline
X_{23}^{2} & 0 & 0 & 1 & 0 & 0 & 0 & 0 & 0 & 1 & 0 & 0 & 0 & 0 & 0 & 0 & 0 & 0 & 1 & 1 & 0 & 1 & 1 & 1 & 1 & 0 \\
X_{34}^{2} & 0 & 0 & 0 & 1 & 0 & 0 & 0 & 0 & 0 & 1 & 0 & 0 & 0 & 0 & 0 & 0 & 0 & 0 & 0 & 1 & 0 & 0 & 1 & 0 & 1 \\
X_{41}^{2} & 0 & 0 & 1 & 0 & 0 & 0 & 0 & 1 & 0 & 0 & 0 & 0 & 1 & 0 & 0 & 1 & 1 & 1 & 0 & 0 & 0 & 0 & 0 & 1 & 1 \\
X_{52}^{2} & 0 & 0 & 0 & 1 & 0 & 1 & 1 & 0 & 0 & 0 & 0 & 0 & 0 & 0 & 0 & 1 & 1 & 0 & 1 & 1 & 0 & 0 & 0 & 1 & 1 \\
X_{67}^{2} & 0 & 0 & 1 & 0 & 0 & 0 & 0 & 1 & 0 & 0 & 0 & 0 & 0 & 1 & 0 & 0 & 0 & 0 & 1 & 1 & 1 & 0 & 0 & 1 & 1 \\
X_{78}^{2} & 0 & 0 & 0 & 1 & 0 & 0 & 0 & 0 & 0 & 0 & 0 & 1 & 0 & 0 & 0 & 0 & 1 & 0 & 0 & 0 & 0 & 0 & 1 & 1 & 0 \\
X_{85}^{2} & 0 & 0 & 1 & 0 & 0 & 0 & 0 & 0 & 0 & 0 & 1 & 0 & 0 & 0 & 0 & 1 & 0 & 1 & 0 & 0 & 1 & 1 & 1 & 0 & 1 \\
\hline
Q_{15} & 0 & 0 & 0 & 0 & 1 & 0 & 0 & 0 & 0 & 0 & 1 & 1 & 0 & 1 & 1 & 0 & 0 & 0 & 0 & 0 & 1 & 1 & 1 & 0 & 0 \\
Q_{26} & 0 & 0 & 0 & 0 & 1 & 0 & 0 & 0 & 1 & 1 & 0 & 0 & 1 & 0 & 1 & 0 & 0 & 1 & 0 & 0 & 0 & 1 & 1 & 0 & 0 \\
Q_{37} & 0 & 0 & 0 & 0 & 1 & 0 & 0 & 1 & 0 & 1 & 0 & 0 & 1 & 1 & 1 & 0 & 0 & 0 & 0 & 1 & 0 & 0 & 0 & 0 & 1 \\
Q_{48} & 0 & 0 & 0 & 0 & 1 & 0 & 0 & 1 & 0 & 0 & 0 & 1 & 1 & 1 & 1 & 0 & 1 & 0 & 0 & 0 & 0 & 0 & 0 & 1 & 0 \\
Q_{53} & 0 & 0 & 0 & 0 & 1 & 0 & 1 & 0 & 1 & 0 & 0 & 0 & 0 & 0 & 0 & 0 & 0 & 0 & 1 & 0 & 0 & 0 & 0 & 1 & 0 \\
Q_{82} & 0 & 0 & 0 & 0 & 1 & 0 & 1 & 0 & 0 & 0 & 1 & 0 & 0 & 0 & 0 & 1 & 0 & 0 & 0 & 0 & 0 & 0 & 0 & 0 & 1 \\
\hline
P_{17} & 0 & 0 & 0 & 0 & 0 & 1 & 0 & 0 & 0 & 0 & 0 & 0 & 0 & 1 & 1 & 0 & 0 & 0 & 1 & 1 & 1 & 1 & 0 & 0 & 0 \\
P_{46} & 0 & 0 & 0 & 0 & 0 & 1 & 0 & 0 & 0 & 0 & 0 & 0 & 1 & 0 & 1 & 1 & 1 & 1 & 0 & 0 & 0 & 1 & 0 & 0 & 0 \\
\end{array}
\right)
$}
}~,~
\eea
and the corresponding $U(1)$-charge matrix under the $J$- and $E$-terms takes the form
 \beal{es0202}
{\footnotesize
 Q_{JE}=
 \resizebox{0.725\textwidth}{!}{$
\left(
\begin{array}{cccccc|ccccccccc|ccccccc|ccc}
 p_1 & p_2 & p_3 & p_4 & p_5 & p_6 & s_1 & s_2 & s_3 & s_4 & s_5 & s_6 & s_7 & s_8 & s_9 
& o_1 & o_2 & o_3 & o_4 & o_5 & o_6 & o_7
& o_8 & o_9 & o_{10}
\\
\hline 
 3 & 1 & 3 & 1 & 0 & 0 & 3 & 0 & -1 & 0 & -1 & 0 & 0 & 0 & 1 & -1 & 0 & 0 & -1 & 0 & 0 & 0 & 0 & -1 & -1 \\
 2 & 1 & 2 & 1 & 1 & 0 & 1 & 0 & -1 & 0 & -1 & 0 & 0 & 0 & 0 & 0 & 0 & 0 & 0 & 0 & 0 & 0 & 0 & -1 & -1 \\
 2 & 1 & 2 & 1 & 0 & 0 & 2 & 0 & 0 & 0 & -1 & 0 & 0 & 1 & 0 & 0 & 0 & 0 & -1 & 0 & 0 & 0 & 0 & -1 & -1 \\
 2 & 1 & 2 & 1 & 0 & 0 & 2 & 0 & -1 & 0 & 0 & 0 & 1 & 0 & 0 & -1 & 0 & 0 & 0 & 0 & 0 & 0 & 0 & -1 & -1 \\
 1 & 1 & 1 & 1 & 0 & 0 & 1 & 1 & 0 & 0 & 0 & 0 & 0 & 0 & 0 & 0 & 0 & 0 & 0 & 0 & 0 & 0 & 0 & -1 & -1 \\
 1 & 0 & 1 & 0 & 0 & 1 & 1 & 0 & 0 & 0 & 0 & 0 & 0 & 0 & 0 & -1 & 0 & 0 & -1 & 0 & 0 & 0 & 0 & 0 & 0 \\
 1 & 0 & 1 & 0 & 0 & 0 & 2 & 0 & -1 & 0 & -1 & 0 & 0 & 0 & 0 & -1 & 0 & 0 & -1 & 0 & 0 & 1 & 0 & 0 & 0 \\
 1 & 0 & 1 & 0 & 0 & 0 & 1 & 0 & 0 & 0 & 0 & 0 & 0 & 0 & 0 & 0 & 0 & 0 & -1 & 1 & 0 & 0 & 0 & 0 & -1 \\
 1 & 0 & 1 & 0 & 0 & 0 & 1 & 0 & 0 & 0 & 0 & 0 & 0 & 0 & 0 & -1 & 1 & 0 & 0 & 0 & 0 & 0 & 0 & -1 & 0 \\
 1 & 0 & 1 & 0 & 0 & 0 & 1 & 0 & 0 & 0 & -1 & 1 & 0 & 0 & 0 & 0 & 0 & 0 & 0 & 0 & 0 & 0 & 0 & -1 & 0 \\
 1 & 0 & 1 & 0 & 0 & 0 & 1 & 0 & -1 & 1 & 0 & 0 & 0 & 0 & 0 & 0 & 0 & 0 & 0 & 0 & 0 & 0 & 0 & 0 & -1 \\
 0 & 0 & 0 & 0 & 0 & 0 & 1 & 0 & 0 & 0 & -1 & 0 & 0 & 0 & 0 & 0 & 0 & 0 & -1 & 0 & 1 & 0 & 0 & 0 & 0 \\
 0 & 0 & 0 & 0 & 0 & 0 & 1 & 0 & -1 & 0 & 0 & 0 & 0 & 0 & 0 & -1 & 0 & 1 & 0 & 0 & 0 & 0 & 0 & 0 & 0 \\
 0 & -1 & 0 & -1 & 0 & 0 & 1 & 0 & -1 & 0 & -1 & 0 & 0 & 0 & 0 & 0 & 0 & 0 & 0 & 0 & 0 & 0 & 1 & 0 & 0 \\
\end{array}
\right)
$}
}~.~
\eea
The charge matrix for the $D$-terms takes the form
 \beal{es0203}
{\footnotesize
 Q_{D}=
 \resizebox{0.725\textwidth}{!}{$
\left(
\begin{array}{cccccc|ccccccccc|ccccccc|ccc}
 p_1 & p_2 & p_3 & p_4 & p_5 & p_6 & s_1 & s_2 & s_3 & s_4 & s_5 & s_6 & s_7 & s_8 & s_9 
& o_1 & o_2 & o_3 & o_4 & o_5 & o_6 & o_7
& o_8 & o_9 & o_{10}
\\
\hline 
 1 & 1 & 1 & 1 & 0 & 0 & 1 & 0 & 0 & 0 & 0 & 0 & 0 & 0 & 0 & 0 & 0 & 0 & -1 & 0 & 0 & 0 & 0 & 0 & -1 \\
 1 & 0 & 1 & 0 & 0 & 0 & 1 & 0 & 0 & 0 & 0 & 0 & 0 & 0 & 0 & 0 & 0 & 0 & 0 & 0 & 0 & 0 & 0 & -1 & 0 \\
 1 & 0 & 1 & 0 & 0 & 0 & 1 & 0 & -1 & 0 & 0 & 0 & 0 & 0 & 0 & -1 & 0 & 0 & 0 & 0 & 0 & 0 & 0 & 0 & 0 \\
 0 & 0 & 0 & 0 & 0 & 0 & 1 & 0 & 0 & 0 & -1 & 0 & 0 & 0 & 0 & 0 & 0 & 0 & 0 & 0 & 0 & 0 & 0 & 0 & 0 \\
 0 & 0 & 0 & 0 & 0 & 0 & -1 & 0 & 1 & 0 & 0 & 0 & 0 & 0 & 0 & 0 & 0 & 0 & 0 & 0 & 0 & 0 & 0 & 0 & 0 \\
 -1 & 0 & -1 & 0 & 0 & 0 & -1 & 0 & 0 & 0 & 0 & 0 & 0 & 0 & 0 & 0 & 0 & 0 & 0 & 0 & 0 & 0 & 0 & 0 & 1 \\
 -1 & -1 & -1 & -1 & 0 & 0 & -1 & 0 & 0 & 0 & 0 & 0 & 0 & 0 & 0 & 1 & 0 & 0 & 0 & 0 & 0 & 0 & 0 & 1 & 0 \\
\end{array}
\right)
$}
}~.~
\eea
We note that the charge matrices are invariant under the following swap of GLSM fields, 
\beal{es0204}
p_1 \leftrightarrow p_3 ~,~
p_2 \leftrightarrow p_4 ~,~
\eea
This further illustrates that the global symmetry is $SU(2)\times SU(2) \times U(1)^2$, which matches with the isometry group on $Y^{2,1}(\mathbb{CP}^1 \times \mathbb{CP}^1)$.

The toric data for the brane brick model is given by
 \beal{es0205}
{\footnotesize
 G_{t}=
 \resizebox{0.725\textwidth}{!}{$
\left(
\begin{array}{cccccc|ccccccccc|ccccccc|ccc}
 p_1 & p_2 & p_3 & p_4 & p_5 & p_6 & s_1 & s_2 & s_3 & s_4 & s_5 & s_6 & s_7 & s_8 & s_9 
& o_1 & o_2 & o_3 & o_4 & o_5 & o_6 & o_7
& o_8 & o_9 & o_{10}
\\
\hline 
 1 & 1 & 1 & 1 & 1 & 1 & 1 & 1 & 1 & 1 & 1 & 1 & 1 & 1 & 1 & 2 & 2 & 2 & 2 & 2 & 2 & 2 & 3 & 3 & 3 \\
 1 & 0 & -1 & 0 & 0 & 0 & 0 & 0 & 0 & 0 & 0 & 0 & 0 & 0 & 0 & 0 & 0 & 0 & 0 & 0 & 0 & 0 & 0 & 0 & 0 \\
 0 & 1 & 0 & -1 & 0 & 0 & 0 & 0 & 0 & 0 & 0 & 0 & 0 & 0 & 0 & 0 & 0 & 0 & 0 & 0 & 0 & 0 & 0 & 0 & 0 \\
 0 & 0 & 1 & 1 & 2 & 0 & 1 & 1 & 1 & 1 & 1 & 1 & 1 & 1 & 1 & 1 & 1 & 1 & 1 & 1 & 1 & 1 & 2 & 2 & 2 \\
\end{array}
\right)
$}
}~,~
\eea
where we identify the $o_k$ as extra GLSM fields \cite{Franco:2015tna}.
The corresponding toric diagram is shown in \fref{f_toric_cp1_cp1_21} and it agrees with the expectation.
We note that $Y^{2,1}(\mathbb{CP}^1 \times \mathbb{CP}^1)$ is one of the Fano 3-folds studied in \cite{Franco:2022gvl}. In fact, to be precise, it corresponds to Model 9 in \cite{Franco:2022gvl}. 
Given the fact that the $2d$ $(0,2)$ theory identified here is different from the one given for the same geometry in \cite{Franco:2022gvl}, we expect both theories to be related by a (sequence of) triality transformation(s) \cite{Franco:2016nwv}.
\\

\begin{figure}[H]
\begin{center}
\resizebox{0.35\hsize}{!}{
\includegraphics[height=6cm]{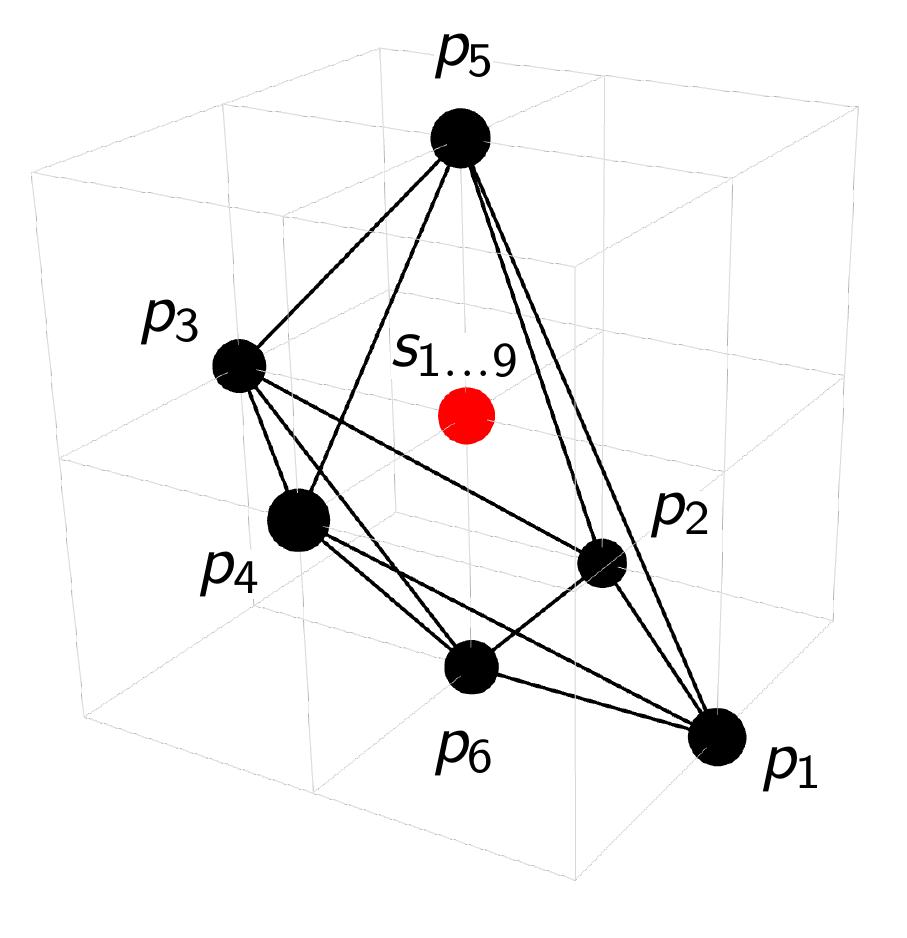} 
}
\caption{
Toric diagram of $Y^{2,1}(\mathbb{CP}^{1}\times\mathbb{CP}^{1})$.
\label{f_toric_cp1_cp1_21}}
 \end{center}
 \end{figure}

\subsubsection{$Y^{2,2}(\mathbb{CP}^1 \times \mathbb{CP}^1)$ \label{sec:cp22}}

\begin{figure}[H]
\begin{center}
\resizebox{0.35\hsize}{!}{
\includegraphics[height=6cm]{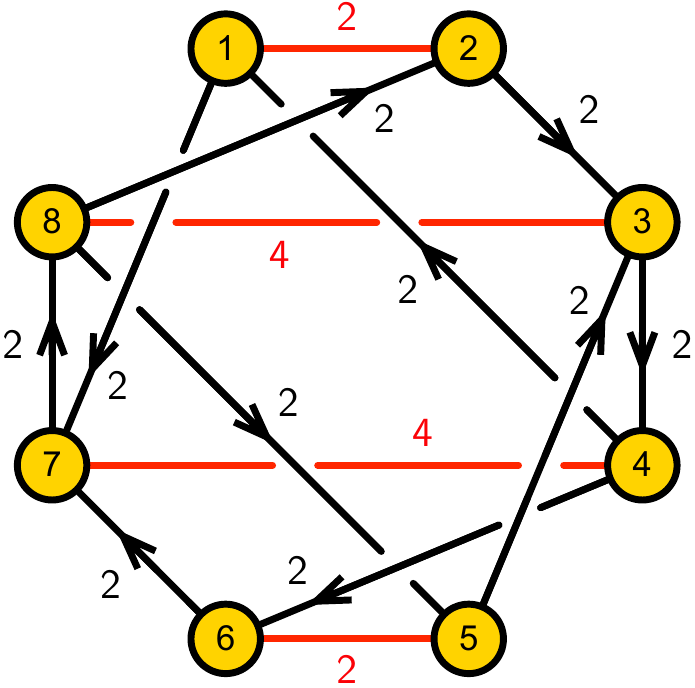} 
}
\caption{
Quiver for $Y^{2,2}(\mathbb{CP}^1 \times \mathbb{CP}^1)$.
\label{f_quiver_cp1_cp1_22}}
 \end{center}
 \end{figure}

The quiver diagram for the $2d$ $(0,2)$ theory corresponding to $Y^{2,2}(\mathbb{CP}^1 \times \mathbb{CP}^1)$ is shown in \fref{f_quiver_cp1_cp1_22}.
The $J$- and $E$-terms for the theory take the following form
\beq
\begin{array}{rccrcl}
 &   J &   & & E &  \\
\Lambda_{12}^{i}  : & \epsilon_{ij} \epsilon_{kl} X^{k}_{23}  X^{j}_{34}  X^{l}_{41} &~&   P_{17}  X^{i}_{78}  Q_{82} &-&  Q_{17}  X^{i}_{78}  P_{82} \\
 \Lambda_{47}^{1i}  : & \epsilon_{ij} \epsilon_{kl} X^{k}_{78}  X^{j}_{85}  Q_{53}  X^{l}_{34} &~&   X^{i}_{41}  P_{17} &-&  P_{46}  X^{i}_{67} \\
 \Lambda_{47}^{2i}  : & \epsilon_{ij} \epsilon_{kl} X^{k}_{78}  X^{j}_{85}  P_{53}  X^{l}_{34} &~&   X^{i}_{41}  Q_{17} &-&  Q_{46}  X^{i}_{67} \\
 \Lambda_{56}^{i}  : & \epsilon_{ij} \epsilon_{kl} X^{k}_{67}  X^{j}_{78}  X^{l}_{85}  &~&   P_{53}  X^{i}_{34}  Q_{46} &-&  Q_{53}  X^{i}_{34}  P_{46} \\
 \Lambda_{83}^{1i}  : & \epsilon_{ij} \epsilon_{kl} X^{k}_{34}  X^{j}_{41}  Q_{17}  X^{l}_{78} &~&   X^{i}_{85}  P_{53} &-&  P_{82}  X^{i}_{23} \\
 \Lambda_{83}^{2i}  : & \epsilon_{ij} \epsilon_{kl} X^{k}_{34}  X^{j}_{41}  P_{17}  X^{l}_{78} &~&   X^{i}_{85}  Q_{53} &-&  Q_{82}  X^{i}_{23} \\
 \end{array} 
~,~
\label{es0300}
 \eeq
 where the global symmetry indices are $i,j,k,l=1,2$.
\newpage

From the $J$- and $E$-terms, we can construct the $P$-matrix which takes the following form
 \beal{es0301}
{\footnotesize
 P=
\resizebox{0.5\textwidth}{!}{$
\left(
\begin{array}{r|cccccc|cccccccccc}
\; & p_1 & p_2 & p_3 & p_4 & p_5 & p_6 &
s_1 & s_2 & s_3 & s_4 & s_5 & s_6 & s_7 & s_8 & s_9 & s_{10}
\\
\hline
X_{23}^{1} & 1 & 0 & 0 & 0 & 0 & 0 & 0 & 0 & 0 & 1 & 0 & 1 & 0 & 0 & 0 & 0 \\
X_{34}^{1} & 0 & 1 & 0 & 0 & 0 & 0 & 0 & 0 & 1 & 0 & 0 & 0 & 0 & 0 & 0 & 0 \\
X_{41}^{1} & 1 & 0 & 0 & 0 & 0 & 0 & 0 & 1 & 0 & 0 & 0 & 0 & 0 & 1 & 0 & 0 \\
X_{67}^{1} & 1 & 0 & 0 & 0 & 0 & 0 & 0 & 1 & 0 & 0 & 0 & 0 & 0 & 0 & 1 & 0 \\
X_{78}^{1} & 0 & 1 & 0 & 0 & 0 & 0 & 1 & 0 & 0 & 0 & 0 & 0 & 0 & 0 & 0 & 0 \\
X_{85}^{1} & 1 & 0 & 0 & 0 & 0 & 0 & 0 & 0 & 0 & 1 & 1 & 0 & 0 & 0 & 0 & 0 \\
\hline
X_{23}^{2} & 0 & 0 & 1 & 0 & 0 & 0 & 0 & 0 & 0 & 1 & 0 & 1 & 0 & 0 & 0 & 0 \\
X_{34}^{2} & 0 & 0 & 0 & 1 & 0 & 0 & 0 & 0 & 1 & 0 & 0 & 0 & 0 & 0 & 0 & 0 \\
X_{41}^{2} & 0 & 0 & 1 & 0 & 0 & 0 & 0 & 1 & 0 & 0 & 0 & 0 & 0 & 1 & 0 & 0 \\
X_{67}^{2} & 0 & 0 & 1 & 0 & 0 & 0 & 0 & 1 & 0 & 0 & 0 & 0 & 0 & 0 & 1 & 0 \\
X_{78}^{2} & 0 & 0 & 0 & 1 & 0 & 0 & 1 & 0 & 0 & 0 & 0 & 0 & 0 & 0 & 0 & 0 \\
X_{85}^{2} & 0 & 0 & 1 & 0 & 0 & 0 & 0 & 0 & 0 & 1 & 1 & 0 & 0 & 0 & 0 & 0 \\
\hline
Q_{17} & 0 & 0 & 0 & 0 & 1 & 0 & 0 & 0 & 0 & 0 & 0 & 0 & 0 & 0 & 1 & 1 \\
Q_{46} & 0 & 0 & 0 & 0 & 1 & 0 & 0 & 0 & 0 & 0 & 0 & 0 & 0 & 1 & 0 & 1 \\
Q_{53} & 0 & 0 & 0 & 0 & 1 & 0 & 0 & 0 & 0 & 0 & 0 & 1 & 1 & 0 & 0 & 0 \\
Q_{82} & 0 & 0 & 0 & 0 & 1 & 0 & 0 & 0 & 0 & 0 & 1 & 0 & 1 & 0 & 0 & 0 \\
\hline
P_{17} & 0 & 0 & 0 & 0 & 0 & 1 & 0 & 0 & 0 & 0 & 0 & 0 & 0 & 0 & 1 & 1 \\
P_{46} & 0 & 0 & 0 & 0 & 0 & 1 & 0 & 0 & 0 & 0 & 0 & 0 & 0 & 1 & 0 & 1 \\
P_{53} & 0 & 0 & 0 & 0 & 0 & 1 & 0 & 0 & 0 & 0 & 0 & 1 & 1 & 0 & 0 & 0 \\
P_{82} & 0 & 0 & 0 & 0 & 0 & 1 & 0 & 0 & 0 & 0 & 1 & 0 & 1 & 0 & 0 & 0 \\
\end{array}
\right)
$}
}~.~
\eea
The corresponding $U(1)$-charge matrix for the $J$- and $E$-terms is given by
\beal{es0302}
{\footnotesize
 Q_{JE}=
 \resizebox{0.5\textwidth}{!}{$
\left(
\begin{array}{cccccc|cccccccccc}
 p_1 & p_2 & p_3 & p_4 & p_5 & p_6 &
s_1 & s_2 & s_3 & s_4 & s_5 & s_6 & s_7 & s_8 & s_9 & s_{10}
\\
\hline 
 1 & 0 & 1 & 0 & 0 & 0 & 0 & -1 & 0 & -1 & 0 & 0 & 0 & 0 & 0 & 0 \\
 0 & 1 & 0 & 1 & 0 & 0 & -1 & 0 & -1 & 0 & 0 & 0 & 0 & 0 & 0 & 0 \\
 0 & 0 & 0 & 0 & 1 & 1 & 0 & 1 & 0 & 1 & -1 & -1 & 0 & -1 & -1 & 0 \\
 0 & 0 & 0 & 0 & 0 & 0 & 0 & 1 & 0 & 0 & 0 & 0 & 0 & -1 & -1 & 1 \\
 0 & 0 & 0 & 0 & 0 & 0 & 0 & 0 & 0 & 1 & -1 & -1 & 1 & 0 & 0 & 0 \\
\end{array}
\right)
$}
}~,~
\eea
and the charge matrix for the $D$-terms is given by
 \beal{es0303}
{\footnotesize
 Q_{D}=
 \resizebox{0.5\textwidth}{!}{$
\left(
\begin{array}{cccccc|cccccccccc}
 p_1 & p_2 & p_3 & p_4 & p_5 & p_6 &
s_1 & s_2 & s_3 & s_4 & s_5 & s_6 & s_7 & s_8 & s_9 & s_{10}
\\
\hline 
 0 & 0 & 0 & 0 & 0 & 0 & 1 & 0 & 0 & 0 & 0 & 0 & 0 & 0 & -1 & 0 \\
 0 & 0 & 0 & 0 & 0 & 0 & 0 & 1 & 0 & 0 & 0 & 0 & 0 & -1 & 0 & 0 \\
 0 & 0 & 0 & 0 & 0 & 0 & 0 & 0 & 1 & 0 & 0 & -1 & 0 & 0 & 0 & 0 \\
 0 & 0 & 0 & 0 & 0 & 0 & 0 & 0 & 0 & 1 & -1 & 0 & 0 & 0 & 0 & 0 \\
 0 & 0 & 0 & 0 & 0 & 0 & 0 & 0 & 0 & -1 & 0 & 1 & 0 & 0 & 0 & 0 \\
 0 & 0 & 0 & 0 & 0 & 0 & 0 & 0 & -1 & 0 & 0 & 0 & 0 & 1 & 0 & 0 \\
 0 & 0 & 0 & 0 & 0 & 0 & -1 & 0 & 0 & 0 & 1 & 0 & 0 & 0 & 0 & 0 \\
\end{array}
\right)
$}
}~.~
\eea
We note that the charge matrices are invariant under
\beal{es304}
p_1 \leftrightarrow p_3 ~,~
p_2 \leftrightarrow p_4 ~.~
\eea
This symmetry further indicates that the global symmetry takes the form $SU(2)\times SU(2) \times U(1)^2$.

\begin{figure}[ht!!]
\begin{center}
\resizebox{0.35\hsize}{!}{
\includegraphics[height=6cm]{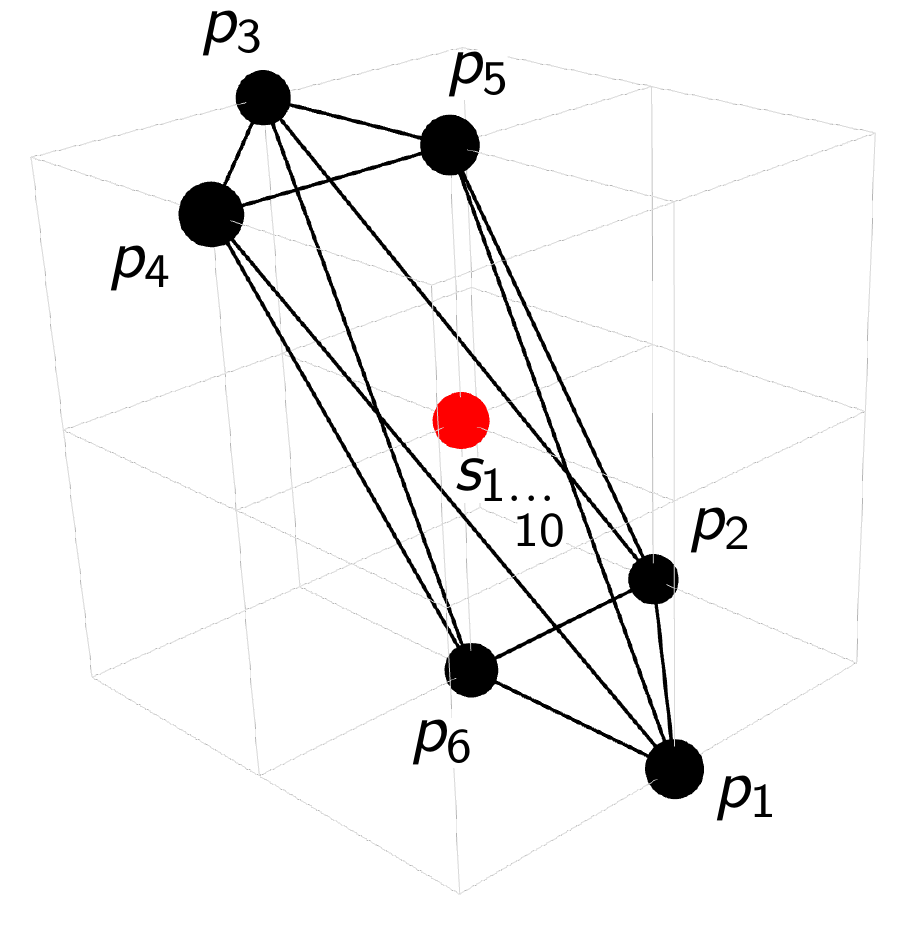} 
}
\caption{
Toric diagram of $Y^{2,2}(\mathbb{CP}^{1}\times\mathbb{CP}^{1})$.
\label{f_toric_cp1_cp1_22}}
 \end{center}
 \end{figure}

The toric data of the $2d$ $(0,2)$ theory is given by
 \beal{es0305}
{\footnotesize
 G_{t}=
 \resizebox{0.5\textwidth}{!}{$
\left(
\begin{array}{cccccc|cccccccccc}
 p_1 & p_2 & p_3 & p_4 & p_5 & p_6 &
s_1 & s_2 & s_3 & s_4 & s_5 & s_6 & s_7 & s_8 & s_9 & s_{10}
\\
\hline 
 1 & 1 & 1 & 1 & 1 & 1 & 1 & 1 & 1 & 1 & 1 & 1 & 1 & 1 & 1 & 1 \\
 1 & 0 & -1 & 0 & 0 & 0 & 0 & 0 & 0 & 0 & 0 & 0 & 0 & 0 & 0 & 0 \\
 0 & 1 & 0 & -1 & 0 & 0 & 0 & 0 & 0 & 0 & 0 & 0 & 0 & 0 & 0 & 0 \\
 0 & 0 & 2 & 2 & 2 & 0 & 1 & 1 & 1 & 1 & 1 & 1 & 1 & 1 & 1 & 1 \\
\end{array}
\right)
$}
}~.~
\eea
The resulting toric diagram is shown in \fref{f_toric_cp1_cp1_22} and it agrees with the expectation. We note that $Y^{2,2}(\mathbb{CP}^1 \times \mathbb{CP}^1)$ corresponds to $Q^{1,1,1}/\mathbb{Z}_2$. A full classification of the toric phases for this geometry was presented in \cite{Franco:2018qsc}. The model constructed above indeed corresponds to phase A in this classification. This theory also appeared in \cite{Franco:2016nwv}.
\\

\section{$Y^{p,k}(\mathbb{CP}^2)$ models \label{sec:cp2}}

\subsection{Construction of the $2d$ $(0,2)$ theories \label{sec:cp2intro}}

The $2d$ $(0,2)$ gauge theories corresponding to the family of $Y^{p,k}(\mathbb{CP}^2)$  manifolds are parameterized in terms of $p,k \in \mathbb{Z}$, where  $0\leq k \leq \frac{3}{2} p$. 

The total number of gauge groups in the $2d$ $(0,2)$ theories is $3p$. 
The notation used on the gauge group labels for the chiral and Fermi fields is defined as follows
\beal{es10a7}
[m] \equiv ((m-1) ~\text{mod}~ 3p) +1 ~.~
\eea
The chiral and Fermi fields form a periodic layered structure with a total number of $p$ layers.
Each layer in the structure contains 3 gauge groups, which is the same number of gauge groups as in the $4d$ $\mathcal{N}=1$ theory corresponding to $dP_0$ \cite{Feng:2001xr}.
In fact, this structure follows from the particular way in which we constructed the gauge theoories.
We made use of orbifold reduction \cite{Franco:2016fxm}, $3d$ printing \cite{Franco:2018qsc} and partial resolution via higgsing \cite{Franco:2015tna}, starting with the $4d$ $\mathcal{N}=1$ quiver gauge theory corresponding to $dP_0$.

Between layers, the gauge group labels are shifted by additional indices given by $a,b,c$ and $d$.
When the parameter $k$ for the $Y^{p,k}(\mathbb{CP}^2)$ models is in the range $0 \leq k \leq p$, 
only Fermi and chiral fields depending on $a$, $b$ and $c$ contribute to the $2d$ $(0,2)$ theory, with the values of the indices being
\beal{es10a5}
a=0,\dots,p-1 ~,~
b=k,\dots,p-1 ~,~
c=0,\dots,k-1 ~.~ 
\eea
When the model parameter $k$ is in the range $p < k  \leq \frac{3}{2} p$, only the Fermi and chiral fields depending on indices $a$, $c$ and $d$ contribute to the theory, with the values of the indices being
\beal{es10a6}
a=0,\dots,2p-k ~,~
c=k-p, \dots, 2p-k ~,~
d=0,\dots,k-p-1 ~.~
\eea
Finally, we also have indices for the $SU(3)\times U(1)^2$ global symmetry of the $2d$ $(0,2)$ theories corresponding to $Y^{p,k}(\mathbb{CP}^2)$. These global symmetry indices are
\beal{es10a20}
i,j,k=1,2,3 ~.~
\eea

Under these indices, the Fermi fields of the $2d$ $(0,2)$ theories are as follows
\beal{es10a21}
&
\Lambda_{[3a+1][3a+2]}^{i}~,~
\Lambda_{[3b+6][3b+1]}^{2i}~,~
\Lambda_{[3b+5][3b+3]}^{2i}~,~
\Lambda_{[3c+6][3c+3]}^{2i}~,~
\Lambda_{[3d+3][3d+6]}^{1i}~.~
&
\eea
The chiral fields of the $2d$ $(0,2)$ theories take the following form
\beal{es10a22}
&
X_{[3a+2][3a+3]}^{i} ~,~
X_{[3a+3][3a+1]}^{i} ~,~
X_{[3b+1][3b+5]}^{i} ~,~
&
\nn\\
&
Q_{[3b+4][3b+1]} ~,~
Q_{[3b+6][3b+3]} ~,~
Q_{[3b+5][3b+2]} ~,~
&
\nn\\
&
Q_{[3c+4][3c+3]} ~,~
Q_{[3c+6][3c+2]} ~,~
P_{[3c+1][3c+5]} ~,~
&
\nn\\
&
Q_{[3d+4][3d+2]} ~,~
P_{[3d+1][3d+6]} ~,~
P_{[3d+3][3d+5]} ~.~
&
\eea

The corresponding $J$-terms are
\beal{es10a0}
J[\Lambda_{[3a+1][3a+2]}^{i}] 
&=& 
\epsilon_{ijk}
X_{[3a+2][3a+3]}^{j} X_{[3a+3][3a+1]}^{k}
~,~
\nn\\
J[\Lambda_{[3b+6][3b+1]}^{2i}] 
&=& 
\epsilon_{ijk}
X_{[3b+1][3b+5]}^{j} X_{[3b+5][3b+6]}^{k}
~,~
\nn\\
J[\Lambda_{[3b+5][3b+3]}^{2i}] 
&=& 
\epsilon_{ijk}
X_{[3b+3][3b+1]}^{j} X_{[3b+1][3b+5]}^{k}
~,~
\nn\\
J[\Lambda_{[3c+6][3c+3]}^{2i}] 
&=& 
\epsilon_{ijk}
X_{[3c+3][3c+1]}^{j} P_{[3c+1][3c+5]} X_{[3c+5][3c+6]}^{k}
~,~
\nn\\
J[\Lambda_{[3d+3][3d+6]}^{1i}] 
&=& 
\epsilon_{ijk}
X_{[3d+6][3d+4]}^{j} Q_{[3d+4][3d+2]} X_{[3d+2][3d+3]}^{k}
~,~
\eea
and the associated $E$-terms take the following form
\beal{es10a1}
E[\Lambda_{[3b+6][3b+1]}^{2i}] 
&=& 
X_{[3b+6][3b+4]}^{i} Q_{[3b+4][3b+1]}
-
Q_{[3b+6][3b+3]} X_{[3b+3][3b+1]}^{i}
~,~
\nn\\
E[\Lambda_{[3b+5][3b+3]}^{2i}] 
&=& 
X_{[3b+5][3b+6]}^{i} Q_{[3b+6][3b+3]}
-
Q_{[3b+5][3b+2]} X_{[3b+2][3b+3]}^{i}
~,~
\nn\\
E[\Lambda_{[3c+6][3c+3]}^{2i}] 
&=& 
X_{[3c+6][3c+4]}^{i} Q_{[3c+4][3c+3]}
-
Q_{[3c+6][3c+2]} X_{[3c+2][3c+3]}^{i}
~,~
\nn\\
E[\Lambda_{[3d+3][3d+6]}^{1i}] 
&=& 
X_{[3d+3][3d+1]}^{i} P_{[3d+1][3d+6]}
-
P_{[3d+3][3d+5]} X_{[3d+5][3d+6]}^{i}
~.~
\eea
Additionally, we have $E$-term plaquettes,
\beal{es10a2}
E_{+}[\Lambda_{[3b+1][3b+2]}^{i}] 
&=& 
X_{[3b+1][3b+5]}^{i} Q_{[3b+5][3b+2]}
~,~
\nn\\
E_{-}[\Lambda_{[3b+4][3b+5]}^{i}] 
&=& 
- Q_{[3b+4][3b+1]} X_{[3b+1][3b+5]}^{i}
~,~
\nn\\
E_{+}[\Lambda_{[3c+1],[3c+2]}^{i}] 
&=& 
P_{[3c+1][3c+5]} X_{[3c+5][3c+6]}^{i} Q_{[3c+6][3c+2]}
~,~
\nn\\
E_{-}[\Lambda_{[3c+4][3c+5]}^{i}] 
&=& 
- Q_{[3c+4][3c+3]} X_{[3c+3][3c+1]}^{i} P_{[3c+1][3c+5]}
~,~
\nn\\
E_{+}[\Lambda_{[3d+1][3d+2]}^{i}] 
&=& 
P_{[3d+1][3d+6]} X_{[3d+6][3d+4]}^{i} Q_{[3d+4][3d+2]}
~,~
\nn\\
E_{-}[\Lambda_{[3d+4][3d+5]}^{i}] 
&=& 
- Q_{[3d+4][3d+2]} X_{[3d+2][3d+3]}^{i} P_{[3d+3][3d+5]}
~,~
\eea
which combine to proper $E$-terms of the form $E=E_+ - E_-$.
The $E$-terms plaquettes are shown here separately because quadratic and cubic plaquettes are combined depending on the value of $k$ in order to form a single $E$-term.
\\

\subsection{Examples \label{sec:cp200}}

In this section, we summarize the construction of $2d$ $(0,2)$ theories corresponding to $Y^{p,k}(\mathbb{CP}^2)$ where $p=3$.
In total there are 5 toric diagrams with $p=3$ whose toric diagrams have precisely 2 internal points. 
In appendix \sref{sec:sap2} and \sref{sec:sap3}, the $J$- and $E$-terms corresponding to $2d$ $(0,2)$ theories with $p=1$ and $p=2$ are summarized for completeness.
\\

\subsubsection{$Y^{3,0}(\mathbb{CP}^2)$ \label{sec:cp20}}

\begin{figure}[H]
\begin{center}
\resizebox{0.4\hsize}{!}{
\includegraphics[height=6cm]{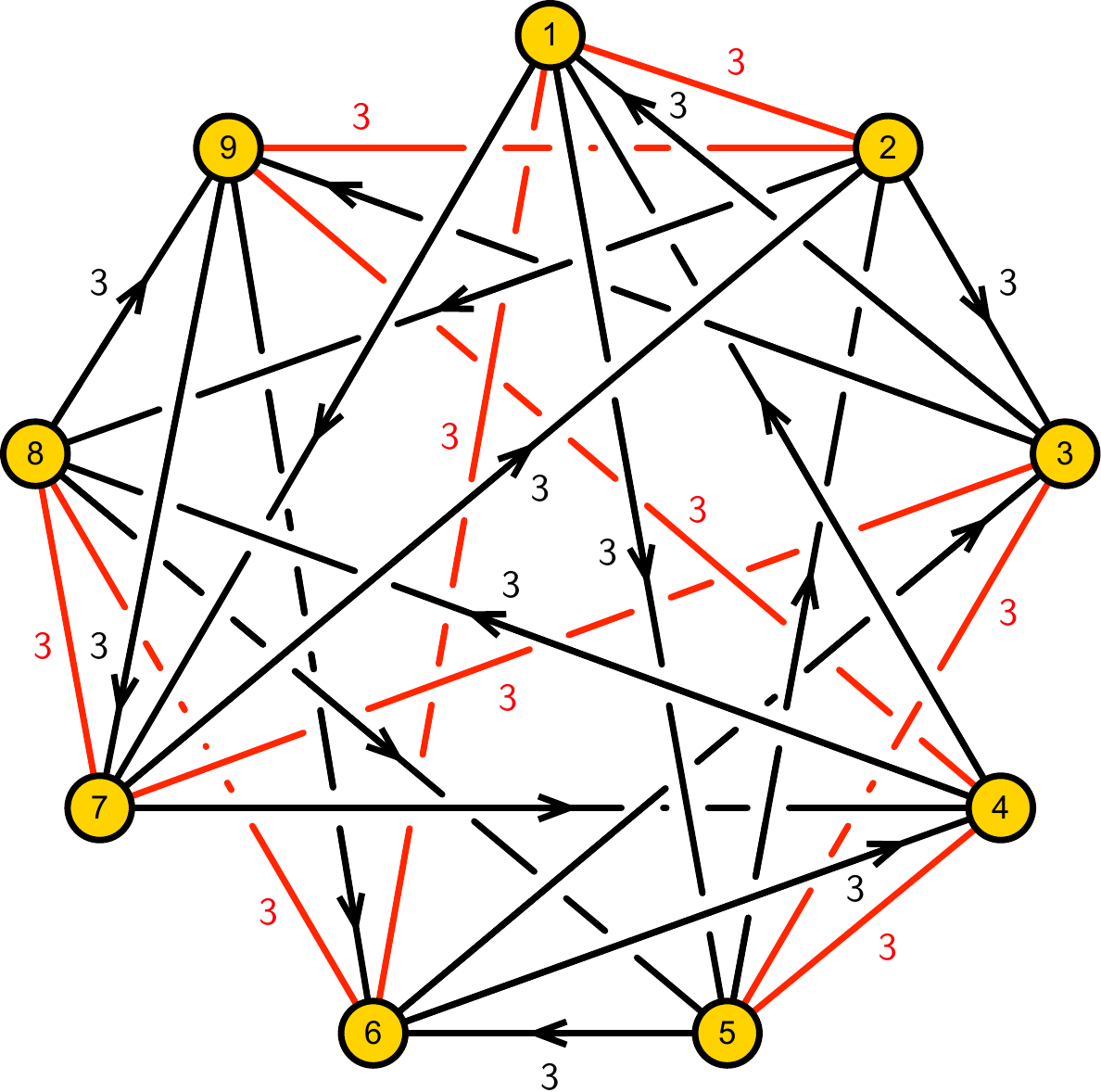} 
}
\caption{
Quiver for $Y^{3,0}(\mathbb{CP}^2)$.
\label{f_quiver_cp2_30}}
 \end{center}
 \end{figure}

The $J$- and $E$-terms of the $2d$ $(0,2)$ theory for $Y^{3,0}(\mathbb{CP}^2)$ take the form
  \beq
\begin{array}{rccrcl}
 &   J &   & & E &  \\
 \Lambda_{12}^{i}  : & \epsilon_{ijk} X^{j}_{23}  X^{k}_{31} &~&  X^{i}_{15}  Q_{52 }&-& Q_{17}  X^{i}_{72 } \\
 \Lambda_{29}^{2i}  : & \epsilon_{ijk} X^{j}_{97}  X^{k}_{72 } &~&  X^{i}_{23}  Q_{39 }&-& Q_{28}  X^{i}_{89} \\
 \Lambda_{37}^{2i}  : & \epsilon_{ijk} X^{j}_{72}  X^{k}_{23 } &~&  X^{i}_{31}  Q_{17 }&-& Q_{39}  X^{i}_{97} \\
 \Lambda_{45}^{i}  : & \epsilon_{ijk} X^{j}_{56}  X^{k}_{64 } &~&  X^{i}_{48}  Q_{85 }&-& Q_{41}  X^{i}_{15} \\
 \Lambda_{53}^{2i}  : & \epsilon_{ijk} X^{j}_{31}  X^{k}_{15 } &~&  X^{i}_{56}  Q_{63 }&-& Q_{52}  X^{i}_{23} \\
 \Lambda_{61}^{2i}  : & \epsilon_{ijk}  X^{j}_{15}  X^{k}_{56 } &~&  X^{i}_{64}  Q_{41 }&-& Q_{63}  X^{i}_{31} \\
 \Lambda_{78}^{i}  : & \epsilon_{ijk} X^{j}_{89}  X^{k}_{97 } &~&  X^{i}_{72}  Q_{28 }&-& Q_{74}  X^{i}_{48} \\
 \Lambda_{86}^{2i}  : & \epsilon_{ijk} X^{j}_{64}  X^{k}_{48 } &~&  X^{i}_{89}  Q_{96 }&-& Q_{85}  X^{i}_{56} \\
 \Lambda_{94}^{2i}  : & \epsilon_{ijk} X^{j}_{48}  X^{k}_{89 } &~&  X^{i}_{97}  Q_{74 }&-& Q_{96}  X^{i}_{64} \\
 \end{array} 
~,~
\label{es0400b}
 \eeq
where the global symmetry indices are $i,j,k=1,2,3$.
The corresponding quiver diagram is shown in \fref{f_quiver_cp2_30}.
 
 From the $J$- and $E$-terms, we can construct the $P$-matrix, which takes the form
 \beal{es0401}
{\footnotesize
 P=
\resizebox{0.92\textwidth}{!}{$
\left(
\begin{array}{r|cccc|ccc|cccccccccccccccccc|ccccccccc|ccccccccc}
\; &
p_1 & p_2 & p_3 & p_4 & 
q_1 & q_2 & q_3 &
u_1 & u_2 & u_3 & u_4 & u_5 & 
u_6 & u_7 & u_8 & u_9 & u_{10} & 
u_{11} & u_{12} & u_{13} & u_{14} & u_{15} & 
u_{16} & u_{17} & u_{18} & 
v_1 & v_2 & v_3 & v_4 & v_5 & v_6 & v_7 & v_8 & v_9 &
o_1 & o_2 & o_3 & o_4 & o_5 & o_6 & o_7 & o_8 & o_9 \\
\hline
X_{15}^{1} & 1 & 0 & 0 & 0 & 0 & 0 & 1 & 1 & 0 & 0 & 0 & 0 & 0 & 0 & 0 & 0 & 0 & 0 & 0 & 1 & 0 & 0 & 1 & 0 & 1 & 0 & 0 & 0 & 0 & 0 & 1 & 0 & 0 & 0 & 1 & 0 & 1 & 0 & 0 & 0 & 1 & 1 & 1 \\
X_{23}^{1} & 1 & 0 & 0 & 0 & 0 & 1 & 0 & 0 & 0 & 0 & 0 & 1 & 0 & 0 & 1 & 1 & 1 & 0 & 0 & 0 & 0 & 0 & 0 & 0 & 0 & 0 & 1 & 0 & 0 & 0 & 0 & 0 & 0 & 0 & 1 & 0 & 0 & 0 & 0 & 1 & 1 & 1 & 1 \\
X_{31}^{1} & 1 & 0 & 0 & 0 & 1 & 0 & 0 & 0 & 0 & 0 & 1 & 0 & 0 & 1 & 0 & 0 & 0 & 1 & 1 & 0 & 0 & 0 & 0 & 0 & 0 & 0 & 0 & 1 & 0 & 0 & 0 & 0 & 0 & 0 & 0 & 1 & 1 & 1 & 1 & 1 & 0 & 0 & 0 \\
X_{48}^{1} & 1 & 0 & 0 & 0 & 0 & 0 & 1 & 0 & 0 & 0 & 0 & 0 & 1 & 0 & 0 & 0 & 1 & 0 & 1 & 0 & 0 & 0 & 0 & 0 & 1 & 0 & 0 & 0 & 0 & 0 & 0 & 0 & 0 & 1 & 0 & 1 & 0 & 0 & 1 & 1 & 0 & 1 & 1 \\
X_{56}^{1} & 1 & 0 & 0 & 0 & 0 & 1 & 0 & 0 & 1 & 0 & 0 & 0 & 0 & 0 & 1 & 0 & 0 & 0 & 0 & 0 & 1 & 0 & 0 & 1 & 0 & 0 & 0 & 0 & 0 & 0 & 0 & 1 & 0 & 0 & 0 & 1 & 0 & 1 & 1 & 1 & 0 & 1 & 0 \\
X_{64}^{1} & 1 & 0 & 0 & 0 & 1 & 0 & 0 & 0 & 0 & 1 & 0 & 0 & 0 & 0 & 0 & 1 & 0 & 1 & 0 & 0 & 0 & 1 & 0 & 0 & 0 & 0 & 0 & 0 & 0 & 0 & 0 & 0 & 1 & 0 & 1 & 0 & 1 & 1 & 0 & 0 & 1 & 0 & 1 \\
X_{72}^{1} & 1 & 0 & 0 & 0 & 0 & 0 & 1 & 1 & 1 & 1 & 0 & 0 & 1 & 0 & 0 & 0 & 0 & 0 & 0 & 0 & 0 & 0 & 0 & 0 & 0 & 1 & 0 & 0 & 0 & 0 & 0 & 0 & 0 & 0 & 1 & 1 & 1 & 1 & 1 & 0 & 0 & 0 & 0 \\
X_{89}^{1} & 1 & 0 & 0 & 0 & 0 & 1 & 0 & 0 & 0 & 0 & 0 & 1 & 0 & 1 & 0 & 0 & 0 & 0 & 0 & 0 & 0 & 0 & 1 & 1 & 0 & 0 & 0 & 0 & 0 & 1 & 0 & 0 & 0 & 0 & 1 & 0 & 1 & 1 & 1 & 0 & 1 & 0 & 0 \\
X_{97}^{1} & 1 & 0 & 0 & 0 & 1 & 0 & 0 & 0 & 0 & 0 & 1 & 0 & 0 & 0 & 0 & 0 & 0 & 0 & 0 & 1 & 1 & 1 & 0 & 0 & 0 & 0 & 0 & 0 & 1 & 0 & 0 & 0 & 0 & 0 & 0 & 1 & 0 & 0 & 0 & 1 & 1 & 1 & 1 \\
\hline
X_{15}^{2} & 0 & 1 & 0 & 0 & 0 & 0 & 1 & 1 & 0 & 0 & 0 & 0 & 0 & 0 & 0 & 0 & 0 & 0 & 0 & 1 & 0 & 0 & 1 & 0 & 1 & 0 & 0 & 0 & 0 & 0 & 1 & 0 & 0 & 0 & 1 & 0 & 1 & 0 & 0 & 0 & 1 & 1 & 1 \\
X_{23}^{2} & 0 & 1 & 0 & 0 & 0 & 1 & 0 & 0 & 0 & 0 & 0 & 1 & 0 & 0 & 1 & 1 & 1 & 0 & 0 & 0 & 0 & 0 & 0 & 0 & 0 & 0 & 1 & 0 & 0 & 0 & 0 & 0 & 0 & 0 & 1 & 0 & 0 & 0 & 0 & 1 & 1 & 1 & 1 \\
X_{31}^{2} & 0 & 1 & 0 & 0 & 1 & 0 & 0 & 0 & 0 & 0 & 1 & 0 & 0 & 1 & 0 & 0 & 0 & 1 & 1 & 0 & 0 & 0 & 0 & 0 & 0 & 0 & 0 & 1 & 0 & 0 & 0 & 0 & 0 & 0 & 0 & 1 & 1 & 1 & 1 & 1 & 0 & 0 & 0 \\
X_{48}^{2} & 0 & 1 & 0 & 0 & 0 & 0 & 1 & 0 & 0 & 0 & 0 & 0 & 1 & 0 & 0 & 0 & 1 & 0 & 1 & 0 & 0 & 0 & 0 & 0 & 1 & 0 & 0 & 0 & 0 & 0 & 0 & 0 & 0 & 1 & 0 & 1 & 0 & 0 & 1 & 1 & 0 & 1 & 1 \\
X_{56}^{2} & 0 & 1 & 0 & 0 & 0 & 1 & 0 & 0 & 1 & 0 & 0 & 0 & 0 & 0 & 1 & 0 & 0 & 0 & 0 & 0 & 1 & 0 & 0 & 1 & 0 & 0 & 0 & 0 & 0 & 0 & 0 & 1 & 0 & 0 & 0 & 1 & 0 & 1 & 1 & 1 & 0 & 1 & 0 \\
X_{64}^{2} & 0 & 1 & 0 & 0 & 1 & 0 & 0 & 0 & 0 & 1 & 0 & 0 & 0 & 0 & 0 & 1 & 0 & 1 & 0 & 0 & 0 & 1 & 0 & 0 & 0 & 0 & 0 & 0 & 0 & 0 & 0 & 0 & 1 & 0 & 1 & 0 & 1 & 1 & 0 & 0 & 1 & 0 & 1 \\
X_{72}^{2} & 0 & 1 & 0 & 0 & 0 & 0 & 1 & 1 & 1 & 1 & 0 & 0 & 1 & 0 & 0 & 0 & 0 & 0 & 0 & 0 & 0 & 0 & 0 & 0 & 0 & 1 & 0 & 0 & 0 & 0 & 0 & 0 & 0 & 0 & 1 & 1 & 1 & 1 & 1 & 0 & 0 & 0 & 0 \\
X_{89}^{2} & 0 & 1 & 0 & 0 & 0 & 1 & 0 & 0 & 0 & 0 & 0 & 1 & 0 & 1 & 0 & 0 & 0 & 0 & 0 & 0 & 0 & 0 & 1 & 1 & 0 & 0 & 0 & 0 & 0 & 1 & 0 & 0 & 0 & 0 & 1 & 0 & 1 & 1 & 1 & 0 & 1 & 0 & 0 \\
X_{97}^{2} & 0 & 1 & 0 & 0 & 1 & 0 & 0 & 0 & 0 & 0 & 1 & 0 & 0 & 0 & 0 & 0 & 0 & 0 & 0 & 1 & 1 & 1 & 0 & 0 & 0 & 0 & 0 & 0 & 1 & 0 & 0 & 0 & 0 & 0 & 0 & 1 & 0 & 0 & 0 & 1 & 1 & 1 & 1 \\
\hline
X_{15}^{3} & 0 & 0 & 1 & 0 & 0 & 0 & 1 & 1 & 0 & 0 & 0 & 0 & 0 & 0 & 0 & 0 & 0 & 0 & 0 & 1 & 0 & 0 & 1 & 0 & 1 & 0 & 0 & 0 & 0 & 0 & 1 & 0 & 0 & 0 & 1 & 0 & 1 & 0 & 0 & 0 & 1 & 1 & 1 \\
X_{23}^{3} & 0 & 0 & 1 & 0 & 0 & 1 & 0 & 0 & 0 & 0 & 0 & 1 & 0 & 0 & 1 & 1 & 1 & 0 & 0 & 0 & 0 & 0 & 0 & 0 & 0 & 0 & 1 & 0 & 0 & 0 & 0 & 0 & 0 & 0 & 1 & 0 & 0 & 0 & 0 & 1 & 1 & 1 & 1 \\
X_{31}^{3} & 0 & 0 & 1 & 0 & 1 & 0 & 0 & 0 & 0 & 0 & 1 & 0 & 0 & 1 & 0 & 0 & 0 & 1 & 1 & 0 & 0 & 0 & 0 & 0 & 0 & 0 & 0 & 1 & 0 & 0 & 0 & 0 & 0 & 0 & 0 & 1 & 1 & 1 & 1 & 1 & 0 & 0 & 0 \\
X_{48}^{3} & 0 & 0 & 1 & 0 & 0 & 0 & 1 & 0 & 0 & 0 & 0 & 0 & 1 & 0 & 0 & 0 & 1 & 0 & 1 & 0 & 0 & 0 & 0 & 0 & 1 & 0 & 0 & 0 & 0 & 0 & 0 & 0 & 0 & 1 & 0 & 1 & 0 & 0 & 1 & 1 & 0 & 1 & 1 \\
X_{56}^{3} & 0 & 0 & 1 & 0 & 0 & 1 & 0 & 0 & 1 & 0 & 0 & 0 & 0 & 0 & 1 & 0 & 0 & 0 & 0 & 0 & 1 & 0 & 0 & 1 & 0 & 0 & 0 & 0 & 0 & 0 & 0 & 1 & 0 & 0 & 0 & 1 & 0 & 1 & 1 & 1 & 0 & 1 & 0 \\
X_{64}^{3} & 0 & 0 & 1 & 0 & 1 & 0 & 0 & 0 & 0 & 1 & 0 & 0 & 0 & 0 & 0 & 1 & 0 & 1 & 0 & 0 & 0 & 1 & 0 & 0 & 0 & 0 & 0 & 0 & 0 & 0 & 0 & 0 & 1 & 0 & 1 & 0 & 1 & 1 & 0 & 0 & 1 & 0 & 1 \\
X_{72}^{3} & 0 & 0 & 1 & 0 & 0 & 0 & 1 & 1 & 1 & 1 & 0 & 0 & 1 & 0 & 0 & 0 & 0 & 0 & 0 & 0 & 0 & 0 & 0 & 0 & 0 & 1 & 0 & 0 & 0 & 0 & 0 & 0 & 0 & 0 & 1 & 1 & 1 & 1 & 1 & 0 & 0 & 0 & 0 \\
X_{89}^{3} & 0 & 0 & 1 & 0 & 0 & 1 & 0 & 0 & 0 & 0 & 0 & 1 & 0 & 1 & 0 & 0 & 0 & 0 & 0 & 0 & 0 & 0 & 1 & 1 & 0 & 0 & 0 & 0 & 0 & 1 & 0 & 0 & 0 & 0 & 1 & 0 & 1 & 1 & 1 & 0 & 1 & 0 & 0 \\
X_{97}^{3} & 0 & 0 & 1 & 0 & 1 & 0 & 0 & 0 & 0 & 0 & 1 & 0 & 0 & 0 & 0 & 0 & 0 & 0 & 0 & 1 & 1 & 1 & 0 & 0 & 0 & 0 & 0 & 0 & 1 & 0 & 0 & 0 & 0 & 0 & 0 & 1 & 0 & 0 & 0 & 1 & 1 & 1 & 1 \\
\hline
Q_{17} & 0 & 0 & 0 & 1 & 0 & 0 & 0 & 0 & 0 & 0 & 0 & 0 & 0 & 0 & 0 & 0 & 0 & 0 & 0 & 1 & 1 & 1 & 1 & 1 & 1 & 0 & 0 & 0 & 1 & 1 & 1 & 1 & 1 & 1 & 0 & 0 & 0 & 0 & 0 & 0 & 1 & 1 & 1 \\
Q_{28} & 0 & 0 & 0 & 1 & 0 & 0 & 0 & 0 & 0 & 0 & 0 & 0 & 0 & 0 & 1 & 1 & 1 & 1 & 1 & 0 & 0 & 0 & 0 & 0 & 1 & 0 & 1 & 1 & 0 & 0 & 1 & 1 & 1 & 1 & 0 & 0 & 0 & 0 & 0 & 1 & 0 & 1 & 1 \\
Q_{39} & 0 & 0 & 0 & 1 & 0 & 0 & 0 & 0 & 0 & 0 & 0 & 0 & 0 & 1 & 0 & 0 & 0 & 1 & 1 & 0 & 0 & 0 & 1 & 1 & 1 & 0 & 0 & 1 & 0 & 1 & 1 & 1 & 1 & 1 & 0 & 0 & 1 & 1 & 1 & 0 & 0 & 0 & 0 \\
Q_{41} & 0 & 0 & 0 & 1 & 0 & 0 & 0 & 0 & 0 & 0 & 1 & 1 & 1 & 1 & 0 & 0 & 1 & 0 & 1 & 0 & 0 & 0 & 0 & 0 & 0 & 1 & 1 & 1 & 1 & 1 & 0 & 0 & 0 & 1 & 0 & 1 & 0 & 0 & 1 & 1 & 0 & 0 & 0 \\
Q_{52} & 0 & 0 & 0 & 1 & 0 & 0 & 0 & 0 & 1 & 1 & 0 & 0 & 1 & 0 & 0 & 0 & 0 & 0 & 0 & 0 & 1 & 1 & 0 & 1 & 0 & 1 & 0 & 0 & 1 & 1 & 0 & 1 & 1 & 1 & 0 & 1 & 0 & 1 & 1 & 0 & 0 & 0 & 0 \\
Q_{63} & 0 & 0 & 0 & 1 & 0 & 0 & 0 & 0 & 0 & 1 & 0 & 1 & 1 & 0 & 0 & 1 & 1 & 0 & 0 & 0 & 0 & 1 & 0 & 0 & 0 & 1 & 1 & 0 & 1 & 1 & 0 & 0 & 1 & 1 & 1 & 0 & 0 & 0 & 0 & 0 & 1 & 0 & 1 \\
Q_{74} & 0 & 0 & 0 & 1 & 0 & 0 & 0 & 1 & 1 & 1 & 0 & 0 & 0 & 0 & 1 & 1 & 0 & 1 & 0 & 0 & 0 & 0 & 0 & 0 & 0 & 1 & 1 & 1 & 0 & 0 & 1 & 1 & 1 & 0 & 1 & 0 & 1 & 1 & 0 & 0 & 0 & 0 & 0 \\
Q_{85} & 0 & 0 & 0 & 1 & 0 & 0 & 0 & 1 & 0 & 0 & 1 & 1 & 0 & 1 & 0 & 0 & 0 & 0 & 0 & 1 & 0 & 0 & 1 & 0 & 0 & 1 & 1 & 1 & 1 & 1 & 1 & 0 & 0 & 0 & 1 & 0 & 1 & 0 & 0 & 0 & 1 & 0 & 0 \\
Q_{96} & 0 & 0 & 0 & 1 & 0 & 0 & 0 & 1 & 1 & 0 & 1 & 0 & 0 & 0 & 1 & 0 & 0 & 0 & 0 & 1 & 1 & 0 & 0 & 0 & 0 & 1 & 1 & 1 & 1 & 0 & 1 & 1 & 0 & 0 & 0 & 1 & 0 & 0 & 0 & 1 & 0 & 1 & 0 \\
\end{array}
\right)
$}
}~.~
\nn\\
\eea

\begin{figure}[ht!!]
\begin{center}
\resizebox{0.35\hsize}{!}{
\includegraphics[height=6cm]{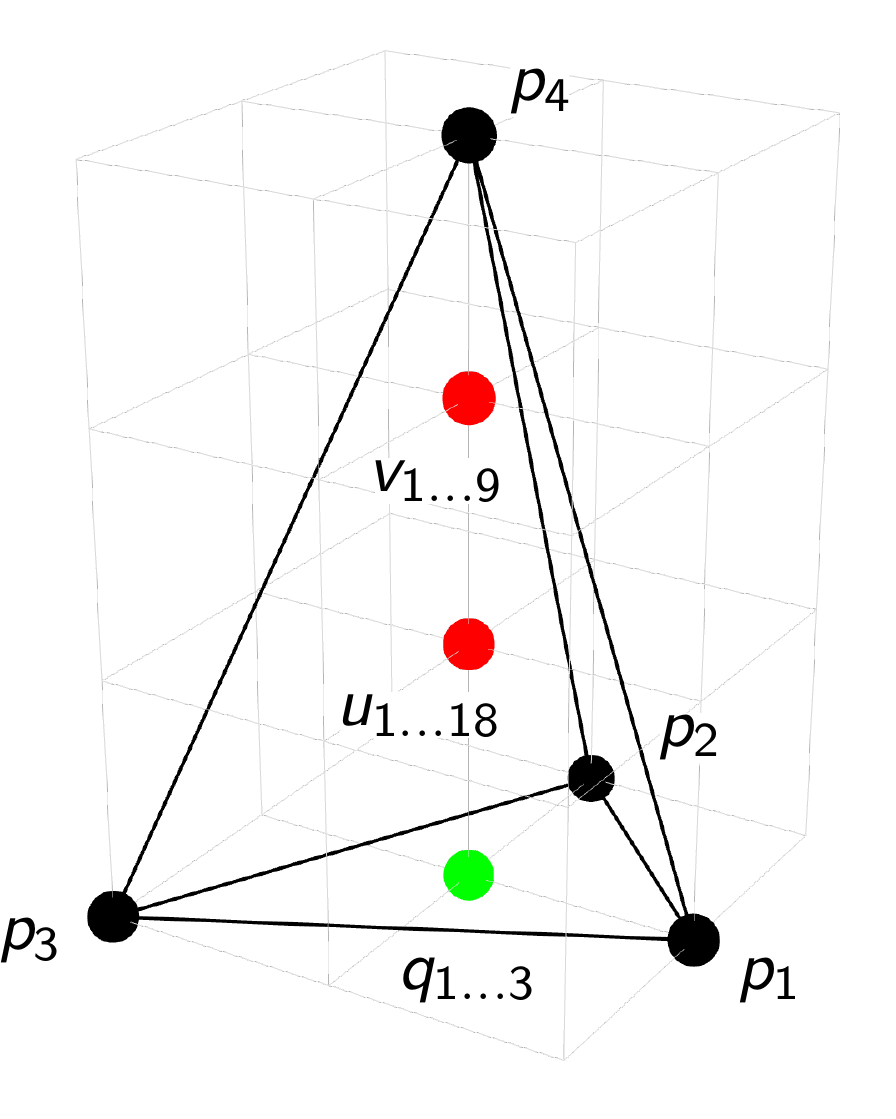} 
}
\caption{
Toric diagram of $Y^{3,0}(\mathbb{CP}^2)$.
\label{f_toric_cp2_20}}
 \end{center}
 \end{figure}

We can also construct the $U(1)$-charge matrix under the $J$- and $E$-terms,
 \beal{es0402}
{\footnotesize
 Q_{JE}=
 \resizebox{0.88\textwidth}{!}{$
\left(
\begin{array}{cccc|ccc|cccccccccccccccccc|ccccccccc|ccccccccc}
p_1 & p_2 & p_3 & p_4 & 
q_1 & q_2 & q_3 &
u_1 & u_2 & u_3 & u_4 & u_5 & 
u_6 & u_7 & u_8 & u_9 & u_{10} & 
u_{11} & u_{12} & u_{13} & u_{14} & u_{15} & 
u_{16} & u_{17} & u_{18} & 
v_1 & v_2 & v_3 & v_4 & v_5 & v_6 & v_7 & v_8 & v_9 &
o_1 & o_2 & o_3 & o_4 & o_5 & o_6 & o_7 & o_8 & o_9 \\
\hline
 1 & 1 & 1 & 0 & -1 & -1 & -1 & 0 & 0 & 0 & 0 & 0 & 0 & 0 & 0 & 0 & 0 & 0 & 0 & 0 & 0 & 0 & 0 & 0 & 0 & 0 & 0 & 0 & 0 & 0 & 0 & 0 & 0 & 0 & 0 & 0 & 0 & 0 & 0 & 0 & 0 & 0 & 0 \\
 0 & 0 & 0 & 1 & 2 & 2 & 0 & 3 & -1 & 0 & -1 & 0 & 0 & 0 & -1 & 0 & 0 & 0 & 0 & -1 & 0 & 0 & 0 & 0 & 0 & 0 & 0 & 0 & 0 & 0 & 0 & 0 & 0 & 0 & -1 & 0 & -1 & 0 & 0 & 0 & 0 & 0 & 0 \\
 0 & 0 & 0 & 0 & 2 & 2 & -1 & 4 & -1 & 0 & -1 & 0 & 0 & 0 & -1 & 0 & 0 & 0 & 0 & -1 & 0 & 0 & 0 & 0 & 0 & 0 & 0 & 0 & 0 & 0 & 0 & 0 & 0 & 1 & -1 & 0 & -1 & 0 & 0 & 0 & 0 & 0 & 0 \\
 0 & 0 & 0 & 0 & 2 & 1 & 0 & 3 & -1 & 0 & -1 & 0 & 0 & 0 & 0 & 0 & 0 & 0 & 0 & -1 & 0 & 0 & 0 & 0 & 0 & 0 & 0 & 0 & 0 & 1 & 0 & 0 & 0 & 0 & -1 & 0 & -1 & 0 & 0 & 0 & 0 & 0 & 0 \\
 0 & 0 & 0 & 0 & 1 & 2 & 0 & 3 & -1 & 0 & 0 & 0 & 0 & 0 & -1 & 0 & 0 & 0 & 0 & -1 & 0 & 0 & 0 & 0 & 0 & 0 & 0 & 0 & 0 & 0 & 0 & 0 & 1 & 0 & -1 & 0 & -1 & 0 & 0 & 0 & 0 & 0 & 0 \\
 0 & 0 & 0 & 0 & 1 & 1 & 0 & 2 & -1 & 0 & 0 & 0 & 0 & 0 & -1 & 0 & 0 & 0 & 0 & -1 & 0 & 0 & 0 & 0 & 0 & 0 & 0 & 0 & 0 & 0 & 0 & 1 & 0 & 0 & 0 & 0 & -1 & 0 & 0 & 0 & 0 & 0 & 0 \\
 0 & 0 & 0 & 0 & 1 & 1 & 0 & 2 & -1 & 0 & -1 & 0 & 0 & 0 & 0 & 0 & 0 & 0 & 0 & -1 & 0 & 0 & 0 & 0 & 0 & 0 & 0 & 0 & 1 & 0 & 0 & 0 & 0 & 0 & -1 & 0 & 0 & 0 & 0 & 0 & 0 & 0 & 0 \\
 0 & 0 & 0 & 0 & 1 & 1 & 0 & 1 & 0 & 0 & 0 & 0 & 0 & 0 & -1 & 0 & 0 & 0 & 0 & -1 & 0 & 0 & 0 & 0 & 0 & 0 & 0 & 0 & 0 & 0 & 1 & 0 & 0 & 0 & 0 & 0 & -1 & 0 & 0 & 0 & 0 & 0 & 0 \\
 0 & 0 & 0 & 0 & 1 & 1 & 0 & 1 & 0 & 0 & -1 & 0 & 0 & 0 & -1 & 0 & 0 & 0 & 0 & 0 & 0 & 0 & 0 & 0 & 0 & 0 & 1 & 0 & 0 & 0 & 0 & 0 & 0 & 0 & -1 & 0 & 0 & 0 & 0 & 0 & 0 & 0 & 0 \\
 0 & 0 & 0 & 0 & 1 & 1 & 0 & 1 & 0 & 0 & -1 & 0 & 0 & 0 & -1 & 0 & 0 & 0 & 0 & 0 & 0 & 0 & 0 & 0 & 0 & 0 & 0 & 1 & 0 & 0 & 0 & 0 & 0 & 0 & 0 & 0 & -1 & 0 & 0 & 0 & 0 & 0 & 0 \\
 0 & 0 & 0 & 0 & 1 & 1 & 0 & 1 & -1 & 0 & -1 & 0 & 0 & 0 & 0 & 0 & 0 & 0 & 0 & 0 & 0 & 0 & 0 & 0 & 0 & 1 & 0 & 0 & 0 & 0 & 0 & 0 & 0 & 0 & -1 & 0 & 0 & 0 & 0 & 0 & 0 & 0 & 0 \\
 0 & 0 & 0 & 0 & 1 & 1 & -1 & 2 & 0 & 0 & 0 & 0 & 0 & 0 & -1 & 0 & 0 & 0 & 0 & -1 & 0 & 0 & 0 & 0 & 1 & 0 & 0 & 0 & 0 & 0 & 0 & 0 & 0 & 0 & 0 & 0 & -1 & 0 & 0 & 0 & 0 & 0 & 0 \\
 0 & 0 & 0 & 0 & 1 & 1 & -1 & 2 & 0 & 0 & -1 & 0 & 0 & 0 & -1 & 0 & 1 & 0 & 0 & 0 & 0 & 0 & 0 & 0 & 0 & 0 & 0 & 0 & 0 & 0 & 0 & 0 & 0 & 0 & -1 & 0 & 0 & 0 & 0 & 0 & 0 & 0 & 0 \\
 0 & 0 & 0 & 0 & 1 & 1 & -1 & 2 & 0 & 0 & -1 & 0 & 0 & 0 & -1 & 0 & 0 & 0 & 1 & 0 & 0 & 0 & 0 & 0 & 0 & 0 & 0 & 0 & 0 & 0 & 0 & 0 & 0 & 0 & 0 & 0 & -1 & 0 & 0 & 0 & 0 & 0 & 0 \\
 0 & 0 & 0 & 0 & 1 & 1 & -1 & 2 & -1 & 0 & -1 & 0 & 1 & 0 & 0 & 0 & 0 & 0 & 0 & 0 & 0 & 0 & 0 & 0 & 0 & 0 & 0 & 0 & 0 & 0 & 0 & 0 & 0 & 0 & -1 & 0 & 0 & 0 & 0 & 0 & 0 & 0 & 0 \\
 0 & 0 & 0 & 0 & 1 & 0 & 0 & 2 & -1 & 0 & 0 & 0 & 0 & 0 & 0 & 0 & 0 & 0 & 0 & -1 & 0 & 0 & 0 & 1 & 0 & 0 & 0 & 0 & 0 & 0 & 0 & 0 & 0 & 0 & 0 & 0 & -1 & 0 & 0 & 0 & 0 & 0 & 0 \\
 0 & 0 & 0 & 0 & 1 & 0 & 0 & 1 & 0 & 0 & 0 & 0 & 0 & 0 & 0 & 0 & 0 & 0 & 0 & -1 & 0 & 0 & 1 & 0 & 0 & 0 & 0 & 0 & 0 & 0 & 0 & 0 & 0 & 0 & 0 & 0 & -1 & 0 & 0 & 0 & 0 & 0 & 0 \\
 0 & 0 & 0 & 0 & 1 & 0 & 0 & 1 & 0 & 0 & -1 & 1 & 0 & 0 & 0 & 0 & 0 & 0 & 0 & 0 & 0 & 0 & 0 & 0 & 0 & 0 & 0 & 0 & 0 & 0 & 0 & 0 & 0 & 0 & -1 & 0 & 0 & 0 & 0 & 0 & 0 & 0 & 0 \\
 0 & 0 & 0 & 0 & 1 & 0 & 0 & 1 & 0 & 0 & -1 & 0 & 0 & 1 & 0 & 0 & 0 & 0 & 0 & 0 & 0 & 0 & 0 & 0 & 0 & 0 & 0 & 0 & 0 & 0 & 0 & 0 & 0 & 0 & 0 & 0 & -1 & 0 & 0 & 0 & 0 & 0 & 0 \\
 0 & 0 & 0 & 0 & 1 & 0 & -1 & 2 & -1 & 0 & -1 & 0 & 0 & 0 & 0 & 0 & 0 & 0 & 0 & 0 & 0 & 0 & 0 & 0 & 0 & 0 & 0 & 0 & 0 & 0 & 0 & 0 & 0 & 0 & 0 & 0 & -1 & 0 & 1 & 0 & 0 & 0 & 0 \\
 0 & 0 & 0 & 0 & 0 & 1 & 0 & 2 & -1 & 0 & 0 & 0 & 0 & 0 & 0 & 0 & 0 & 0 & 0 & -1 & 0 & 1 & 0 & 0 & 0 & 0 & 0 & 0 & 0 & 0 & 0 & 0 & 0 & 0 & -1 & 0 & 0 & 0 & 0 & 0 & 0 & 0 & 0 \\
 0 & 0 & 0 & 0 & 0 & 1 & 0 & 1 & 0 & 0 & 0 & 0 & 0 & 0 & -1 & 1 & 0 & 0 & 0 & 0 & 0 & 0 & 0 & 0 & 0 & 0 & 0 & 0 & 0 & 0 & 0 & 0 & 0 & 0 & -1 & 0 & 0 & 0 & 0 & 0 & 0 & 0 & 0 \\
 0 & 0 & 0 & 0 & 0 & 1 & 0 & 1 & 0 & 0 & 0 & 0 & 0 & 0 & -1 & 0 & 0 & 1 & 0 & 0 & 0 & 0 & 0 & 0 & 0 & 0 & 0 & 0 & 0 & 0 & 0 & 0 & 0 & 0 & 0 & 0 & -1 & 0 & 0 & 0 & 0 & 0 & 0 \\
 0 & 0 & 0 & 0 & 0 & 1 & 0 & 1 & -1 & 1 & 0 & 0 & 0 & 0 & 0 & 0 & 0 & 0 & 0 & 0 & 0 & 0 & 0 & 0 & 0 & 0 & 0 & 0 & 0 & 0 & 0 & 0 & 0 & 0 & -1 & 0 & 0 & 0 & 0 & 0 & 0 & 0 & 0 \\
 0 & 0 & 0 & 0 & 0 & 1 & -1 & 2 & 0 & 0 & 0 & 0 & 0 & 0 & -1 & 0 & 0 & 0 & 0 & -1 & 0 & 0 & 0 & 0 & 0 & 0 & 0 & 0 & 0 & 0 & 0 & 0 & 0 & 0 & -1 & 0 & 0 & 0 & 0 & 0 & 0 & 0 & 1 \\
 0 & 0 & 0 & 0 & 0 & 0 & 0 & 1 & 0 & 0 & 0 & 0 & 0 & 0 & 0 & 0 & 0 & 0 & 0 & -1 & 0 & 0 & 0 & 0 & 0 & 0 & 0 & 0 & 0 & 0 & 0 & 0 & 0 & 0 & -1 & 0 & 0 & 0 & 0 & 0 & 1 & 0 & 0 \\
 0 & 0 & 0 & 0 & 0 & 0 & 0 & 1 & -1 & 0 & 0 & 0 & 0 & 0 & 0 & 0 & 0 & 0 & 0 & 0 & 0 & 0 & 0 & 0 & 0 & 0 & 0 & 0 & 0 & 0 & 0 & 0 & 0 & 0 & 0 & 0 & -1 & 1 & 0 & 0 & 0 & 0 & 0 \\
 0 & 0 & 0 & 0 & 0 & 0 & 0 & 1 & -1 & 0 & 0 & 0 & 0 & 0 & 0 & 0 & 0 & 0 & 0 & -1 & 1 & 0 & 0 & 0 & 0 & 0 & 0 & 0 & 0 & 0 & 0 & 0 & 0 & 0 & 0 & 0 & 0 & 0 & 0 & 0 & 0 & 0 & 0 \\
 0 & 0 & 0 & 0 & 0 & 0 & -1 & 1 & 0 & 0 & 0 & 0 & 0 & 0 & -1 & 0 & 0 & 0 & 0 & -1 & 0 & 0 & 0 & 0 & 0 & 0 & 0 & 0 & 0 & 0 & 0 & 0 & 0 & 0 & 0 & 0 & 0 & 0 & 0 & 0 & 0 & 1 & 0 \\
 0 & 0 & 0 & 0 & 0 & 0 & -1 & 1 & 0 & 0 & -1 & 0 & 0 & 0 & -1 & 0 & 0 & 0 & 0 & 0 & 0 & 0 & 0 & 0 & 0 & 0 & 0 & 0 & 0 & 0 & 0 & 0 & 0 & 0 & 0 & 0 & 0 & 0 & 0 & 1 & 0 & 0 & 0 \\
 0 & 0 & 0 & 0 & 0 & 0 & -1 & 1 & -1 & 0 & -1 & 0 & 0 & 0 & 0 & 0 & 0 & 0 & 0 & 0 & 0 & 0 & 0 & 0 & 0 & 0 & 0 & 0 & 0 & 0 & 0 & 0 & 0 & 0 & 0 & 1 & 0 & 0 & 0 & 0 & 0 & 0 & 0 \\
\end{array}
\right)
$}
}~,~
\nn\\
\eea
and also the charge matrix for the $D$-terms,
 \beal{es0403}
{\footnotesize
 Q_{D}=
 \resizebox{0.88\textwidth}{!}{$
\left(
\begin{array}{cccc|ccc|cccccccccccccccccc|ccccccccc|ccccccccc}
p_1 & p_2 & p_3 & p_4 & 
q_1 & q_2 & q_3 &
u_1 & u_2 & u_3 & u_4 & u_5 & 
u_6 & u_7 & u_8 & u_9 & u_{10} & 
u_{11} & u_{12} & u_{13} & u_{14} & u_{15} & 
u_{16} & u_{17} & u_{18} & 
v_1 & v_2 & v_3 & v_4 & v_5 & v_6 & v_7 & v_8 & v_9 &
o_1 & o_2 & o_3 & o_4 & o_5 & o_6 & o_7 & o_8 & o_9 \\
\hline
 0 & 0 & 0 & 0 & 1 & 0 & 0 & 1 & 0 & 0 & 0 & 0 & 0 & 0 & 0 & 0 & 0 & 0 & 0 & 0 & 0 & 0 & 0 & 0 & 0 & 0 & 0 & 0 & 0 & 0 & 0 & 0 & 0 & 0 & 0 & 0 & -1 & 0 & 0 & 0 & 0 & 0 & 0 \\
 0 & 0 & 0 & 0 & 0 & 1 & -1 & 1 & 0 & 0 & 0 & 0 & 0 & 0 & -1 & 0 & 0 & 0 & 0 & 0 & 0 & 0 & 0 & 0 & 0 & 0 & 0 & 0 & 0 & 0 & 0 & 0 & 0 & 0 & 0 & 0 & 0 & 0 & 0 & 0 & 0 & 0 & 0 \\
 0 & 0 & 0 & 0 & 0 & 0 & 0 & 1 & 0 & 0 & 0 & 0 & 0 & 0 & 0 & 0 & 0 & 0 & 0 & -1 & 0 & 0 & 0 & 0 & 0 & 0 & 0 & 0 & 0 & 0 & 0 & 0 & 0 & 0 & 0 & 0 & 0 & 0 & 0 & 0 & 0 & 0 & 0 \\
 0 & 0 & 0 & 0 & 0 & 0 & 0 & 0 & 0 & 0 & 0 & 0 & 0 & 0 & 0 & 0 & 0 & 0 & 0 & 0 & 0 & 0 & 0 & 0 & 0 & 0 & 0 & 0 & 0 & 0 & 0 & 0 & 0 & 0 & -1 & 0 & 1 & 0 & 0 & 0 & 0 & 0 & 0 \\
 0 & 0 & 0 & 0 & 0 & 0 & 0 & 0 & -1 & 0 & 0 & 0 & 0 & 0 & 1 & 0 & 0 & 0 & 0 & 0 & 0 & 0 & 0 & 0 & 0 & 0 & 0 & 0 & 0 & 0 & 0 & 0 & 0 & 0 & 0 & 0 & 0 & 0 & 0 & 0 & 0 & 0 & 0 \\
 0 & 0 & 0 & 0 & 0 & 0 & 0 & -1 & 1 & 0 & 0 & 0 & 0 & 0 & 0 & 0 & 0 & 0 & 0 & 0 & 0 & 0 & 0 & 0 & 0 & 0 & 0 & 0 & 0 & 0 & 0 & 0 & 0 & 0 & 0 & 0 & 0 & 0 & 0 & 0 & 0 & 0 & 0 \\
 0 & 0 & 0 & 0 & 0 & -1 & 0 & -1 & 0 & 0 & 0 & 0 & 0 & 0 & 0 & 0 & 0 & 0 & 0 & 0 & 0 & 0 & 0 & 0 & 0 & 0 & 0 & 0 & 0 & 0 & 0 & 0 & 0 & 0 & 1 & 0 & 0 & 0 & 0 & 0 & 0 & 0 & 0 \\
 0 & 0 & 0 & 0 & -1 & 0 & 1 & -1 & 0 & 0 & 1 & 0 & 0 & 0 & 0 & 0 & 0 & 0 & 0 & 0 & 0 & 0 & 0 & 0 & 0 & 0 & 0 & 0 & 0 & 0 & 0 & 0 & 0 & 0 & 0 & 0 & 0 & 0 & 0 & 0 & 0 & 0 & 0 \\
\end{array}
\right)
$}
}~.~
\nn\\
\eea
It can be seen that the charge matrices are invariant under permutations of the extremal GLSM fields $(p_1,p_2,p_3)$.
This is the result of the $SU(3)$ non-abelian factor in the global symmetry $SU(3)\times U(1)^2$ of the $2d$ $(0,2)$ theory.

The toric data is summarized by
 \beal{es0405}
{\footnotesize
 G_{t}=
 \resizebox{0.88\textwidth}{!}{$
\left(
\begin{array}{cccc|ccc|cccccccccccccccccc|ccccccccc|ccccccccc}
p_1 & p_2 & p_3 & p_4 & 
q_1 & q_2 & q_3 &
u_1 & u_2 & u_3 & u_4 & u_5 & 
u_6 & u_7 & u_8 & u_9 & u_{10} & 
u_{11} & u_{12} & u_{13} & u_{14} & u_{15} & 
u_{16} & u_{17} & u_{18} & 
v_1 & v_2 & v_3 & v_4 & v_5 & v_6 & v_7 & v_8 & v_9 &
o_1 & o_2 & o_3 & o_4 & o_5 & o_6 & o_7 & o_8 & o_9 \\
\hline
 1 & 1 & 1 & 1 & 1 & 1 & 1 & 1 & 1 & 1 & 1 & 1 & 1 & 1 & 1 & 1 & 1 & 1 & 1 & 1 & 1 & 1 & 1 & 1 & 1 & 1 & 1 & 1 & 1 & 1 & 1 & 1 & 1 & 1 & 2 & 2 & 2 & 2 & 2 & 2 & 2 & 2 & 2 \\
 1 & 0 & -1 & 0 & 0 & 0 & 0 & 0 & 0 & 0 & 0 & 0 & 0 & 0 & 0 & 0 & 0 & 0 & 0 & 0 & 0 & 0 & 0 & 0 & 0 & 0 & 0 & 0 & 0 & 0 & 0 & 0 & 0 & 0 & 0 & 0 & 0 & 0 & 0 & 0 & 0 & 0 & 0 \\
 0 & 1 & -1 & 0 & 0 & 0 & 0 & 0 & 0 & 0 & 0 & 0 & 0 & 0 & 0 & 0 & 0 & 0 & 0 & 0 & 0 & 0 & 0 & 0 & 0 & 0 & 0 & 0 & 0 & 0 & 0 & 0 & 0 & 0 & 0 & 0 & 0 & 0 & 0 & 0 & 0 & 0 & 0 \\
 0 & 0 & 0 & 3 & 0 & 0 & 0 & 1 & 1 & 1 & 1 & 1 & 1 & 1 & 1 & 1 & 1 & 1 & 1 & 1 & 1 & 1 & 1 & 1 & 1 & 2 & 2 & 2 & 2 & 2 & 2 & 2 & 2 & 2 & 1 & 1 & 1 & 1 & 1 & 1 & 1 & 1 & 1 \\
\end{array}
\right)
$}
}~,~
\nn\\
\eea
where $o_k$ correspond to extra GLSM fields \cite{Franco:2015tna}.
The corresponding toric diagram is shown in \fref{f_toric_cp2_20}.
We note that the $2d$ $(0,2)$ theories for $Y^{3,0}(\mathbb{CP}^2)$ can be obtained by orbifold reduction \cite{Franco:2016fxm} of the $4d$ $\mathcal{N}=1$ theory for $dP_0$. 
\\

\subsubsection{$Y^{3,1}(\mathbb{CP}^2)$ \label{sec:cp21}}

\begin{figure}[H]
\begin{center}
\resizebox{0.4\hsize}{!}{
\includegraphics[height=6cm]{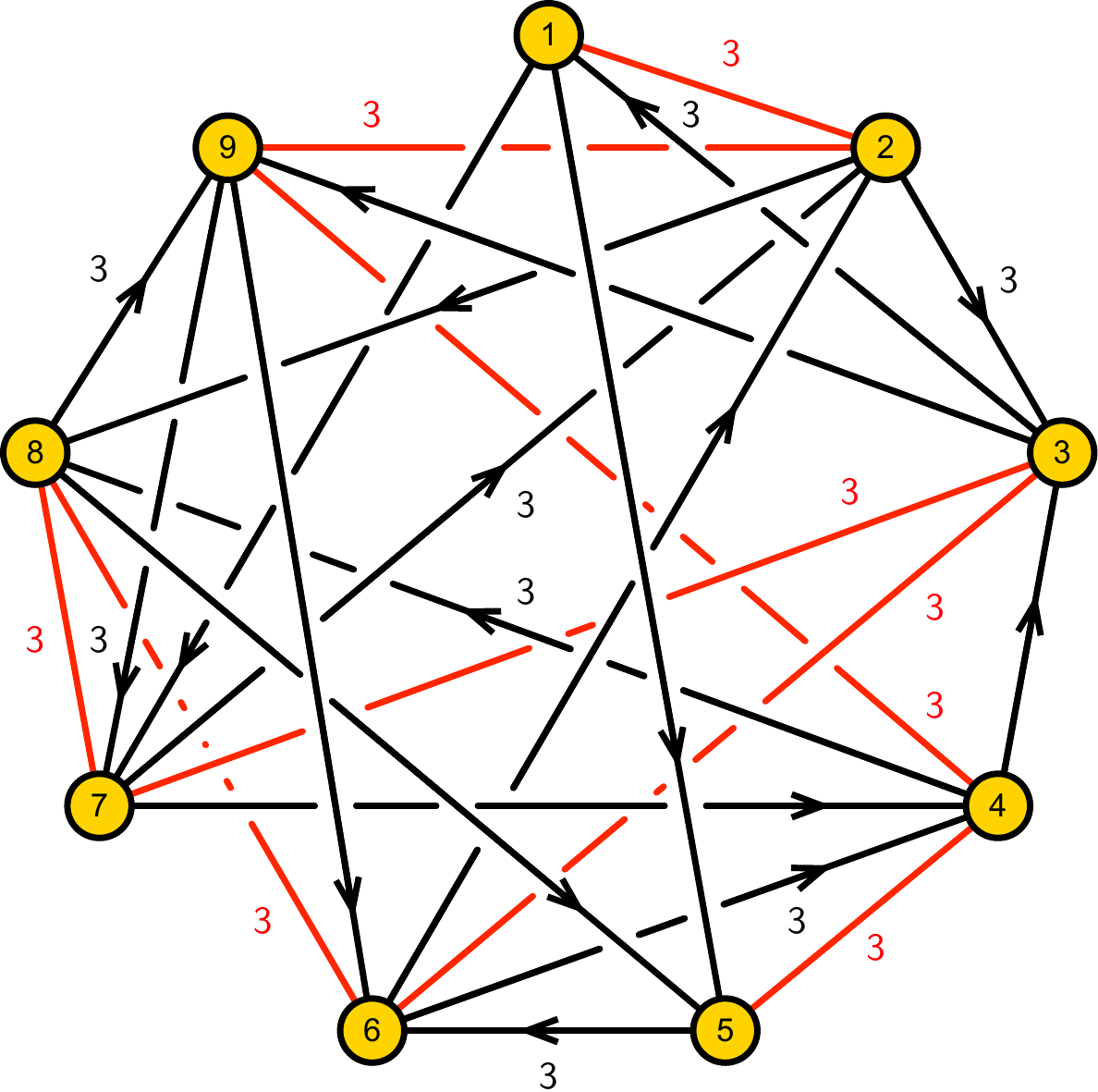} 
}
\caption{
Quiver for $Y^{3,1}(\mathbb{CP}^2)$.
\label{f_quiver_cp2_31}}
 \end{center}
 \end{figure}

The quiver for the $2d$ $(0,2)$ theory corresponding to $Y^{3,1}(\mathbb{CP}^2)$ is shown in \fref{f_quiver_cp2_31}.

\noindent
The corresponding $J$- and $E$-terms take the following form
  \beq
\begin{array}{rccrcl}
 &   J &   & & E &  \\
 \Lambda_{12}^{i}  : & \epsilon_{ijk} X^{j}_{23} X^{k}_{31 } &~&  P_{15} X^{i}_{56} Q_{62 } &-&  Q_{17} X^{i}_{72 } \\
 \Lambda_{29}^{2i}  : & \epsilon_{ijk} X^{j}_{97} X^{k}_{72 } &~&  X^{i}_{23} Q_{39 } &-&  Q_{28} X^{i}_{89 } \\
 \Lambda_{37}^{2i}  : & \epsilon_{ijk}  X^{j}_{72} X^{k}_{23 } &~&  X^{i}_{31} Q_{17 } &-&  Q_{39} X^{i}_{97} \\
 \Lambda_{45}^{i}  : & \epsilon_{ijk} X^{j}_{56} X^{k}_{64 } &~&  X^{i}_{48} Q_{85 } &-&   Q_{43} X^{i}_{31} P_{15} \\
 \Lambda_{63}^{2i}  : & \epsilon_{ijk} X^{j}_{31} P_{15} X^{k}_{56 } &~&  X^{i}_{64} Q_{43 } &-&  Q_{62} X^{i}_{23} \\
 \Lambda_{78}^{i}  : & \epsilon_{ijk} X^{j}_{89} X^{k}_{97 } &~&  X^{i}_{72} Q_{28 } &-&  Q_{74} X^{i}_{48 } \\
 \Lambda_{86}^{2i}  : & \epsilon_{ijk} X^{j}_{64} X^{k}_{48 } &~&  X^{i}_{89} Q_{96 } &-&  Q_{85} X^{i}_{56} \\
 \Lambda_{94}^{2i}  : & \epsilon_{ijk} X^{j}_{48} X^{k}_{89 } &~&  X^{i}_{97} Q_{74 } &-&  Q_{96} X^{i}_{64} \\
 \end{array} 
~,~
\label{es0500}
 \eeq
where the global symmetry indices are $i,j,k=1,2,3$.
 
 The $J$- and $E$-terms can be used to obtain the $P$-matrix for the theory, which takes the following form
 \beal{es0501}
{\footnotesize
 P=
\resizebox{0.92\textwidth}{!}{$
\left(
\begin{array}{r|ccccc|cccccccccccccccc|ccccccccc|ccccccccccc}
\; & p_1 & p_2 & p_3 & p_4 & p_5 &
u_1 & u_2 & u_3 & u_4 & u_5 & u_6 & u_7 & u_8 & u_9 & u_{10} &
u_{11} & u_{12} & u_{13} & u_{14} & u_{15} & u_{16} & 
v_1 & v_2 & v_3 & v_4 & v_5 & v_6 & v_7 & v_8 & v_9 &
o_1 & o_2 & o_3 & o_4 & o_5 & o_6 & o_7 & o_8 & o_9 & o_{10} & o_{11}
\\
\hline
X_{23}^{1} & 1 & 0 & 0 & 0 & 0 & 0 & 0 & 0 & 1 & 0 & 0 & 1 & 1 & 0 & 0 & 0 & 0 & 0 & 0 & 0 & 0 & 0 & 1 & 0 & 0 & 0 & 0 & 0 & 0 & 0 & 0 & 1 & 1 & 0 & 1 & 0 & 1 & 1 & 1 & 1 & 1 \\
X_{31}^{1} & 1 & 0 & 0 & 0 & 0 & 1 & 0 & 0 & 0 & 0 & 1 & 0 & 0 & 1 & 0 & 0 & 0 & 0 & 0 & 0 & 0 & 0 & 0 & 1 & 0 & 0 & 0 & 0 & 0 & 0 & 1 & 1 & 0 & 1 & 0 & 1 & 1 & 0 & 0 & 0 & 0 \\
X_{48}^{1} & 1 & 0 & 0 & 0 & 1 & 0 & 0 & 0 & 0 & 1 & 0 & 0 & 1 & 1 & 0 & 0 & 0 & 0 & 0 & 0 & 1 & 0 & 0 & 0 & 0 & 0 & 0 & 0 & 1 & 0 & 1 & 0 & 0 & 1 & 1 & 1 & 1 & 0 & 0 & 0 & 1 \\
X_{56}^{1} & 1 & 0 & 0 & 0 & 0 & 0 & 1 & 0 & 0 & 0 & 0 & 0 & 0 & 0 & 1 & 0 & 1 & 0 & 0 & 0 & 0 & 0 & 0 & 0 & 0 & 0 & 1 & 0 & 0 & 0 & 1 & 0 & 1 & 0 & 1 & 1 & 0 & 1 & 0 & 0 & 0 \\
X_{64}^{1} & 1 & 0 & 0 & 0 & 0 & 0 & 0 & 1 & 0 & 0 & 0 & 1 & 0 & 0 & 0 & 1 & 0 & 0 & 0 & 0 & 0 & 0 & 0 & 0 & 0 & 0 & 0 & 1 & 0 & 0 & 0 & 1 & 1 & 1 & 0 & 0 & 1 & 1 & 1 & 1 & 1 \\
X_{72}^{1} & 1 & 0 & 0 & 0 & 1 & 0 & 1 & 1 & 0 & 1 & 0 & 0 & 0 & 0 & 0 & 0 & 0 & 1 & 0 & 0 & 0 & 1 & 0 & 0 & 0 & 0 & 0 & 0 & 0 & 0 & 1 & 0 & 1 & 1 & 1 & 1 & 0 & 0 & 1 & 0 & 0 \\
X_{89}^{1} & 1 & 0 & 0 & 0 & 0 & 0 & 0 & 0 & 1 & 0 & 1 & 0 & 0 & 0 & 0 & 0 & 1 & 0 & 0 & 1 & 0 & 0 & 0 & 0 & 0 & 1 & 0 & 0 & 0 & 0 & 0 & 1 & 1 & 0 & 1 & 1 & 0 & 1 & 1 & 1 & 0 \\
X_{97}^{1} & 1 & 0 & 0 & 0 & 0 & 1 & 0 & 0 & 0 & 0 & 0 & 0 & 0 & 0 & 1 & 1 & 0 & 0 & 1 & 0 & 0 & 0 & 0 & 0 & 1 & 0 & 0 & 0 & 0 & 0 & 1 & 1 & 0 & 1 & 0 & 0 & 1 & 1 & 0 & 1 & 1 \\
\hline
X_{23}^{2} & 0 & 1 & 0 & 0 & 0 & 0 & 0 & 0 & 1 & 0 & 0 & 1 & 1 & 0 & 0 & 0 & 0 & 0 & 0 & 0 & 0 & 0 & 1 & 0 & 0 & 0 & 0 & 0 & 0 & 0 & 0 & 1 & 1 & 0 & 1 & 0 & 1 & 1 & 1 & 1 & 1 \\
X_{31}^{2} & 0 & 1 & 0 & 0 & 0 & 1 & 0 & 0 & 0 & 0 & 1 & 0 & 0 & 1 & 0 & 0 & 0 & 0 & 0 & 0 & 0 & 0 & 0 & 1 & 0 & 0 & 0 & 0 & 0 & 0 & 1 & 1 & 0 & 1 & 0 & 1 & 1 & 0 & 0 & 0 & 0 \\
X_{48}^{2} & 0 & 1 & 0 & 0 & 1 & 0 & 0 & 0 & 0 & 1 & 0 & 0 & 1 & 1 & 0 & 0 & 0 & 0 & 0 & 0 & 1 & 0 & 0 & 0 & 0 & 0 & 0 & 0 & 1 & 0 & 1 & 0 & 0 & 1 & 1 & 1 & 1 & 0 & 0 & 0 & 1 \\
X_{56}^{2} & 0 & 1 & 0 & 0 & 0 & 0 & 1 & 0 & 0 & 0 & 0 & 0 & 0 & 0 & 1 & 0 & 1 & 0 & 0 & 0 & 0 & 0 & 0 & 0 & 0 & 0 & 1 & 0 & 0 & 0 & 1 & 0 & 1 & 0 & 1 & 1 & 0 & 1 & 0 & 0 & 0 \\
X_{64}^{2} & 0 & 1 & 0 & 0 & 0 & 0 & 0 & 1 & 0 & 0 & 0 & 1 & 0 & 0 & 0 & 1 & 0 & 0 & 0 & 0 & 0 & 0 & 0 & 0 & 0 & 0 & 0 & 1 & 0 & 0 & 0 & 1 & 1 & 1 & 0 & 0 & 1 & 1 & 1 & 1 & 1 \\
X_{72}^{2} & 0 & 1 & 0 & 0 & 1 & 0 & 1 & 1 & 0 & 1 & 0 & 0 & 0 & 0 & 0 & 0 & 0 & 1 & 0 & 0 & 0 & 1 & 0 & 0 & 0 & 0 & 0 & 0 & 0 & 0 & 1 & 0 & 1 & 1 & 1 & 1 & 0 & 0 & 1 & 0 & 0 \\
X_{89}^{2} & 0 & 1 & 0 & 0 & 0 & 0 & 0 & 0 & 1 & 0 & 1 & 0 & 0 & 0 & 0 & 0 & 1 & 0 & 0 & 1 & 0 & 0 & 0 & 0 & 0 & 1 & 0 & 0 & 0 & 0 & 0 & 1 & 1 & 0 & 1 & 1 & 0 & 1 & 1 & 1 & 0 \\
X_{97}^{2} & 0 & 1 & 0 & 0 & 0 & 1 & 0 & 0 & 0 & 0 & 0 & 0 & 0 & 0 & 1 & 1 & 0 & 0 & 1 & 0 & 0 & 0 & 0 & 0 & 1 & 0 & 0 & 0 & 0 & 0 & 1 & 1 & 0 & 1 & 0 & 0 & 1 & 1 & 0 & 1 & 1 \\
\hline
X_{23}^{3} & 0 & 0 & 1 & 0 & 0 & 0 & 0 & 0 & 1 & 0 & 0 & 1 & 1 & 0 & 0 & 0 & 0 & 0 & 0 & 0 & 0 & 0 & 1 & 0 & 0 & 0 & 0 & 0 & 0 & 0 & 0 & 1 & 1 & 0 & 1 & 0 & 1 & 1 & 1 & 1 & 1 \\
X_{31}^{3} & 0 & 0 & 1 & 0 & 0 & 1 & 0 & 0 & 0 & 0 & 1 & 0 & 0 & 1 & 0 & 0 & 0 & 0 & 0 & 0 & 0 & 0 & 0 & 1 & 0 & 0 & 0 & 0 & 0 & 0 & 1 & 1 & 0 & 1 & 0 & 1 & 1 & 0 & 0 & 0 & 0 \\
X_{48}^{3} & 0 & 0 & 1 & 0 & 1 & 0 & 0 & 0 & 0 & 1 & 0 & 0 & 1 & 1 & 0 & 0 & 0 & 0 & 0 & 0 & 1 & 0 & 0 & 0 & 0 & 0 & 0 & 0 & 1 & 0 & 1 & 0 & 0 & 1 & 1 & 1 & 1 & 0 & 0 & 0 & 1 \\
X_{56}^{3} & 0 & 0 & 1 & 0 & 0 & 0 & 1 & 0 & 0 & 0 & 0 & 0 & 0 & 0 & 1 & 0 & 1 & 0 & 0 & 0 & 0 & 0 & 0 & 0 & 0 & 0 & 1 & 0 & 0 & 0 & 1 & 0 & 1 & 0 & 1 & 1 & 0 & 1 & 0 & 0 & 0 \\
X_{64}^{3} & 0 & 0 & 1 & 0 & 0 & 0 & 0 & 1 & 0 & 0 & 0 & 1 & 0 & 0 & 0 & 1 & 0 & 0 & 0 & 0 & 0 & 0 & 0 & 0 & 0 & 0 & 0 & 1 & 0 & 0 & 0 & 1 & 1 & 1 & 0 & 0 & 1 & 1 & 1 & 1 & 1 \\
X_{72}^{3} & 0 & 0 & 1 & 0 & 1 & 0 & 1 & 1 & 0 & 1 & 0 & 0 & 0 & 0 & 0 & 0 & 0 & 1 & 0 & 0 & 0 & 1 & 0 & 0 & 0 & 0 & 0 & 0 & 0 & 0 & 1 & 0 & 1 & 1 & 1 & 1 & 0 & 0 & 1 & 0 & 0 \\
X_{89}^{3} & 0 & 0 & 1 & 0 & 0 & 0 & 0 & 0 & 1 & 0 & 1 & 0 & 0 & 0 & 0 & 0 & 1 & 0 & 0 & 1 & 0 & 0 & 0 & 0 & 0 & 1 & 0 & 0 & 0 & 0 & 0 & 1 & 1 & 0 & 1 & 1 & 0 & 1 & 1 & 1 & 0 \\
X_{97}^{3} & 0 & 0 & 1 & 0 & 0 & 1 & 0 & 0 & 0 & 0 & 0 & 0 & 0 & 0 & 1 & 1 & 0 & 0 & 1 & 0 & 0 & 0 & 0 & 0 & 1 & 0 & 0 & 0 & 0 & 0 & 1 & 1 & 0 & 1 & 0 & 0 & 1 & 1 & 0 & 1 & 1 \\
\hline
Q_{17} & 0 & 0 & 0 & 1 & 0 & 0 & 0 & 0 & 0 & 0 & 0 & 0 & 0 & 0 & 1 & 1 & 1 & 0 & 1 & 1 & 1 & 0 & 0 & 0 & 1 & 1 & 1 & 1 & 1 & 1 & 0 & 0 & 0 & 0 & 0 & 0 & 0 & 1 & 0 & 1 & 1 \\
Q_{28} & 0 & 0 & 0 & 1 & 0 & 0 & 0 & 0 & 0 & 0 & 0 & 1 & 1 & 1 & 0 & 0 & 0 & 0 & 0 & 0 & 1 & 0 & 1 & 1 & 0 & 0 & 1 & 1 & 1 & 1 & 0 & 0 & 0 & 0 & 0 & 0 & 1 & 0 & 0 & 0 & 1 \\
Q_{39} & 0 & 0 & 0 & 1 & 0 & 0 & 0 & 0 & 0 & 0 & 1 & 0 & 0 & 1 & 0 & 0 & 1 & 0 & 0 & 1 & 1 & 0 & 0 & 1 & 0 & 1 & 1 & 1 & 1 & 1 & 0 & 0 & 0 & 0 & 0 & 1 & 0 & 0 & 0 & 0 & 0 \\
Q_{43} & 0 & 0 & 0 & 1 & 0 & 0 & 0 & 0 & 1 & 1 & 0 & 0 & 1 & 0 & 0 & 0 & 0 & 0 & 0 & 0 & 0 & 1 & 1 & 0 & 1 & 1 & 0 & 0 & 1 & 0 & 0 & 0 & 0 & 0 & 1 & 0 & 0 & 0 & 0 & 0 & 0 \\
Q_{62} & 0 & 0 & 0 & 1 & 0 & 0 & 0 & 1 & 0 & 1 & 0 & 0 & 0 & 0 & 0 & 1 & 0 & 0 & 0 & 0 & 0 & 1 & 0 & 0 & 1 & 1 & 0 & 1 & 1 & 0 & 0 & 0 & 0 & 1 & 0 & 0 & 0 & 0 & 0 & 0 & 0 \\
Q_{74} & 0 & 0 & 0 & 1 & 0 & 0 & 1 & 1 & 0 & 0 & 0 & 1 & 0 & 0 & 0 & 0 & 0 & 1 & 0 & 0 & 0 & 1 & 1 & 1 & 0 & 0 & 1 & 1 & 0 & 1 & 0 & 0 & 1 & 0 & 0 & 0 & 0 & 0 & 1 & 0 & 0 \\
Q_{85} & 0 & 0 & 0 & 1 & 0 & 1 & 0 & 0 & 1 & 0 & 1 & 0 & 0 & 0 & 0 & 0 & 0 & 1 & 1 & 1 & 0 & 1 & 1 & 1 & 1 & 1 & 0 & 0 & 0 & 1 & 0 & 1 & 0 & 0 & 0 & 0 & 0 & 0 & 1 & 1 & 0 \\
Q_{96} & 0 & 0 & 0 & 1 & 0 & 1 & 1 & 0 & 0 & 0 & 0 & 0 & 0 & 0 & 1 & 0 & 0 & 1 & 1 & 0 & 0 & 1 & 1 & 1 & 1 & 0 & 1 & 0 & 0 & 1 & 1 & 0 & 0 & 0 & 0 & 0 & 0 & 0 & 0 & 0 & 0 \\
\hline
P_{15} & 0 & 0 & 0 & 0 & 1 & 0 & 0 & 0 & 0 & 0 & 0 & 0 & 0 & 0 & 0 & 0 & 0 & 1 & 1 & 1 & 1 & 0 & 0 & 0 & 0 & 0 & 0 & 0 & 0 & 1 & 0 & 0 & 0 & 0 & 0 & 0 & 0 & 0 & 1 & 1 & 1 \\
\end{array}
\right)
$}
}~.~
\nn\\
\eea

\begin{figure}[ht!!]
\begin{center}
\resizebox{0.35\hsize}{!}{
\includegraphics[height=6cm]{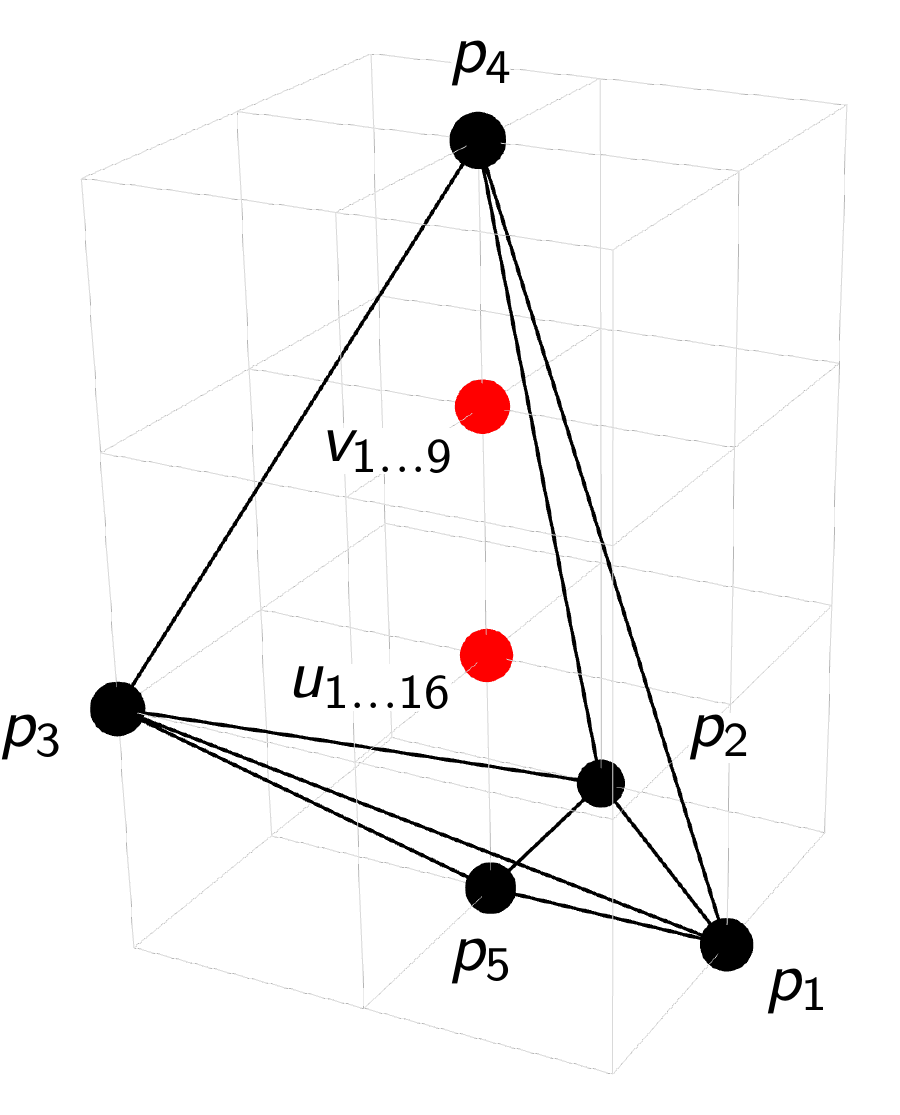} 
}
\caption{
Toric diagram of $Y^{3,1}(\mathbb{CP}^2)$.
\label{f_toric_cp2_31}}
 \end{center}
 \end{figure}

The $U(1)$-charges under the $J$- and $E$-terms are summarized in
 \beal{es0502}
{\footnotesize
 Q_{JE}=
 \resizebox{0.88\textwidth}{!}{$
\left(
\begin{array}{ccccc|cccccccccccccccc|ccccccccc|ccccccccccc}
p_1 & p_2 & p_3 & p_4 & p_5 &
u_1 & u_2 & u_3 & u_4 & u_5 & u_6 & u_7 & u_8 & u_9 & u_{10} &
u_{11} & u_{12} & u_{13} & u_{14} & u_{15} & u_{16} & 
v_1 & v_2 & v_3 & v_4 & v_5 & v_6 & v_7 & v_8 & v_9 &
o_1 & o_2 & o_3 & o_4 & o_5 & o_6 & o_7 & o_8 & o_9 & o_{10} & o_{11}
\\
\hline
5 & 5 & 5 & 1 & 0 & 0 & 0 & 0 & 0 & 0 & 0 & 0 & 0 & 0 & 0 & 0 & 0 & 0 & 0 & 0 & 0 & 0 & 0 & 0 & 0 & 0 & 0 & 0 & 0 & 0 & -1 & -1 & -1 & -1 & -1 & -1 & -1 & -1 & 0 & 0 & 0 \\
 4 & 4 & 4 & 0 & -1 & 0 & 0 & 0 & 0 & 0 & 0 & 0 & 0 & 0 & 0 & 0 & 0 & 0 & 0 & 0 & 0 & 0 & 0 & 0 & 0 & 0 & 0 & 0 & 0 & 1 & -1 & -1 & -1 & 0 & 0 & -1 & -1 & -1 & 0 & 0 & 0 \\
 3 & 3 & 3 & 0 & 0 & 0 & 0 & 0 & 0 & 0 & 0 & 0 & 0 & 0 & 0 & 0 & 0 & 0 & 0 & 0 & 0 & 1 & 0 & 0 & 0 & 0 & 0 & 0 & 0 & 0 & -1 & -1 & -1 & -1 & -1 & 0 & 0 & 0 & 0 & 0 & 0 \\
 3 & 3 & 3 & 0 & 0 & 0 & 0 & 0 & 0 & 0 & 0 & 0 & 0 & 0 & 0 & 0 & 0 & 0 & 0 & 0 & 0 & 0 & 1 & 0 & 0 & 0 & 0 & 0 & 0 & 0 & -1 & -1 & -1 & 0 & -1 & 0 & -1 & 0 & 0 & 0 & 0 \\
 3 & 3 & 3 & 0 & 0 & 0 & 0 & 0 & 0 & 0 & 0 & 0 & 0 & 0 & 0 & 0 & 0 & 0 & 0 & 0 & 0 & 0 & 0 & 1 & 0 & 0 & 0 & 0 & 0 & 0 & -1 & -1 & -1 & 0 & 0 & -1 & -1 & 0 & 0 & 0 & 0 \\
 3 & 3 & 3 & 0 & 0 & 0 & 0 & 0 & 0 & 0 & 0 & 0 & 0 & 0 & 0 & 0 & 0 & 0 & 0 & 0 & 0 & 0 & 0 & 0 & 1 & 0 & 0 & 0 & 0 & 0 & -1 & -1 & 0 & -1 & -1 & 0 & 0 & -1 & 0 & 0 & 0 \\
 3 & 3 & 3 & 0 & 0 & 0 & 0 & 0 & 0 & 0 & 0 & 0 & 0 & 0 & 0 & 0 & 0 & 0 & 0 & 0 & 0 & 0 & 0 & 0 & 0 & 1 & 0 & 0 & 0 & 0 & 0 & -1 & 0 & -1 & -1 & -1 & 0 & -1 & 0 & 0 & 0 \\
 3 & 3 & 3 & 0 & 0 & 0 & 0 & 0 & 0 & 0 & 0 & 0 & 0 & 0 & 0 & 0 & 0 & 0 & 0 & 0 & 0 & 0 & 0 & 0 & 0 & 0 & 1 & 0 & 0 & 0 & -1 & 0 & -1 & 0 & 0 & -1 & -1 & -1 & 0 & 0 & 0 \\
 3 & 3 & 3 & 0 & 0 & 0 & 0 & 0 & 0 & 0 & 0 & 0 & 0 & 0 & 0 & 0 & 0 & 0 & 0 & 0 & 0 & 0 & 0 & 0 & 0 & 0 & 0 & 1 & 0 & 0 & 0 & 0 & -1 & -1 & 0 & -1 & -1 & -1 & 0 & 0 & 0 \\
 3 & 3 & 3 & 0 & 0 & 0 & 0 & 0 & 0 & 0 & 0 & 0 & 0 & 0 & 0 & 0 & 0 & 0 & 0 & 0 & 0 & 0 & 0 & 0 & 0 & 0 & 0 & 0 & 1 & 0 & 0 & 0 & 0 & -1 & -1 & -1 & -1 & -1 & 0 & 0 & 0 \\
 2 & 2 & 2 & 0 & -1 & 0 & 0 & 0 & 0 & 0 & 0 & 0 & 0 & 0 & 0 & 0 & 0 & 1 & 0 & 0 & 0 & 0 & 0 & 0 & 0 & 0 & 0 & 0 & 0 & 0 & -1 & -1 & -1 & 0 & 0 & 0 & 0 & 0 & 0 & 0 & 0 \\
 2 & 2 & 2 & 0 & -1 & 0 & 0 & 0 & 0 & 0 & 0 & 0 & 0 & 0 & 0 & 0 & 0 & 0 & 1 & 0 & 0 & 0 & 0 & 0 & 0 & 0 & 0 & 0 & 0 & 0 & -1 & -1 & 0 & 0 & 0 & 0 & 0 & -1 & 0 & 0 & 0 \\
 2 & 2 & 2 & 0 & -1 & 0 & 0 & 0 & 0 & 0 & 0 & 0 & 0 & 0 & 0 & 0 & 0 & 0 & 0 & 1 & 0 & 0 & 0 & 0 & 0 & 0 & 0 & 0 & 0 & 0 & 0 & -1 & 0 & 0 & 0 & -1 & 0 & -1 & 0 & 0 & 0 \\
 2 & 2 & 2 & 0 & -1 & 0 & 0 & 0 & 0 & 0 & 0 & 0 & 0 & 0 & 0 & 0 & 0 & 0 & 0 & 0 & 1 & 0 & 0 & 0 & 0 & 0 & 0 & 0 & 0 & 0 & 0 & 0 & 0 & 0 & 0 & -1 & -1 & -1 & 0 & 0 & 0 \\
 1 & 1 & 1 & 0 & 0 & 1 & 0 & 0 & 0 & 0 & 0 & 0 & 0 & 0 & 0 & 0 & 0 & 0 & 0 & 0 & 0 & 0 & 0 & 0 & 0 & 0 & 0 & 0 & 0 & 0 & -1 & -1 & 0 & 0 & 0 & 0 & 0 & 0 & 0 & 0 & 0 \\
 1 & 1 & 1 & 0 & 0 & 0 & 1 & 0 & 0 & 0 & 0 & 0 & 0 & 0 & 0 & 0 & 0 & 0 & 0 & 0 & 0 & 0 & 0 & 0 & 0 & 0 & 0 & 0 & 0 & 0 & -1 & 0 & -1 & 0 & 0 & 0 & 0 & 0 & 0 & 0 & 0 \\
 1 & 1 & 1 & 0 & 0 & 0 & 0 & 1 & 0 & 0 & 0 & 0 & 0 & 0 & 0 & 0 & 0 & 0 & 0 & 0 & 0 & 0 & 0 & 0 & 0 & 0 & 0 & 0 & 0 & 0 & 0 & 0 & -1 & -1 & 0 & 0 & 0 & 0 & 0 & 0 & 0 \\
 1 & 1 & 1 & 0 & 0 & 0 & 0 & 0 & 1 & 0 & 0 & 0 & 0 & 0 & 0 & 0 & 0 & 0 & 0 & 0 & 0 & 0 & 0 & 0 & 0 & 0 & 0 & 0 & 0 & 0 & 0 & -1 & 0 & 0 & -1 & 0 & 0 & 0 & 0 & 0 & 0 \\
 1 & 1 & 1 & 0 & 0 & 0 & 0 & 0 & 0 & 1 & 0 & 0 & 0 & 0 & 0 & 0 & 0 & 0 & 0 & 0 & 0 & 0 & 0 & 0 & 0 & 0 & 0 & 0 & 0 & 0 & 0 & 0 & 0 & -1 & -1 & 0 & 0 & 0 & 0 & 0 & 0 \\
 1 & 1 & 1 & 0 & 0 & 0 & 0 & 0 & 0 & 0 & 1 & 0 & 0 & 0 & 0 & 0 & 0 & 0 & 0 & 0 & 0 & 0 & 0 & 0 & 0 & 0 & 0 & 0 & 0 & 0 & 0 & -1 & 0 & 0 & 0 & -1 & 0 & 0 & 0 & 0 & 0 \\
 1 & 1 & 1 & 0 & 0 & 0 & 0 & 0 & 0 & 0 & 0 & 1 & 0 & 0 & 0 & 0 & 0 & 0 & 0 & 0 & 0 & 0 & 0 & 0 & 0 & 0 & 0 & 0 & 0 & 0 & 0 & 0 & -1 & 0 & 0 & 0 & -1 & 0 & 0 & 0 & 0 \\
 1 & 1 & 1 & 0 & 0 & 0 & 0 & 0 & 0 & 0 & 0 & 0 & 1 & 0 & 0 & 0 & 0 & 0 & 0 & 0 & 0 & 0 & 0 & 0 & 0 & 0 & 0 & 0 & 0 & 0 & 0 & 0 & 0 & 0 & -1 & 0 & -1 & 0 & 0 & 0 & 0 \\
 1 & 1 & 1 & 0 & 0 & 0 & 0 & 0 & 0 & 0 & 0 & 0 & 0 & 1 & 0 & 0 & 0 & 0 & 0 & 0 & 0 & 0 & 0 & 0 & 0 & 0 & 0 & 0 & 0 & 0 & 0 & 0 & 0 & 0 & 0 & -1 & -1 & 0 & 0 & 0 & 0 \\
 1 & 1 & 1 & 0 & 0 & 0 & 0 & 0 & 0 & 0 & 0 & 0 & 0 & 0 & 1 & 0 & 0 & 0 & 0 & 0 & 0 & 0 & 0 & 0 & 0 & 0 & 0 & 0 & 0 & 0 & -1 & 0 & 0 & 0 & 0 & 0 & 0 & -1 & 0 & 0 & 0 \\
 1 & 1 & 1 & 0 & 0 & 0 & 0 & 0 & 0 & 0 & 0 & 0 & 0 & 0 & 0 & 1 & 0 & 0 & 0 & 0 & 0 & 0 & 0 & 0 & 0 & 0 & 0 & 0 & 0 & 0 & 0 & 0 & 0 & -1 & 0 & 0 & 0 & -1 & 0 & 0 & 0 \\
 1 & 1 & 1 & 0 & 0 & 0 & 0 & 0 & 0 & 0 & 0 & 0 & 0 & 0 & 0 & 0 & 1 & 0 & 0 & 0 & 0 & 0 & 0 & 0 & 0 & 0 & 0 & 0 & 0 & 0 & 0 & 0 & 0 & 0 & 0 & -1 & 0 & -1 & 0 & 0 & 0 \\
 1 & 1 & 1 & 0 & -1 & 0 & 0 & 0 & 0 & 0 & 0 & 0 & 0 & 0 & 0 & 0 & 0 & 0 & 0 & 0 & 0 & 0 & 0 & 0 & 0 & 0 & 0 & 0 & 0 & 0 & 0 & 0 & 0 & 0 & 0 & 0 & -1 & -1 & 0 & 0 & 1 \\
 1 & 1 & 1 & 0 & -1 & 0 & 0 & 0 & 0 & 0 & 0 & 0 & 0 & 0 & 0 & 0 & 0 & 0 & 0 & 0 & 0 & 0 & 0 & 0 & 0 & 0 & 0 & 0 & 0 & 0 & 0 & -1 & 0 & 0 & 0 & 0 & 0 & -1 & 0 & 1 & 0 \\
 1 & 1 & 1 & 0 & -1 & 0 & 0 & 0 & 0 & 0 & 0 & 0 & 0 & 0 & 0 & 0 & 0 & 0 & 0 & 0 & 0 & 0 & 0 & 0 & 0 & 0 & 0 & 0 & 0 & 0 & 0 & -1 & -1 & 0 & 0 & 0 & 0 & 0 & 1 & 0 & 0 \\
\end{array}
\right)
$}
}~,~
\nn\\
\eea
and the $U(1)$-charges for the $D$-terms are summarized in 
 \beal{es0503}
{\footnotesize
 Q_{D}=
 \resizebox{0.88\textwidth}{!}{$
\left(
\begin{array}{ccccc|cccccccccccccccc|ccccccccc|ccccccccccc}
p_1 & p_2 & p_3 & p_4 & p_5 &
u_1 & u_2 & u_3 & u_4 & u_5 & u_6 & u_7 & u_8 & u_9 & u_{10} &
u_{11} & u_{12} & u_{13} & u_{14} & u_{15} & u_{16} & 
v_1 & v_2 & v_3 & v_4 & v_5 & v_6 & v_7 & v_8 & v_9 &
o_1 & o_2 & o_3 & o_4 & o_5 & o_6 & o_7 & o_8 & o_9 & o_{10} & o_{11}
\\
\hline
 1 & 1 & 1 & 0 & -1 & 0 & 0 & 0 & 0 & 0 & 0 & 0 & 0 & 0 & 0 & 0 & 0 & 0 & 0 & 0 & 0 & 0 & 0 & 0 & 0 & 0 & 0 & 0 & 0 & 0 & 0 & -1 & 0 & 0 & 0 & 0 & 0 & 0 & 0 & 0 & 0 \\
 0 & 0 & 0 & 0 & 0 & 0 & 0 & 0 & 0 & 0 & 0 & 0 & 0 & 0 & 0 & 0 & 0 & 0 & 0 & 0 & 0 & 0 & 0 & 0 & 0 & 0 & 0 & 0 & 0 & 0 & 1 & 0 & 0 & 0 & 0 & -1 & 0 & 0 & 0 & 0 & 0 \\
 0 & 0 & 0 & 0 & 0 & 0 & 0 & 0 & 0 & 0 & 0 & 0 & 0 & 0 & 0 & 0 & 0 & 0 & 0 & 0 & 0 & 0 & 0 & 0 & 0 & 0 & 0 & 0 & 0 & 0 & 0 & 1 & 0 & 0 & 0 & 0 & -1 & 0 & 0 & 0 & 0 \\
 0 & 0 & 0 & 0 & 0 & 0 & 0 & 0 & 0 & 0 & 0 & 0 & 0 & 0 & 0 & 0 & 0 & 0 & 0 & 0 & 0 & 0 & 0 & 0 & 0 & 0 & 0 & 0 & 0 & 0 & 0 & 0 & 1 & 0 & 0 & 0 & 0 & -1 & 0 & 0 & 0 \\
 0 & 0 & 0 & 0 & 0 & 0 & 0 & 0 & 0 & 0 & 0 & 0 & 0 & 0 & 0 & 0 & 0 & 0 & 0 & 0 & 0 & 0 & 0 & 0 & 0 & 0 & 0 & 0 & 0 & 0 & 0 & 0 & 0 & 0 & -1 & 1 & 0 & 0 & 0 & 0 & 0 \\
 0 & 0 & 0 & 0 & 0 & 0 & 0 & 0 & 0 & 0 & 0 & 0 & 0 & 0 & 0 & 0 & 0 & 0 & 0 & 0 & 0 & 0 & 0 & 0 & 0 & 0 & 0 & 0 & 0 & 0 & 0 & 0 & 0 & -1 & 0 & 0 & 1 & 0 & 0 & 0 & 0 \\
 0 & 0 & 0 & 0 & 0 & 0 & 0 & 0 & 0 & 0 & 0 & 0 & 0 & 0 & 0 & 0 & 0 & 0 & 0 & 0 & 0 & 0 & 0 & 0 & 0 & 0 & 0 & 0 & 0 & 0 & 0 & 0 & -1 & 0 & 1 & 0 & 0 & 0 & 0 & 0 & 0 \\
 0 & 0 & 0 & 0 & 0 & 0 & 0 & 0 & 0 & 0 & 0 & 0 & 0 & 0 & 0 & 0 & 0 & 0 & 0 & 0 & 0 & 0 & 0 & 0 & 0 & 0 & 0 & 0 & 0 & 0 & -1 & 0 & 0 & 1 & 0 & 0 & 0 & 0 & 0 & 0 & 0 \\
\end{array}
\right)
$}
}~.~
\nn\\
\eea
The charge matrices are invariant under permutations of the extremal GLSM fields $(p_1,p_2,p_3)$. This indicates that the global symmetry of the $2d$ $(0,2)$ theories is enhanced to $SU(3)\times U(1)^2$.

The charge matrices are used to obtain the toric data for the $2d$ $(0,2)$ theories, which is summarized in 
 \beal{es0505}
{\footnotesize
 G_{t}=
 \resizebox{0.88\textwidth}{!}{$
\left(
\begin{array}{ccccc|cccccccccccccccc|ccccccccc|ccccccccccc}
p_1 & p_2 & p_3 & p_4 & p_5 &
u_1 & u_2 & u_3 & u_4 & u_5 & u_6 & u_7 & u_8 & u_9 & u_{10} &
u_{11} & u_{12} & u_{13} & u_{14} & u_{15} & u_{16} & 
v_1 & v_2 & v_3 & v_4 & v_5 & v_6 & v_7 & v_8 & v_9 &
o_1 & o_2 & o_3 & o_4 & o_5 & o_6 & o_7 & o_8 & o_9 & o_{10} & o_{11}
\\
\hline
 1 & 1 & 1 & 1 & 1 & 1 & 1 & 1 & 1 & 1 & 1 & 1 & 1 & 1 & 1 & 1 & 1 & 1 & 1 & 1 & 1 & 1 & 1 & 1 & 1 & 1 & 1 & 1 & 1 & 1 & 2 & 2 & 2 & 2 & 2 & 2 & 2 & 2 & 2 & 2 & 2 \\
 1 & 0 & -1 & 0 & 0 & 0 & 0 & 0 & 0 & 0 & 0 & 0 & 0 & 0 & 0 & 0 & 0 & 0 & 0 & 0 & 0 & 0 & 0 & 0 & 0 & 0 & 0 & 0 & 0 & 0 & 0 & 0 & 0 & 0 & 0 & 0 & 0 & 0 & 0 & 0 & 0 \\
 0 & 1 & -1 & 0 & 0 & 0 & 0 & 0 & 0 & 0 & 0 & 0 & 0 & 0 & 0 & 0 & 0 & 0 & 0 & 0 & 0 & 0 & 0 & 0 & 0 & 0 & 0 & 0 & 0 & 0 & 0 & 0 & 0 & 0 & 0 & 0 & 0 & 0 & 0 & 0 & 0 \\
 0 & 0 & 1 & 3 & 0 & 1 & 1 & 1 & 1 & 1 & 1 & 1 & 1 & 1 & 1 & 1 & 1 & 1 & 1 & 1 & 1 & 2 & 2 & 2 & 2 & 2 & 2 & 2 & 2 & 2 & 1 & 1 & 1 & 1 & 1 & 1 & 1 & 1 & 1 & 1 & 1 \\
\end{array}
\right)
$}
}~,~
\nn\\
\eea
where $o_k$ correspond to extra GLSM fields \cite{Franco:2015tna}.
The corresponding toric diagram is shown in \fref{f_toric_cp2_31}.
\\

\subsubsection{$Y^{3,2}(\mathbb{CP}^2)$ \label{sec:cp22}}

\begin{figure}[H]
\begin{center}
\resizebox{0.4\hsize}{!}{
\includegraphics[height=6cm]{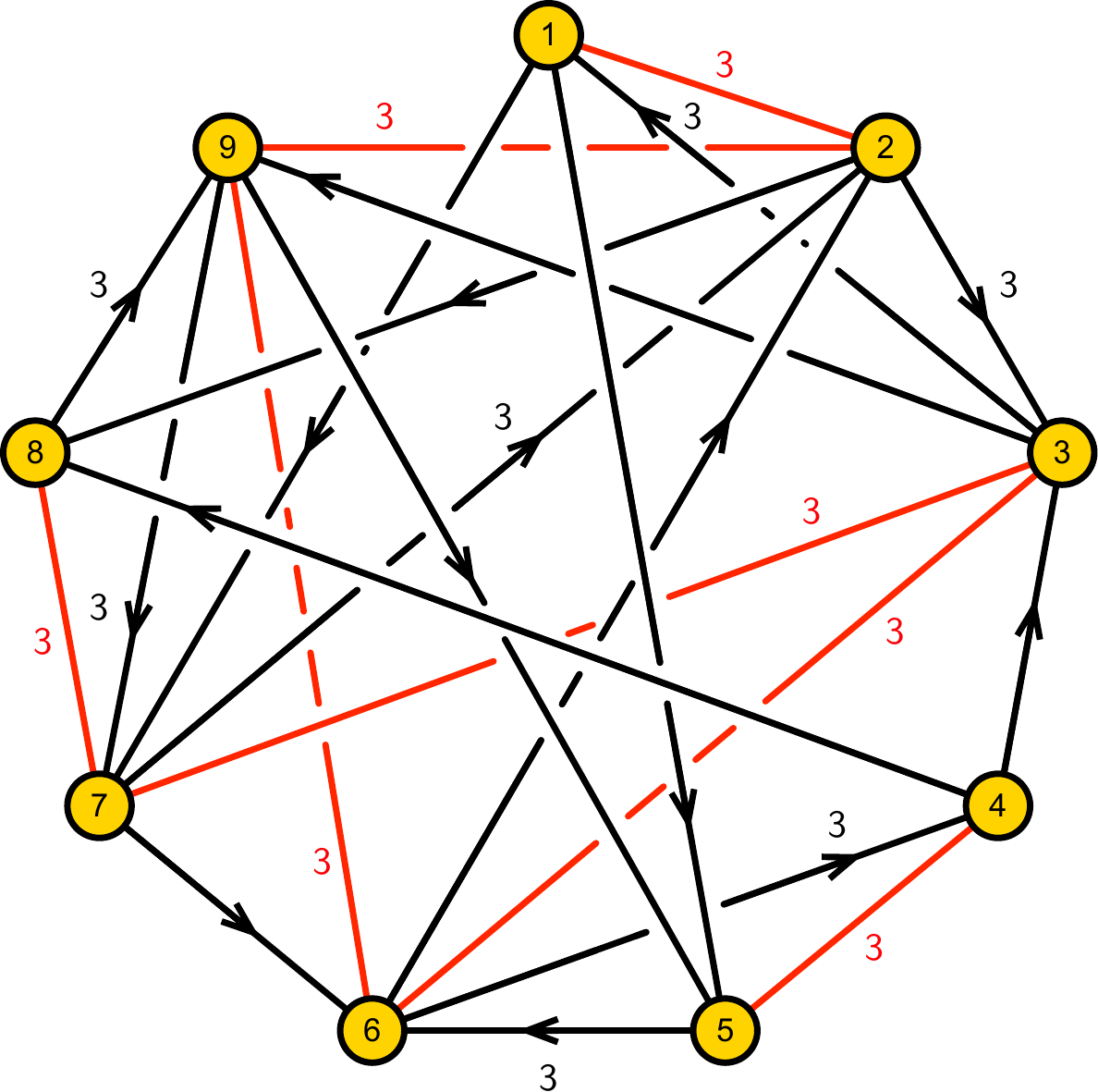} 
}
\caption{
Quiver for $Y^{3,2}(\mathbb{CP}^2)$.
\label{f_quiver_cp2_32}}
 \end{center}
 \end{figure}

The $2d$ $(0,2)$ theories for $Y^{3,2}(\mathbb{CP}^2)$ have the following $J$- and $E$-terms
  \beq
\begin{array}{rccrcl}
 &   J &   & & E &  \\
 \Lambda_{12}^{i}  : & \epsilon_{ijk} X^{j}_{23}  X^{k}_{31 } &~&    P_{15}  X^{i}_{56}  Q_{62 } &-& Q_{17}  X^{i}_{72 } \\    
 \Lambda_{29}^{2i}  : &\epsilon_{ijk} X^{j}_{97}  X^{k}_{72 } &~&    X^{i}_{23}  Q_{39 } &-& Q_{28}  X^{i}_{89 } \\    
 \Lambda_{37}^{2i}  : & \epsilon_{ijk} X^{j}_{72}  X^{k}_{23 } &~&    X^{i}_{31}  Q_{17 } &-& Q_{39}  X^{i}_{97 } \\    
 \Lambda_{45}^{i}  : & \epsilon_{ijk} X^{j}_{56}  X^{k}_{64 } &~&    P_{48}  X^{i}_{89}  Q_{95 } &-&  Q_{43}  X^{i}_{31}  P_{15} \\    
 \Lambda_{63}^{2i}  : & \epsilon_{ijk} X^{j}_{31}  P_{15}  X^{k}_{56 } &~&    X^{i}_{64}  Q_{43 } &-& Q_{62}  X^{i}_{23 } \\    
 \Lambda_{78}^{i}  : & \epsilon_{ijk} X^{j}_{89}  X^{k}_{97 } &~&    X^{i}_{72}  Q_{28 } &-&  Q_{76}  X^{i}_{64}  P_{48} \\    
 \Lambda_{96}^{2i}  : &\epsilon_{ijk}  X^{j}_{64}  P_{48}  X^{k}_{89 }  &~&    X^{i}_{97}  Q_{76 } &-& Q_{95}  X^{i}_{56 } \\    
 \end{array} 
~,~
\label{E_J_C_+-}
 \eeq
with the corresponding quiver diagram shown in \fref{f_quiver_cp2_32}. The global symmetry indices are $i,j,k=1,2,3$.
 
The $P$-matrix can be obtained as follows
\beal{es0601}
{\footnotesize
 P=
\resizebox{0.85\textwidth}{!}{$
\left(
\begin{array}{r|ccccc|cccccccccccccc|ccccccccc|cc}
\; & p_1 & p_2 & p_3 & p_4 & p_5&
u_1 & u_2 & u_3 & u_4 & u_5 & u_6 & u_7 & u_8 & u_9 & u_{10} &
u_{11} & u_{12} & u_{13} & u_{14} & 
v_1 & v_2 & v_3 & v_4 & v_5 & v_6 & v_7 & v_8 & v_9 &
o_1 & o_2
\\
\hline
X_{23}^{1} & 1 & 0 & 0 & 0 & 0 & 0 & 0 & 0 & 1 & 0 & 1 & 0 & 0 & 1 & 0 & 0 & 0 & 0 & 0 & 0 & 1 & 0 & 0 & 0 & 0 & 0 & 0 & 0 & 0 & 1 \\
X_{31}^{1} & 1 & 0 & 0 & 0 & 0 & 1 & 0 & 0 & 0 & 1 & 0 & 0 & 0 & 0 & 1 & 0 & 0 & 0 & 0 & 0 & 0 & 1 & 0 & 0 & 0 & 0 & 0 & 0 & 1 & 0 \\
X_{56}^{1} & 1 & 0 & 0 & 0 & 0 & 0 & 1 & 0 & 0 & 0 & 0 & 1 & 0 & 0 & 0 & 0 & 0 & 0 & 0 & 0 & 0 & 0 & 0 & 0 & 1 & 0 & 0 & 0 & 1 & 0 \\
X_{64}^{1} & 1 & 0 & 0 & 0 & 0 & 0 & 0 & 1 & 0 & 0 & 1 & 0 & 0 & 0 & 0 & 0 & 0 & 0 & 0 & 0 & 0 & 0 & 0 & 0 & 0 & 1 & 0 & 0 & 0 & 1 \\
X_{72}^{1} & 1 & 0 & 0 & 0 & 1 & 0 & 1 & 1 & 0 & 0 & 0 & 0 & 1 & 0 & 0 & 1 & 0 & 0 & 0 & 1 & 0 & 0 & 0 & 0 & 0 & 0 & 0 & 0 & 1 & 1 \\
X_{89}^{1} & 1 & 0 & 0 & 0 & 0 & 0 & 0 & 0 & 1 & 1 & 0 & 0 & 0 & 0 & 0 & 0 & 0 & 1 & 0 & 0 & 0 & 0 & 0 & 1 & 0 & 0 & 0 & 0 & 0 & 1 \\
X_{97}^{1} & 1 & 0 & 0 & 0 & 0 & 1 & 0 & 0 & 0 & 0 & 0 & 1 & 0 & 0 & 0 & 0 & 1 & 0 & 0 & 0 & 0 & 0 & 1 & 0 & 0 & 0 & 0 & 0 & 1 & 0 \\
\hline
X_{23}^{2} & 0 & 1 & 0 & 0 & 0 & 0 & 0 & 0 & 1 & 0 & 1 & 0 & 0 & 1 & 0 & 0 & 0 & 0 & 0 & 0 & 1 & 0 & 0 & 0 & 0 & 0 & 0 & 0 & 0 & 1 \\
X_{31}^{2} & 0 & 1 & 0 & 0 & 0 & 1 & 0 & 0 & 0 & 1 & 0 & 0 & 0 & 0 & 1 & 0 & 0 & 0 & 0 & 0 & 0 & 1 & 0 & 0 & 0 & 0 & 0 & 0 & 1 & 0 \\
X_{56}^{2} & 0 & 1 & 0 & 0 & 0 & 0 & 1 & 0 & 0 & 0 & 0 & 1 & 0 & 0 & 0 & 0 & 0 & 0 & 0 & 0 & 0 & 0 & 0 & 0 & 1 & 0 & 0 & 0 & 1 & 0 \\
X_{64}^{2} & 0 & 1 & 0 & 0 & 0 & 0 & 0 & 1 & 0 & 0 & 1 & 0 & 0 & 0 & 0 & 0 & 0 & 0 & 0 & 0 & 0 & 0 & 0 & 0 & 0 & 1 & 0 & 0 & 0 & 1 \\
X_{72}^{2} & 0 & 1 & 0 & 0 & 1 & 0 & 1 & 1 & 0 & 0 & 0 & 0 & 1 & 0 & 0 & 1 & 0 & 0 & 0 & 1 & 0 & 0 & 0 & 0 & 0 & 0 & 0 & 0 & 1 & 1 \\
X_{89}^{2} & 0 & 1 & 0 & 0 & 0 & 0 & 0 & 0 & 1 & 1 & 0 & 0 & 0 & 0 & 0 & 0 & 0 & 1 & 0 & 0 & 0 & 0 & 0 & 1 & 0 & 0 & 0 & 0 & 0 & 1 \\
X_{97}^{2} & 0 & 1 & 0 & 0 & 0 & 1 & 0 & 0 & 0 & 0 & 0 & 1 & 0 & 0 & 0 & 0 & 1 & 0 & 0 & 0 & 0 & 0 & 1 & 0 & 0 & 0 & 0 & 0 & 1 & 0 \\
\hline
X_{23}^{3} & 0 & 0 & 1 & 0 & 0 & 0 & 0 & 0 & 1 & 0 & 1 & 0 & 0 & 1 & 0 & 0 & 0 & 0 & 0 & 0 & 1 & 0 & 0 & 0 & 0 & 0 & 0 & 0 & 0 & 1 \\
X_{31}^{3} & 0 & 0 & 1 & 0 & 0 & 1 & 0 & 0 & 0 & 1 & 0 & 0 & 0 & 0 & 1 & 0 & 0 & 0 & 0 & 0 & 0 & 1 & 0 & 0 & 0 & 0 & 0 & 0 & 1 & 0 \\
X_{56}^{3} & 0 & 0 & 1 & 0 & 0 & 0 & 1 & 0 & 0 & 0 & 0 & 1 & 0 & 0 & 0 & 0 & 0 & 0 & 0 & 0 & 0 & 0 & 0 & 0 & 1 & 0 & 0 & 0 & 1 & 0 \\
X_{64}^{3} & 0 & 0 & 1 & 0 & 0 & 0 & 0 & 1 & 0 & 0 & 1 & 0 & 0 & 0 & 0 & 0 & 0 & 0 & 0 & 0 & 0 & 0 & 0 & 0 & 0 & 1 & 0 & 0 & 0 & 1 \\
X_{72}^{3} & 0 & 0 & 1 & 0 & 1 & 0 & 1 & 1 & 0 & 0 & 0 & 0 & 1 & 0 & 0 & 1 & 0 & 0 & 0 & 1 & 0 & 0 & 0 & 0 & 0 & 0 & 0 & 0 & 1 & 1 \\
X_{89}^{3} & 0 & 0 & 1 & 0 & 0 & 0 & 0 & 0 & 1 & 1 & 0 & 0 & 0 & 0 & 0 & 0 & 0 & 1 & 0 & 0 & 0 & 0 & 0 & 1 & 0 & 0 & 0 & 0 & 0 & 1 \\
X_{97}^{3} & 0 & 0 & 1 & 0 & 0 & 1 & 0 & 0 & 0 & 0 & 0 & 1 & 0 & 0 & 0 & 0 & 1 & 0 & 0 & 0 & 0 & 0 & 1 & 0 & 0 & 0 & 0 & 0 & 1 & 0 \\
\hline
Q_{17} & 0 & 0 & 0 & 1 & 0 & 0 & 0 & 0 & 0 & 0 & 0 & 1 & 0 & 0 & 0 & 0 & 1 & 1 & 1 & 0 & 0 & 0 & 1 & 1 & 1 & 1 & 1 & 1 & 0 & 0 \\
Q_{28} & 0 & 0 & 0 & 1 & 0 & 0 & 0 & 0 & 0 & 0 & 1 & 0 & 0 & 1 & 1 & 0 & 0 & 0 & 1 & 0 & 1 & 1 & 0 & 0 & 1 & 1 & 1 & 1 & 0 & 0 \\
Q_{39} & 0 & 0 & 0 & 1 & 0 & 0 & 0 & 0 & 0 & 1 & 0 & 0 & 0 & 0 & 1 & 0 & 0 & 1 & 1 & 0 & 0 & 1 & 0 & 1 & 1 & 1 & 1 & 1 & 0 & 0 \\
Q_{43} & 0 & 0 & 0 & 1 & 0 & 0 & 0 & 0 & 1 & 0 & 0 & 0 & 1 & 1 & 0 & 0 & 0 & 0 & 0 & 1 & 1 & 0 & 1 & 1 & 0 & 0 & 1 & 0 & 0 & 0 \\
Q_{62} & 0 & 0 & 0 & 1 & 0 & 0 & 0 & 1 & 0 & 0 & 0 & 0 & 1 & 0 & 0 & 0 & 0 & 0 & 0 & 1 & 0 & 0 & 1 & 1 & 0 & 1 & 1 & 0 & 0 & 0 \\
Q_{76} & 0 & 0 & 0 & 1 & 0 & 0 & 1 & 0 & 0 & 0 & 0 & 0 & 0 & 0 & 0 & 1 & 0 & 0 & 0 & 1 & 1 & 1 & 0 & 0 & 1 & 0 & 0 & 1 & 0 & 0 \\
Q_{95} & 0 & 0 & 0 & 1 & 0 & 1 & 0 & 0 & 0 & 0 & 0 & 0 & 0 & 0 & 0 & 1 & 1 & 0 & 0 & 1 & 1 & 1 & 1 & 0 & 0 & 0 & 0 & 1 & 0 & 0 \\
\hline
P_{15} & 0 & 0 & 0 & 0 & 1 & 0 & 0 & 0 & 0 & 0 & 0 & 0 & 0 & 0 & 0 & 1 & 1 & 1 & 1 & 0 & 0 & 0 & 0 & 0 & 0 & 0 & 0 & 1 & 0 & 1 \\
P_{48} & 0 & 0 & 0 & 0 & 1 & 0 & 0 & 0 & 0 & 0 & 0 & 0 & 1 & 1 & 1 & 0 & 0 & 0 & 1 & 0 & 0 & 0 & 0 & 0 & 0 & 0 & 1 & 0 & 1 & 0 \\
\end{array}
\right)
$}
}~.~
\nn\\
\eea
The corresponding $U(1)$-charges under the $J$- and $E$-terms take the form
\beal{es0602}
{\footnotesize
 Q_{JE}=
 \resizebox{0.85\textwidth}{!}{$
\left(
\begin{array}{ccccc|cccccccccccccc|ccccccccc|cc}
p_1 & p_2 & p_3 & p_4 & p_5&
u_1 & u_2 & u_3 & u_4 & u_5 & u_6 & u_7 & u_8 & u_9 & u_{10} &
u_{11} & u_{12} & u_{13} & u_{14} & 
v_1 & v_2 & v_3 & v_4 & v_5 & v_6 & v_7 & v_8 & v_9 &
o_1 & o_2
\\
\hline
 2 & 2 & 2 & 1 & 0 & -1 & -1 & -1 & -1 & -1 & -1 & -1 & 0 & 0 & 0 & 0 & 0 & 0 & 0 & 0 & 0 & 0 & 0 & 0 & 0 & 0 & 0 & 0 & 0 & 0 \\
 2 & 2 & 2 & 0 & 0 & 0 & 0 & 0 & 0 & -1 & -1 & -1 & 0 & 0 & 0 & 0 & 0 & 0 & 1 & 0 & 0 & 0 & 0 & 0 & 0 & 0 & 0 & 0 & -1 & -1 \\
 2 & 2 & 2 & 0 & 0 & 0 & 0 & -1 & -1 & -1 & -1 & -1 & 0 & 0 & 0 & 0 & 0 & 0 & 0 & 0 & 0 & 0 & 0 & 0 & 0 & 0 & 1 & 0 & -1 & 0 \\
 2 & 2 & 2 & 0 & 0 & -1 & -1 & 0 & 0 & -1 & -1 & -1 & 0 & 0 & 0 & 0 & 0 & 0 & 0 & 0 & 0 & 0 & 0 & 0 & 0 & 0 & 0 & 1 & 0 & -1 \\
 1 & 1 & 1 & 0 & 1 & 0 & 0 & 0 & 0 & 0 & 0 & 0 & 0 & 0 & 0 & 0 & 0 & 0 & 0 & 0 & 0 & 0 & 0 & 0 & 0 & 0 & 0 & 0 & -1 & -1 \\
 1 & 1 & 1 & 0 & 0 & 0 & 0 & 0 & 0 & -1 & 0 & -1 & 0 & 0 & 0 & 0 & 0 & 1 & 0 & 0 & 0 & 0 & 0 & 0 & 0 & 0 & 0 & 0 & 0 & -1 \\
 1 & 1 & 1 & 0 & 0 & 0 & 0 & 0 & 0 & -1 & -1 & 0 & 0 & 0 & 1 & 0 & 0 & 0 & 0 & 0 & 0 & 0 & 0 & 0 & 0 & 0 & 0 & 0 & -1 & 0 \\
 1 & 1 & 1 & 0 & 0 & 0 & 0 & 0 & -1 & 0 & -1 & 0 & 0 & 1 & 0 & 0 & 0 & 0 & 0 & 0 & 0 & 0 & 0 & 0 & 0 & 0 & 0 & 0 & -1 & 0 \\
 1 & 1 & 1 & 0 & 0 & 0 & 0 & -1 & 0 & -1 & -1 & -1 & 0 & 0 & 0 & 0 & 0 & 0 & 0 & 0 & 0 & 0 & 0 & 0 & 0 & 1 & 0 & 0 & 0 & 0 \\
 1 & 1 & 1 & 0 & 0 & 0 & 0 & -1 & -1 & 0 & 0 & 0 & 1 & 0 & 0 & 0 & 0 & 0 & 0 & 0 & 0 & 0 & 0 & 0 & 0 & 0 & 0 & 0 & -1 & 0 \\
 1 & 1 & 1 & 0 & 0 & 0 & 0 & -1 & -1 & -1 & 0 & -1 & 0 & 0 & 0 & 0 & 0 & 0 & 0 & 0 & 0 & 0 & 0 & 1 & 0 & 0 & 0 & 0 & 0 & 0 \\
 1 & 1 & 1 & 0 & 0 & 0 & -1 & 0 & 0 & -1 & -1 & -1 & 0 & 0 & 0 & 0 & 0 & 0 & 0 & 0 & 0 & 0 & 0 & 0 & 1 & 0 & 0 & 0 & 0 & 0 \\
 1 & 1 & 1 & 0 & 0 & -1 & 0 & 0 & 0 & 0 & 0 & -1 & 0 & 0 & 0 & 0 & 1 & 0 & 0 & 0 & 0 & 0 & 0 & 0 & 0 & 0 & 0 & 0 & 0 & -1 \\
 1 & 1 & 1 & 0 & 0 & -1 & 0 & -1 & -1 & 0 & 0 & -1 & 0 & 0 & 0 & 0 & 0 & 0 & 0 & 0 & 0 & 0 & 1 & 0 & 0 & 0 & 0 & 0 & 0 & 0 \\
 1 & 1 & 1 & 0 & 0 & -1 & -1 & 0 & 0 & 0 & 0 & 0 & 0 & 0 & 0 & 1 & 0 & 0 & 0 & 0 & 0 & 0 & 0 & 0 & 0 & 0 & 0 & 0 & 0 & -1 \\
 1 & 1 & 1 & 0 & 0 & -1 & -1 & 0 & 0 & -1 & -1 & 0 & 0 & 0 & 0 & 0 & 0 & 0 & 0 & 0 & 0 & 1 & 0 & 0 & 0 & 0 & 0 & 0 & 0 & 0 \\
 1 & 1 & 1 & 0 & 0 & -1 & -1 & 0 & -1 & 0 & -1 & 0 & 0 & 0 & 0 & 0 & 0 & 0 & 0 & 0 & 1 & 0 & 0 & 0 & 0 & 0 & 0 & 0 & 0 & 0 \\
 1 & 1 & 1 & 0 & 0 & -1 & -1 & -1 & -1 & 0 & 0 & 0 & 0 & 0 & 0 & 0 & 0 & 0 & 0 & 1 & 0 & 0 & 0 & 0 & 0 & 0 & 0 & 0 & 0 & 0 \\
\end{array}
\right)
$}
}~.~
\nn\\
\eea

\begin{figure}[ht!!]
\begin{center}
\resizebox{0.35\hsize}{!}{
\includegraphics[height=6cm]{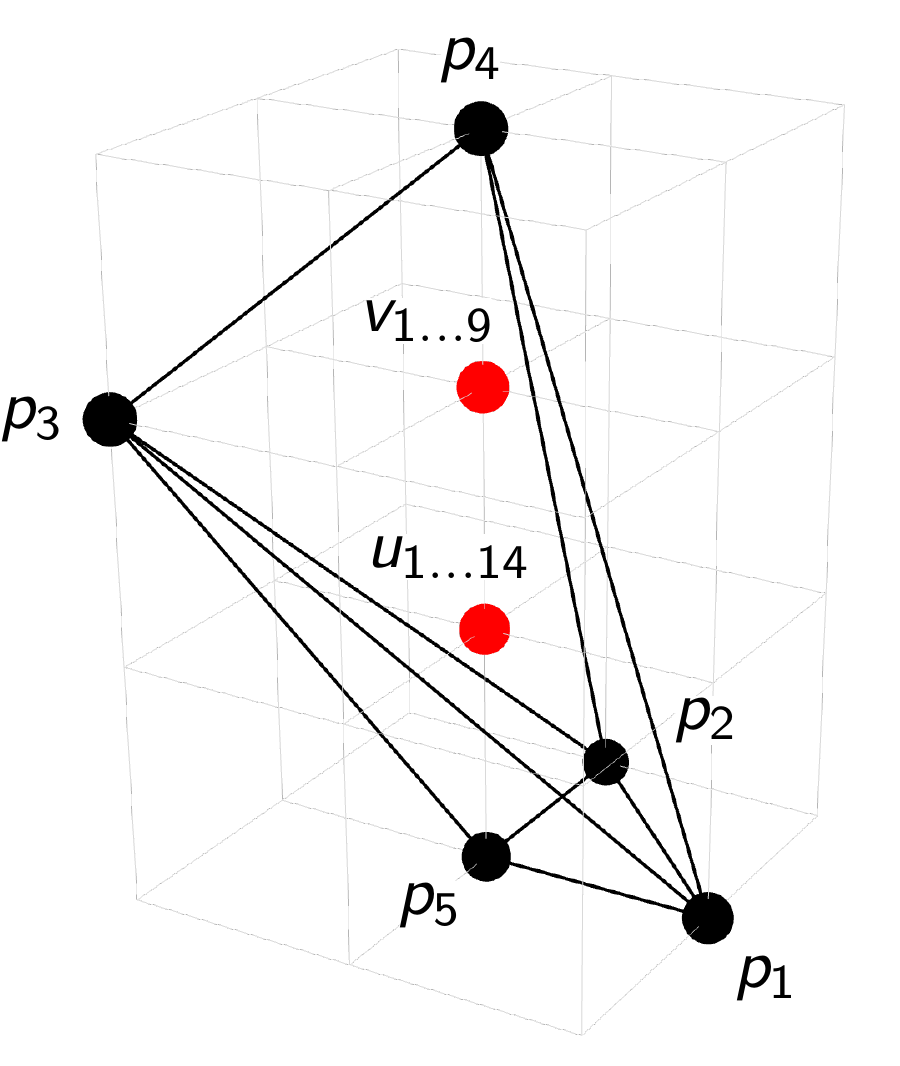} 
}
\caption{
Toric diagram of $Y^{3,2}(\mathbb{CP}^2)$.
\label{f_toric_cp2_32}}
 \end{center}
 \end{figure}

Furthermore, the $U(1)$-charges for the $D$-terms are summarized in 
\beal{es0603}
{\footnotesize
 Q_{D}=
 \resizebox{0.85\textwidth}{!}{$
\left(
\begin{array}{ccccc|cccccccccccccc|ccccccccc|cc}
p_1 & p_2 & p_3 & p_4 & p_5&
u_1 & u_2 & u_3 & u_4 & u_5 & u_6 & u_7 & u_8 & u_9 & u_{10} &
u_{11} & u_{12} & u_{13} & u_{14} & 
v_1 & v_2 & v_3 & v_4 & v_5 & v_6 & v_7 & v_8 & v_9 &
o_1 & o_2
\\
\hline
 1 & 1 & 1 & 0 & 0 & 0 & 0 & 0 & 0 & 0 & -1 & 0 & 0 & 0 & 0 & 0 & 0 & 0 & 0 & 0 & 0 & 0 & 0 & 0 & 0 & 0 & 0 & 0 & -1 & 0 \\
 1 & 1 & 1 & 0 & 0 & -1 & 0 & 0 & 0 & 0 & 0 & 0 & 0 & 0 & 0 & 0 & 0 & 0 & 0 & 0 & 0 & 0 & 0 & 0 & 0 & 0 & 0 & 0 & 0 & -1 \\
 0 & 0 & 0 & 0 & 0 & 1 & 0 & 0 & 0 & -1 & 0 & 0 & 0 & 0 & 0 & 0 & 0 & 0 & 0 & 0 & 0 & 0 & 0 & 0 & 0 & 0 & 0 & 0 & 0 & 0 \\
 0 & 0 & 0 & 0 & 0 & 0 & 1 & 0 & 0 & 0 & 0 & -1 & 0 & 0 & 0 & 0 & 0 & 0 & 0 & 0 & 0 & 0 & 0 & 0 & 0 & 0 & 0 & 0 & 0 & 0 \\
 0 & 0 & 0 & 0 & 0 & 0 & 0 & 0 & -1 & 1 & 0 & 0 & 0 & 0 & 0 & 0 & 0 & 0 & 0 & 0 & 0 & 0 & 0 & 0 & 0 & 0 & 0 & 0 & 0 & 0 \\
 0 & 0 & 0 & 0 & 0 & 0 & 0 & -1 & 0 & 0 & 1 & 0 & 0 & 0 & 0 & 0 & 0 & 0 & 0 & 0 & 0 & 0 & 0 & 0 & 0 & 0 & 0 & 0 & 0 & 0 \\
 0 & 0 & 0 & 0 & 0 & 0 & -1 & 1 & 0 & 0 & 0 & 0 & 0 & 0 & 0 & 0 & 0 & 0 & 0 & 0 & 0 & 0 & 0 & 0 & 0 & 0 & 0 & 0 & 0 & 0 \\
 -1 & -1 & -1 & 0 & 0 & 0 & 0 & 0 & 1 & 0 & 0 & 0 & 0 & 0 & 0 & 0 & 0 & 0 & 0 & 0 & 0 & 0 & 0 & 0 & 0 & 0 & 0 & 0 & 1 & 0 \\
\end{array}
\right)
$}
}~.~
\nn\\
\eea
We note that the $U(1)$-charges under the $J$- and $E$-terms and the $D$-terms remain invariant under any permutation of $(p_1,p_2,p_3)$. 
This further illustrates that the global symmetry of the $2d$ $(0,2)$ theories is $SU(3)\times U(1)^2$.

The corresponding toric data is given by
 \beal{es0605}
{\footnotesize
 G_{t}=
 \resizebox{0.85\textwidth}{!}{$
\left(
\begin{array}{ccccc|cccccccccccccc|ccccccccc|cc}
p_1 & p_2 & p_3 & p_4 & p_5&
u_1 & u_2 & u_3 & u_4 & u_5 & u_6 & u_7 & u_8 & u_9 & u_{10} &
u_{11} & u_{12} & u_{13} & u_{14} & 
v_1 & v_2 & v_3 & v_4 & v_5 & v_6 & v_7 & v_8 & v_9 &
o_1 & o_2
\\
\hline
 1 & 1 & 1 & 1 & 1 & 1 & 1 & 1 & 1 & 1 & 1 & 1 & 1 & 1 & 1 & 1 & 1 & 1 & 1 & 1 & 1 & 1 & 1 & 1 & 1 & 1 & 1 & 1 & 2 & 2 \\
 1 & 0 & -1 & 0 & 0 & 0 & 0 & 0 & 0 & 0 & 0 & 0 & 0 & 0 & 0 & 0 & 0 & 0 & 0 & 0 & 0 & 0 & 0 & 0 & 0 & 0 & 0 & 0 & 0 & 0 \\
 0 & 1 & -1 & 0 & 0 & 0 & 0 & 0 & 0 & 0 & 0 & 0 & 0 & 0 & 0 & 0 & 0 & 0 & 0 & 0 & 0 & 0 & 0 & 0 & 0 & 0 & 0 & 0 & 0 & 0 \\
 0 & 0 & 2 & 3 & 0 & 1 & 1 & 1 & 1 & 1 & 1 & 1 & 1 & 1 & 1 & 1 & 1 & 1 & 1 & 2 & 2 & 2 & 2 & 2 & 2 & 2 & 2 & 2 & 1 & 1 \\
\end{array}
\right)
$}
}~,~
\eea
where $o_1,o_2$ correspond to extra GLSM fields \cite{Franco:2015tna}. 
The toric diagram is shown in \fref{f_toric_cp2_32} and it agrees with the expectation.
\\

\subsubsection{$Y^{3,3}(\mathbb{CP}^2)$ \label{sec:cp22}}

\begin{figure}[H]
\begin{center}
\resizebox{0.4\hsize}{!}{
\includegraphics[height=6cm]{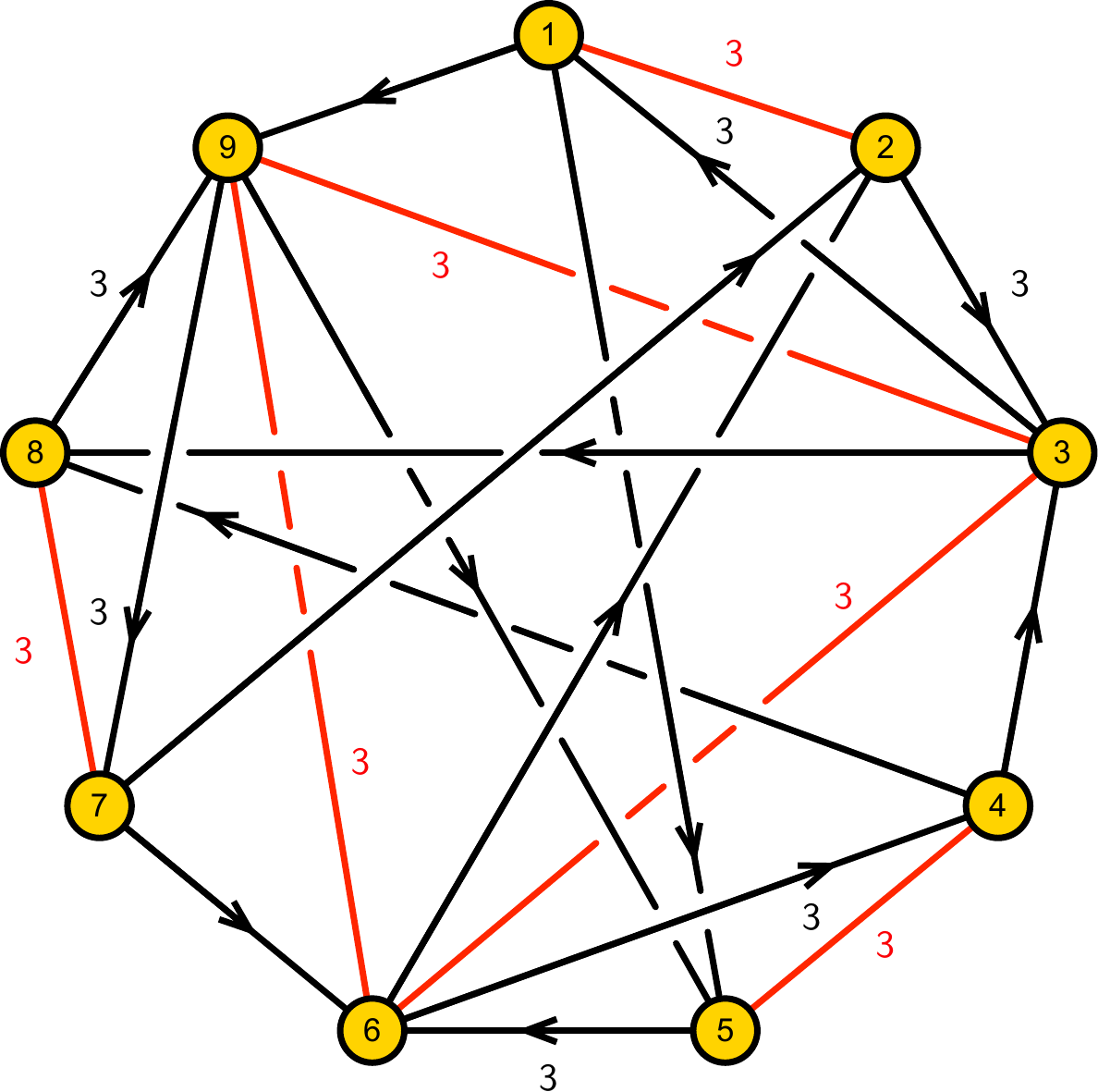} 
}
\caption{
Quiver for $Y^{3,3}(\mathbb{CP}^2)$.
\label{f_quiver_cp2_33}}
 \end{center}
 \end{figure}

The $J$- and $E$-terms of the $2d$ $(0,2)$ theories for $Y^{3,3}(\mathbb{CP}^2)$ take the following form
  \beq
\begin{array}{rccrcl}
 &   J &   & & E &  \\
 \Lambda_{12}^{i}  : &\epsilon_{ijk} X^{j}_{23} X^{k}_{31 } &~&   P_{15} X^{i}_{56}  Q_{62 } &-&  Q_{19} X^{i}_{97}  P_{72} \\   
 \Lambda_{39}^{2i}  : &\epsilon_{ijk} X^{j}_{97}  P_{72} X^{k}_{23 } &~&  X^{i}_{31}  Q_{19 } &-& Q_{38} X^{i}_{89 } \\   
 \Lambda_{45}^{i}  : &\epsilon_{ijk} X^{j}_{56} X^{k}_{64 } &~&   P_{48} X^{i}_{89}  Q_{95 } &-&  Q_{43} X^{i}_{31}  P_{15} \\   
 \Lambda_{63}^{2i}  : &\epsilon_{ijk} X^{j}_{31}  P_{15} X^{k}_{56 } &~&  X^{i}_{64}  Q_{43 } &-& Q_{62} X^{i}_{23 } \\   
 \Lambda_{78}^{i}  : &\epsilon_{ijk} X^{j}_{89} X^{k}_{97 } &~&   P_{72} X^{i}_{23}  Q_{38 } &-&  Q_{76} X^{i}_{64}  P_{48} \\   
 \Lambda_{96}^{2i}  : &\epsilon_{ijk} X^{j}_{64}  P_{48} X^{k}_{89 } &~&  X^{i}_{97}  Q_{76 } &-& Q_{95} X^{i}_{56 } \\ 
 \end{array} 
~,~
\label{E_J_C_+-}
 \eeq
where the global symmetry indices are $i,j,k=1,2,3$.
The corresponding quiver diagram is shown in \fref{f_quiver_cp2_33}.
 
The corresponding $P$-matrix takes the form
\beal{es0701}
{\footnotesize
 P=
\resizebox{0.8\textwidth}{!}{$
\left(
\begin{array}{r|ccccc|cccccccccccc|ccccccccc|cc}
\; & p_1 & p_2 & p_3 & p_4 & p_5&
u_1 & u_2 & u_3 & u_4 & u_5 & u_6 & u_7 & u_8 & u_9 & u_{10} &
u_{11} & u_{12} & 
v_1 & v_2 & v_3 & v_4 & v_5 & v_6 & v_7 & v_8 & v_9 &
o_1 & o_2
\\
\hline
 X_{23}^{1} & 1 & 0 & 0 & 0 & 0 & 1 & 0 & 0 & 1 & 0 & 0 & 0 & 0 & 0 & 0 & 0 & 0 & 1 & 0 & 0 & 0 & 0 & 0 & 0 & 0 & 0 & 0 & 1 \\
 X_{31}^{1} & 1 & 0 & 0 & 0 & 0 & 0 & 1 & 0 & 0 & 1 & 0 & 0 & 0 & 0 & 0 & 0 & 0 & 0 & 0 & 1 & 0 & 0 & 0 & 0 & 0 & 0 & 1 & 0 \\
 X_{56}^{1} & 1 & 0 & 0 & 0 & 0 & 0 & 0 & 1 & 0 & 0 & 1 & 0 & 0 & 0 & 0 & 0 & 0 & 0 & 0 & 0 & 0 & 1 & 0 & 0 & 0 & 0 & 0 & 1 \\
 X_{64}^{1} & 1 & 0 & 0 & 0 & 0 & 1 & 0 & 0 & 0 & 0 & 0 & 1 & 0 & 0 & 0 & 0 & 0 & 0 & 0 & 0 & 0 & 0 & 1 & 0 & 0 & 0 & 1 & 0 \\
 X_{89}^{1} & 1 & 0 & 0 & 0 & 0 & 0 & 1 & 0 & 0 & 0 & 0 & 0 & 0 & 0 & 1 & 0 & 0 & 0 & 0 & 0 & 1 & 0 & 0 & 0 & 0 & 0 & 0 & 1 \\
 X_{97}^{1} & 1 & 0 & 0 & 0 & 0 & 0 & 0 & 1 & 0 & 0 & 0 & 0 & 0 & 1 & 0 & 0 & 0 & 0 & 1 & 0 & 0 & 0 & 0 & 0 & 0 & 0 & 1 & 0 \\
 \hline
 X_{23}^{2} & 0 & 1 & 0 & 0 & 0 & 1 & 0 & 0 & 1 & 0 & 0 & 0 & 0 & 0 & 0 & 0 & 0 & 1 & 0 & 0 & 0 & 0 & 0 & 0 & 0 & 0 & 0 & 1 \\
 X_{31}^{2} & 0 & 1 & 0 & 0 & 0 & 0 & 1 & 0 & 0 & 1 & 0 & 0 & 0 & 0 & 0 & 0 & 0 & 0 & 0 & 1 & 0 & 0 & 0 & 0 & 0 & 0 & 1 & 0 \\
 X_{56}^{2} & 0 & 1 & 0 & 0 & 0 & 0 & 0 & 1 & 0 & 0 & 1 & 0 & 0 & 0 & 0 & 0 & 0 & 0 & 0 & 0 & 0 & 1 & 0 & 0 & 0 & 0 & 0 & 1 \\
 X_{64}^{2} & 0 & 1 & 0 & 0 & 0 & 1 & 0 & 0 & 0 & 0 & 0 & 1 & 0 & 0 & 0 & 0 & 0 & 0 & 0 & 0 & 0 & 0 & 1 & 0 & 0 & 0 & 1 & 0 \\
 X_{89}^{2} & 0 & 1 & 0 & 0 & 0 & 0 & 1 & 0 & 0 & 0 & 0 & 0 & 0 & 0 & 1 & 0 & 0 & 0 & 0 & 0 & 1 & 0 & 0 & 0 & 0 & 0 & 0 & 1 \\
 X_{97}^{2} & 0 & 1 & 0 & 0 & 0 & 0 & 0 & 1 & 0 & 0 & 0 & 0 & 0 & 1 & 0 & 0 & 0 & 0 & 1 & 0 & 0 & 0 & 0 & 0 & 0 & 0 & 1 & 0 \\
 \hline
 X_{23}^{3} & 0 & 0 & 1 & 0 & 0 & 1 & 0 & 0 & 1 & 0 & 0 & 0 & 0 & 0 & 0 & 0 & 0 & 1 & 0 & 0 & 0 & 0 & 0 & 0 & 0 & 0 & 0 & 1 \\
 X_{31}^{3} & 0 & 0 & 1 & 0 & 0 & 0 & 1 & 0 & 0 & 1 & 0 & 0 & 0 & 0 & 0 & 0 & 0 & 0 & 0 & 1 & 0 & 0 & 0 & 0 & 0 & 0 & 1 & 0 \\
 X_{56}^{3} & 0 & 0 & 1 & 0 & 0 & 0 & 0 & 1 & 0 & 0 & 1 & 0 & 0 & 0 & 0 & 0 & 0 & 0 & 0 & 0 & 0 & 1 & 0 & 0 & 0 & 0 & 0 & 1 \\
 X_{64}^{3} & 0 & 0 & 1 & 0 & 0 & 1 & 0 & 0 & 0 & 0 & 0 & 1 & 0 & 0 & 0 & 0 & 0 & 0 & 0 & 0 & 0 & 0 & 1 & 0 & 0 & 0 & 1 & 0 \\
 X_{89}^{3} & 0 & 0 & 1 & 0 & 0 & 0 & 1 & 0 & 0 & 0 & 0 & 0 & 0 & 0 & 1 & 0 & 0 & 0 & 0 & 0 & 1 & 0 & 0 & 0 & 0 & 0 & 0 & 1 \\
 X_{97}^{3} & 0 & 0 & 1 & 0 & 0 & 0 & 0 & 1 & 0 & 0 & 0 & 0 & 0 & 1 & 0 & 0 & 0 & 0 & 1 & 0 & 0 & 0 & 0 & 0 & 0 & 0 & 1 & 0 \\
 \hline
Q_{19} & 0 & 0 & 0 & 1 & 0 & 0 & 0 & 0 & 0 & 0 & 0 & 0 & 0 & 0 & 1 & 0 & 1 & 0 & 0 & 0 & 1 & 1 & 1 & 0 & 1 & 1 & 0 & 1 \\
Q_{38} & 0 & 0 & 0 & 1 & 0 & 0 & 0 & 0 & 0 & 1 & 0 & 0 & 0 & 0 & 0 & 0 & 1 & 0 & 0 & 1 & 0 & 1 & 1 & 0 & 1 & 1 & 1 & 0 \\
Q_{43} & 0 & 0 & 0 & 1 & 0 & 0 & 0 & 0 & 1 & 0 & 0 & 0 & 1 & 0 & 0 & 0 & 0 & 1 & 1 & 0 & 1 & 0 & 0 & 1 & 1 & 0 & 0 & 1 \\
Q_{62} & 0 & 0 & 0 & 1 & 0 & 0 & 0 & 0 & 0 & 0 & 0 & 1 & 1 & 0 & 0 & 0 & 0 & 0 & 1 & 0 & 1 & 0 & 1 & 1 & 1 & 0 & 1 & 0 \\
Q_{76} & 0 & 0 & 0 & 1 & 0 & 0 & 0 & 0 & 0 & 0 & 1 & 0 & 0 & 0 & 0 & 1 & 0 & 1 & 0 & 1 & 0 & 1 & 0 & 1 & 0 & 1 & 0 & 1 \\
Q_{95} & 0 & 0 & 0 & 1 & 0 & 0 & 0 & 0 & 0 & 0 & 0 & 0 & 0 & 1 & 0 & 1 & 0 & 1 & 1 & 1 & 0 & 0 & 0 & 1 & 0 & 1 & 1 & 0 \\
\hline
P_{15} & 0 & 0 & 0 & 0 & 1 & 0 & 0 & 0 & 0 & 0 & 0 & 0 & 0 & 1 & 1 & 1 & 1 & 0 & 0 & 0 & 0 & 0 & 0 & 0 & 0 & 1 & 0 & 0 \\
P_{48} & 0 & 0 & 0 & 0 & 1 & 0 & 0 & 0 & 1 & 1 & 0 & 0 & 1 & 0 & 0 & 0 & 1 & 0 & 0 & 0 & 0 & 0 & 0 & 0 & 1 & 0 & 0 & 0 \\
P_{72} & 0 & 0 & 0 & 0 & 1 & 0 & 0 & 0 & 0 & 0 & 1 & 1 & 1 & 0 & 0 & 1 & 0 & 0 & 0 & 0 & 0 & 0 & 0 & 1 & 0 & 0 & 0 & 0 \\
\end{array}
\right)
$}
}~.~
\eea
The $U(1)$-charges under $J$- and $E$-terms are summarized in 
\beal{es0702}
{\footnotesize
 Q_{JE}=
 \resizebox{0.8\textwidth}{!}{$
\left(
\begin{array}{ccccc|cccccccccccc|ccccccccc|cc}
p_1 & p_2 & p_3 & p_4 & p_5&
u_1 & u_2 & u_3 & u_4 & u_5 & u_6 & u_7 & u_8 & u_9 & u_{10} &
u_{11} & u_{12} & 
v_1 & v_2 & v_3 & v_4 & v_5 & v_6 & v_7 & v_8 & v_9 &
o_1 & o_2
\\
\hline
 1 & 1 & 1 & 0 & 0 & -1 & -1 & -1 & 0 & 0 & 0 & 0 & 0 & 0 & 0 & 0 & 0 & 0 & 0 & 0 & 0 & 0 & 0 & 0 & 0 & 0 & 0 & 0 \\
 0 & 0 & 0 & 1 & 0 & 0 & 1 & 1 & 0 & 0 & 0 & 0 & 0 & 0 & 0 & 0 & 0 & 1 & -1 & -1 & 0 & 0 & 0 & 0 & 0 & 0 & 0 & -1 \\
 0 & 0 & 0 & 0 & 1 & 0 & 0 & 1 & -1 & 0 & -1 & 0 & 0 & -1 & 0 & 0 & 0 & 1 & 0 & 0 & 0 & 0 & 0 & 0 & 0 & 0 & 0 & 0 \\
 0 & 0 & 0 & 0 & 0 & 0 & 1 & 1 & 0 & 0 & 0 & 0 & 0 & -1 & 0 & 0 & 0 & 1 & 0 & -1 & 0 & 0 & 0 & 0 & 0 & 1 & 0 & -1 \\
 0 & 0 & 0 & 0 & 0 & 0 & 1 & 1 & -1 & 0 & 0 & 0 & 0 & 0 & 0 & 0 & 0 & 2 & -1 & -1 & 0 & 0 & 0 & 0 & 1 & 0 & 0 & -1 \\
 0 & 0 & 0 & 0 & 0 & 0 & 1 & 1 & -1 & 0 & 0 & 0 & 0 & -1 & 0 & 0 & 1 & 2 & 0 & -1 & 0 & 0 & 0 & 0 & 0 & 0 & 0 & -1 \\
 0 & 0 & 0 & 0 & 0 & 0 & 1 & 0 & 0 & 0 & 0 & 0 & 0 & 0 & 0 & 0 & 0 & 1 & 0 & -1 & 0 & 1 & 0 & 0 & 0 & 0 & 0 & -1 \\
 0 & 0 & 0 & 0 & 0 & 0 & 0 & 1 & 0 & 0 & 0 & 0 & 0 & 0 & 0 & 0 & 0 & 1 & -1 & 0 & 1 & 0 & 0 & 0 & 0 & 0 & 0 & -1 \\
 0 & 0 & 0 & 0 & 0 & 0 & 0 & 1 & 0 & 0 & 0 & 0 & 0 & -1 & 1 & 0 & 0 & 1 & 0 & 0 & 0 & 0 & 0 & 0 & 0 & 0 & 0 & -1 \\
 0 & 0 & 0 & 0 & 0 & 0 & 0 & 1 & 0 & 0 & -1 & 0 & 0 & 0 & 0 & 0 & 0 & 0 & -1 & 0 & 0 & 0 & 0 & 1 & 0 & 0 & 0 & 0 \\
 0 & 0 & 0 & 0 & 0 & 0 & 0 & 1 & 0 & 0 & -1 & 0 & 0 & -1 & 0 & 1 & 0 & 0 & 0 & 0 & 0 & 0 & 0 & 0 & 0 & 0 & 0 & 0 \\
 0 & 0 & 0 & 0 & 0 & 0 & 0 & 1 & -1 & 0 & -1 & 0 & 1 & 0 & 0 & 0 & 0 & 1 & -1 & 0 & 0 & 0 & 0 & 0 & 0 & 0 & 0 & 0 \\
 0 & 0 & 0 & 0 & 0 & 0 & 0 & 0 & -1 & 1 & 0 & 0 & 0 & 0 & 0 & 0 & 0 & 1 & 0 & -1 & 0 & 0 & 0 & 0 & 0 & 0 & 0 & 0 \\
 0 & 0 & 0 & 0 & 0 & -1 & 1 & 1 & 0 & 0 & 0 & 0 & 0 & 0 & 0 & 0 & 0 & 2 & -1 & -1 & 0 & 0 & 1 & 0 & 0 & 0 & 0 & -1 \\
 0 & 0 & 0 & 0 & 0 & -1 & 0 & 1 & 0 & 0 & -1 & 1 & 0 & 0 & 0 & 0 & 0 & 1 & -1 & 0 & 0 & 0 & 0 & 0 & 0 & 0 & 0 & 0 \\
 0 & 0 & 0 & 0 & 0 & -1 & 0 & 0 & 0 & 0 & 0 & 0 & 0 & 0 & 0 & 0 & 0 & 1 & -1 & -1 & 0 & 0 & 0 & 0 & 0 & 0 & 1 & 0 \\
\end{array}
\right)
$}
}~,~
\nn\\
\eea
and the $U(1)$-charges under $D$-terms are summarized in 
\beal{es0703}
{\footnotesize
 Q_{D}=
 \resizebox{0.8\textwidth}{!}{$
\left(
\begin{array}{ccccc|cccccccccccc|ccccccccc|cc}
p_1 & p_2 & p_3 & p_4 & p_5&
u_1 & u_2 & u_3 & u_4 & u_5 & u_6 & u_7 & u_8 & u_9 & u_{10} &
u_{11} & u_{12} & 
v_1 & v_2 & v_3 & v_4 & v_5 & v_6 & v_7 & v_8 & v_9 &
o_1 & o_2
\\
\hline
 0 & 0 & 0 & 0 & 0 & 1 & 0 & -1 & 0 & 0 & 0 & 0 & 0 & 0 & 0 & 0 & 0 & -1 & 1 & 0 & 0 & 0 & 0 & 0 & 0 & 0 & 0 & 0 \\
 0 & 0 & 0 & 0 & 0 & 0 & 1 & 0 & -1 & 0 & 0 & 0 & 0 & 0 & 0 & 0 & 0 & 1 & 0 & -1 & 0 & 0 & 0 & 0 & 0 & 0 & 0 & 0 \\
 0 & 0 & 0 & 0 & 0 & 0 & 0 & 1 & 0 & 0 & 0 & 0 & 0 & 0 & 0 & 0 & 0 & 1 & 0 & 0 & 0 & 0 & 0 & 0 & 0 & 0 & 0 & -1 \\
 0 & 0 & 0 & 0 & 0 & 0 & 0 & 1 & 0 & 0 & 0 & 0 & 0 & -1 & 0 & 0 & 0 & 0 & 0 & 0 & 0 & 0 & 0 & 0 & 0 & 0 & 0 & 0 \\
 0 & 0 & 0 & 0 & 0 & 0 & 0 & 1 & 0 & 0 & -1 & 0 & 0 & 0 & 0 & 0 & 0 & 1 & -1 & 0 & 0 & 0 & 0 & 0 & 0 & 0 & 0 & 0 \\
 0 & 0 & 0 & 0 & 0 & 0 & 0 & 0 & 0 & 0 & 0 & 0 & 0 & 0 & 0 & 0 & 0 & -1 & 0 & 1 & 0 & 0 & 0 & 0 & 0 & 0 & 0 & 0 \\
 0 & 0 & 0 & 0 & 0 & 0 & 0 & -1 & 0 & 0 & 1 & 0 & 0 & 0 & 0 & 0 & 0 & 0 & 0 & 0 & 0 & 0 & 0 & 0 & 0 & 0 & 0 & 0 \\
 0 & 0 & 0 & 0 & 0 & -1 & 0 & 0 & 1 & 0 & 0 & 0 & 0 & 0 & 0 & 0 & 0 & 0 & 0 & 0 & 0 & 0 & 0 & 0 & 0 & 0 & 0 & 0 \\
\end{array}
\right)
$}
}~.~
\nn\\
\eea
We note that the charge matrices are invariant under permutations of $(p_1,p_2,p_3)$.
This illustrates that the global symmetry for the $2d$ $(0,2)$ theory is $SU(3)\times U(1)^2$.

The toric data is given by
 \beal{es0705}
{\footnotesize
 G_{t}=
 \resizebox{0.8\textwidth}{!}{$
\left(
\begin{array}{ccccc|cccccccccccc|ccccccccc|cc}
p_1 & p_2 & p_3 & p_4 & p_5&
u_1 & u_2 & u_3 & u_4 & u_5 & u_6 & u_7 & u_8 & u_9 & u_{10} &
u_{11} & u_{12} & 
v_1 & v_2 & v_3 & v_4 & v_5 & v_6 & v_7 & v_8 & v_9 &
o_1 & o_2
\\
\hline
 1 & 1 & 1 & 1 & 1 & 1 & 1 & 1 & 1 & 1 & 1 & 1 & 1 & 1 & 1 & 1 & 1 & 1 & 1 & 1 & 1 & 1 & 1 & 1 & 1 & 1 & 2 & 2 \\
 1 & 0 & -1 & 0 & 0 & 0 & 0 & 0 & 0 & 0 & 0 & 0 & 0 & 0 & 0 & 0 & 0 & 0 & 0 & 0 & 0 & 0 & 0 & 0 & 0 & 0 & 0 & 0 \\
 0 & 1 & -1 & 0 & 0 & 0 & 0 & 0 & 0 & 0 & 0 & 0 & 0 & 0 & 0 & 0 & 0 & 0 & 0 & 0 & 0 & 0 & 0 & 0 & 0 & 0 & 0 & 0 \\
 0 & 0 & 3 & 3 & 0 & 1 & 1 & 1 & 1 & 1 & 1 & 1 & 1 & 1 & 1 & 1 & 1 & 2 & 2 & 2 & 2 & 2 & 2 & 2 & 2 & 2 & 3 & 3 \\
\end{array}
\right)
$}
}~,~
\eea
where $o_1,o_2$ correspond to extra GLSM fields \cite{Franco:2015tna}. 
The toric diagram is shown in \fref{f_toric_cp2_33} and it agrees with the expectation.

\begin{figure}[H]
\begin{center}
\resizebox{0.35\hsize}{!}{
\includegraphics[height=6cm]{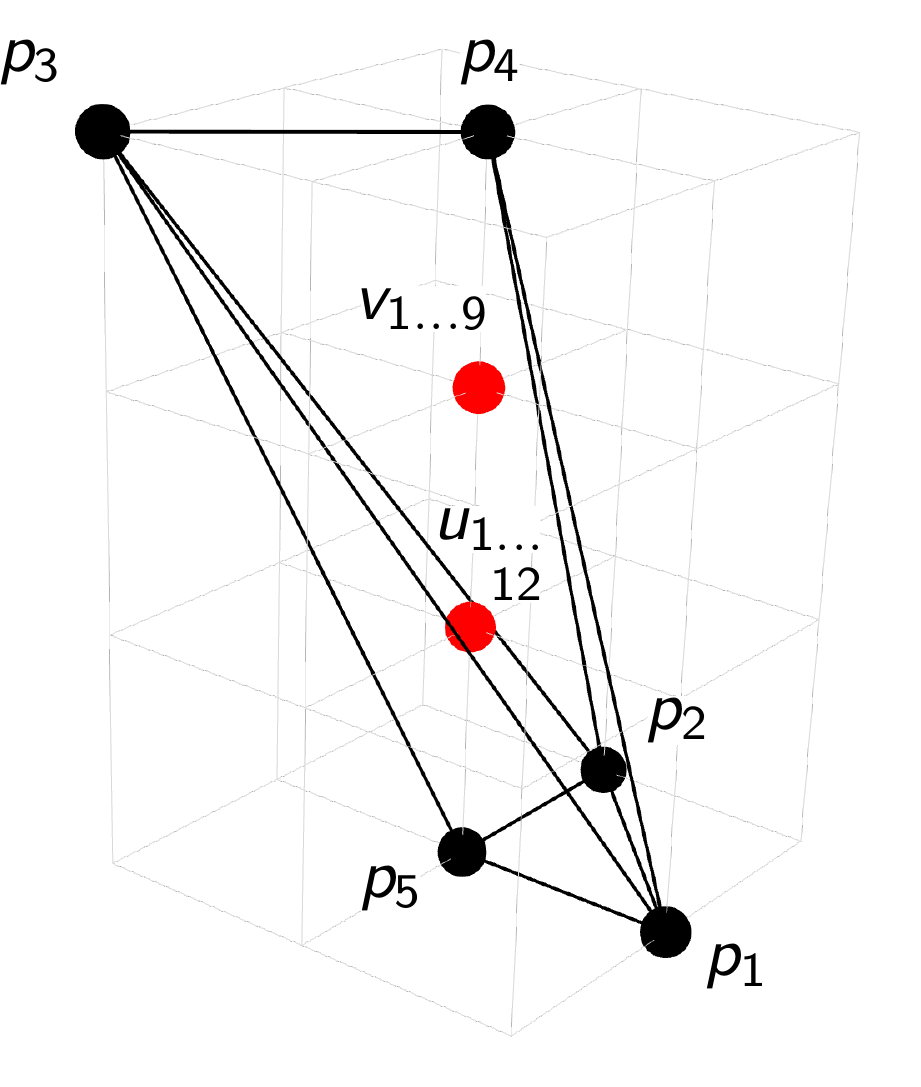} 
}
\caption{
Toric diagram of $Y^{3,3}(\mathbb{CP}^2)$.
\label{f_toric_cp2_33}}
 \end{center}
 \end{figure}

\subsubsection{$Y^{3,4}(\mathbb{CP}^2)$ \label{sec:cp22}}

\begin{figure}[H]
\begin{center}
\resizebox{0.4\hsize}{!}{
\includegraphics[height=6cm]{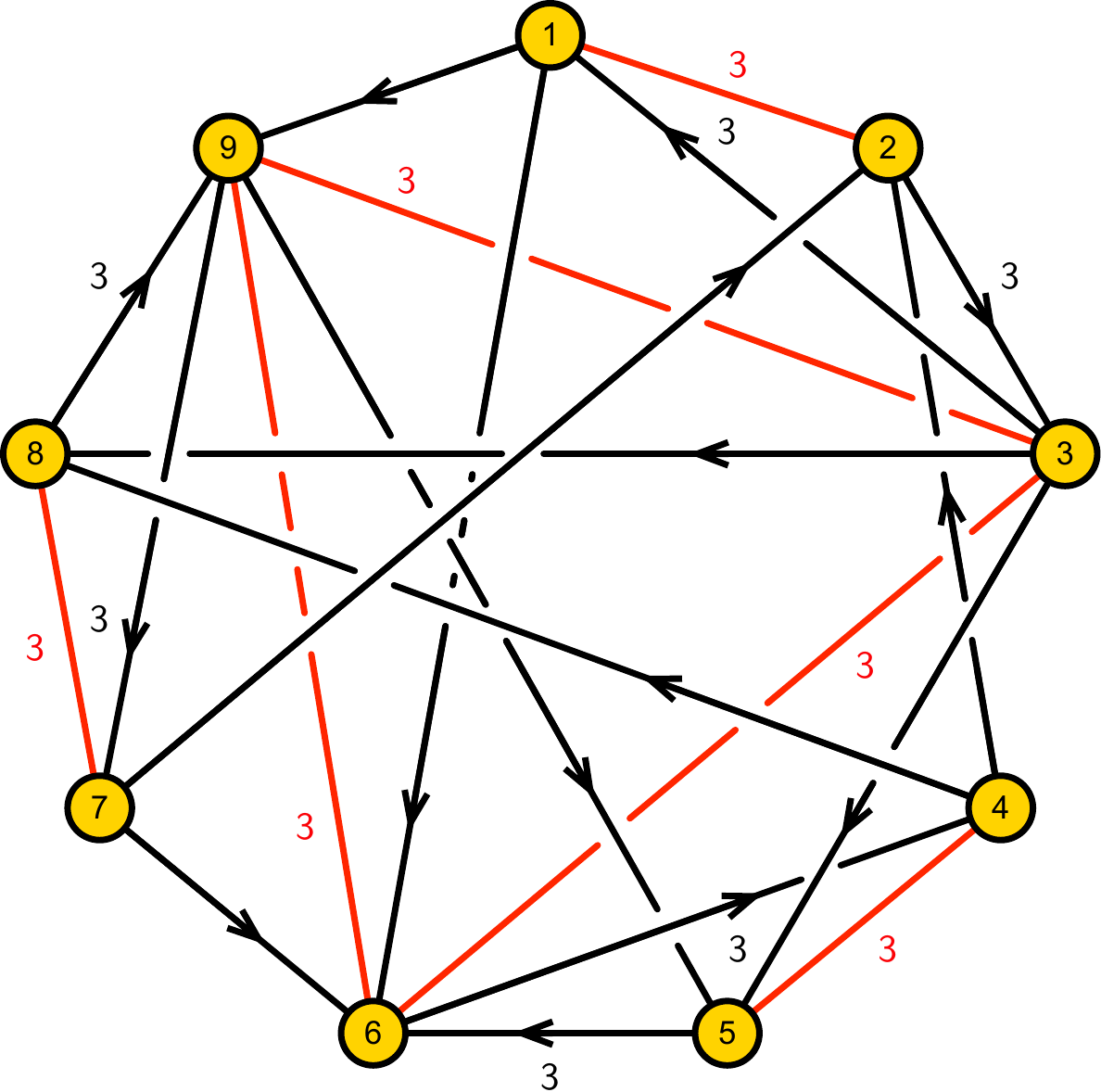} 
}
\caption{
Quiver for $Y^{3,4}(\mathbb{CP}^2)$.
\label{f_quiver_cp2_34}}
 \end{center}
 \end{figure}

The model for $Y^{3,4}(\mathbb{CP}^2)$ has a quiver which is shown in \fref{f_quiver_cp2_34}  and the corresponding $J$- and $E$-terms are given by
  \beq
\begin{array}{rccrcl}
 &   J &   & & E &  \\
 \Lambda_{12}^{i}  : &\epsilon_{ijk} X^{j}_{23}  X^{k}_{31 } &~&   P_{16}  X^{i}_{64}  Q_{42 } &-&   Q_{19}  X^{i}_{97}  P_{72} \\   
 \Lambda_{36}^{1i}  : &\epsilon_{ijk} X^{j}_{64}  Q_{42}  X^{k}_{23 } &~&   X^{i}_{31}  P_{16 } &-&  P_{35}  X^{i}_{56 } \\  
 \Lambda_{39}^{2i}  : &\epsilon_{ijk} X^{j}_{97}  P_{72}  X^{k}_{23 } &~&   X^{i}_{31}  Q_{19 } &-&  Q_{38}  X^{i}_{89} \\   
 \Lambda_{45}^{i}  : &\epsilon_{ijk} X^{j}_{56}  X^{k}_{64 } &~&   P_{48}  X^{i}_{89}  Q_{95 } &-&   Q_{42}  X^{i}_{23}  P_{35} \\   
 \Lambda_{78}^{i}  : &\epsilon_{ijk} X^{j}_{89}  X^{k}_{97 } &~&   P_{72}  X^{i}_{23}  Q_{38 } &-&   Q_{76}  X^{i}_{64}  P_{48} \\   
 \Lambda_{96}^{2i}  : &\epsilon_{ijk} X^{j}_{64}  P_{48}  X^{k}_{89 } &~&   X^{i}_{97}  Q_{76 } &-&  Q_{95}  X^{i}_{56 } \\  
 \end{array} 
~,~
\label{es0800}
 \eeq
where the global symmetry indices are $i,j,k=1,2,3$.
 
 The $P$-matrix takes the form
\beal{es0801}
{\footnotesize
 P=
\resizebox{0.85\textwidth}{!}{$
\left(
\begin{array}{r|ccccc|ccccccccccc|cccccccccc|ccccccccc}
\; & p_1 & p_2 & p_3 & p_4 & p_5&
u_1 & u_2 & u_3 & u_4 & u_5 & u_6 & u_7 & u_8 & u_9 & u_{10} &
u_{11} &
v_1 & v_2 & v_3 & v_4 & v_5 & v_6 & v_7 & v_8 & v_9 & v_{10} &
o_1 & o_2 & o_3 & o_4 & o_5 & o_6 & o_7 & o_8 & o_9
\\
\hline
X_{23}^{1} & 1 & 0 & 0 & 0 & 0 & 0 & 0 & 0 & 1 & 0 & 0 & 0 & 0 & 0 & 0 & 0 & 1 & 0 & 0 & 0 & 0 & 0 & 0 & 0 & 0 & 0 & 1 & 1 & 0 & 0 & 1 & 0 & 0 & 0 & 1 \\
X_{31}^{1} & 1 & 0 & 0 & 0 & 0 & 0 & 1 & 0 & 0 & 1 & 0 & 0 & 0 & 0 & 0 & 0 & 0 & 0 & 1 & 0 & 0 & 0 & 0 & 1 & 0 & 0 & 0 & 1 & 1 & 1 & 0 & 1 & 1 & 1 & 0 \\
X_{56}^{1} & 1 & 0 & 0 & 0 & 0 & 0 & 0 & 0 & 0 & 0 & 1 & 1 & 0 & 0 & 0 & 0 & 0 & 0 & 1 & 0 & 0 & 0 & 0 & 0 & 1 & 0 & 0 & 1 & 1 & 1 & 0 & 1 & 1 & 0 & 1 \\
X_{64}^{1} & 1 & 0 & 0 & 0 & 0 & 0 & 0 & 1 & 0 & 0 & 0 & 0 & 0 & 0 & 0 & 0 & 0 & 0 & 0 & 0 & 1 & 0 & 0 & 0 & 0 & 0 & 1 & 0 & 0 & 1 & 1 & 0 & 0 & 1 & 0 \\
X_{89}^{1} & 1 & 0 & 0 & 0 & 0 & 0 & 1 & 0 & 0 & 0 & 0 & 0 & 0 & 1 & 0 & 0 & 0 & 0 & 0 & 1 & 0 & 0 & 0 & 0 & 0 & 0 & 0 & 1 & 1 & 0 & 1 & 1 & 0 & 0 & 1 \\
X_{97}^{1} & 1 & 0 & 0 & 0 & 0 & 0 & 0 & 0 & 0 & 0 & 1 & 0 & 1 & 0 & 0 & 0 & 0 & 1 & 0 & 0 & 0 & 0 & 0 & 0 & 0 & 0 & 1 & 0 & 1 & 1 & 0 & 0 & 1 & 1 & 0 \\
\hline
X_{23}^{2} & 0 & 1 & 0 & 0 & 0 & 0 & 0 & 0 & 1 & 0 & 0 & 0 & 0 & 0 & 0 & 0 & 1 & 0 & 0 & 0 & 0 & 0 & 0 & 0 & 0 & 0 & 1 & 1 & 0 & 0 & 1 & 0 & 0 & 0 & 1 \\
X_{31}^{2} & 0 & 1 & 0 & 0 & 0 & 0 & 1 & 0 & 0 & 1 & 0 & 0 & 0 & 0 & 0 & 0 & 0 & 0 & 1 & 0 & 0 & 0 & 0 & 1 & 0 & 0 & 0 & 1 & 1 & 1 & 0 & 1 & 1 & 1 & 0 \\
X_{56}^{2} & 0 & 1 & 0 & 0 & 0 & 0 & 0 & 0 & 0 & 0 & 1 & 1 & 0 & 0 & 0 & 0 & 0 & 0 & 1 & 0 & 0 & 0 & 0 & 0 & 1 & 0 & 0 & 1 & 1 & 1 & 0 & 1 & 1 & 0 & 1 \\
X_{64}^{2} & 0 & 1 & 0 & 0 & 0 & 0 & 0 & 1 & 0 & 0 & 0 & 0 & 0 & 0 & 0 & 0 & 0 & 0 & 0 & 0 & 1 & 0 & 0 & 0 & 0 & 0 & 1 & 0 & 0 & 1 & 1 & 0 & 0 & 1 & 0 \\
X_{89}^{2} & 0 & 1 & 0 & 0 & 0 & 0 & 1 & 0 & 0 & 0 & 0 & 0 & 0 & 1 & 0 & 0 & 0 & 0 & 0 & 1 & 0 & 0 & 0 & 0 & 0 & 0 & 0 & 1 & 1 & 0 & 1 & 1 & 0 & 0 & 1 \\
X_{97}^{2} & 0 & 1 & 0 & 0 & 0 & 0 & 0 & 0 & 0 & 0 & 1 & 0 & 1 & 0 & 0 & 0 & 0 & 1 & 0 & 0 & 0 & 0 & 0 & 0 & 0 & 0 & 1 & 0 & 1 & 1 & 0 & 0 & 1 & 1 & 0 \\
\hline
X_{23}^{3} & 0 & 0 & 1 & 0 & 0 & 0 & 0 & 0 & 1 & 0 & 0 & 0 & 0 & 0 & 0 & 0 & 1 & 0 & 0 & 0 & 0 & 0 & 0 & 0 & 0 & 0 & 1 & 1 & 0 & 0 & 1 & 0 & 0 & 0 & 1 \\
X_{31}^{3} & 0 & 0 & 1 & 0 & 0 & 0 & 1 & 0 & 0 & 1 & 0 & 0 & 0 & 0 & 0 & 0 & 0 & 0 & 1 & 0 & 0 & 0 & 0 & 1 & 0 & 0 & 0 & 1 & 1 & 1 & 0 & 1 & 1 & 1 & 0 \\
X_{56}^{3} & 0 & 0 & 1 & 0 & 0 & 0 & 0 & 0 & 0 & 0 & 1 & 1 & 0 & 0 & 0 & 0 & 0 & 0 & 1 & 0 & 0 & 0 & 0 & 0 & 1 & 0 & 0 & 1 & 1 & 1 & 0 & 1 & 1 & 0 & 1 \\
X_{64}^{3} & 0 & 0 & 1 & 0 & 0 & 0 & 0 & 1 & 0 & 0 & 0 & 0 & 0 & 0 & 0 & 0 & 0 & 0 & 0 & 0 & 1 & 0 & 0 & 0 & 0 & 0 & 1 & 0 & 0 & 1 & 1 & 0 & 0 & 1 & 0 \\
X_{89}^{3} & 0 & 0 & 1 & 0 & 0 & 0 & 1 & 0 & 0 & 0 & 0 & 0 & 0 & 1 & 0 & 0 & 0 & 0 & 0 & 1 & 0 & 0 & 0 & 0 & 0 & 0 & 0 & 1 & 1 & 0 & 1 & 1 & 0 & 0 & 1 \\
X_{97}^{3} & 0 & 0 & 1 & 0 & 0 & 0 & 0 & 0 & 0 & 0 & 1 & 0 & 1 & 0 & 0 & 0 & 0 & 1 & 0 & 0 & 0 & 0 & 0 & 0 & 0 & 0 & 1 & 0 & 1 & 1 & 0 & 0 & 1 & 1 & 0 \\
\hline
Q_{19} & 0 & 0 & 0 & 1 & 0 & 0 & 0 & 0 & 0 & 0 & 0 & 0 & 0 & 1 & 0 & 1 & 0 & 0 & 0 & 1 & 1 & 0 & 1 & 0 & 1 & 1 & 0 & 0 & 0 & 0 & 1 & 0 & 0 & 0 & 1 \\
Q_{38} & 0 & 0 & 0 & 1 & 0 & 0 & 0 & 0 & 0 & 1 & 0 & 0 & 0 & 0 & 0 & 1 & 0 & 0 & 1 & 0 & 1 & 0 & 1 & 1 & 1 & 1 & 0 & 0 & 0 & 1 & 0 & 0 & 1 & 1 & 0 \\
Q_{42} & 0 & 0 & 0 & 1 & 0 & 1 & 0 & 0 & 0 & 0 & 0 & 0 & 0 & 0 & 0 & 0 & 0 & 1 & 0 & 1 & 0 & 1 & 1 & 0 & 0 & 0 & 0 & 0 & 1 & 0 & 0 & 1 & 1 & 0 & 0 \\
Q_{76} & 0 & 0 & 0 & 1 & 0 & 0 & 0 & 0 & 0 & 0 & 0 & 1 & 0 & 0 & 1 & 0 & 1 & 0 & 1 & 0 & 0 & 1 & 0 & 1 & 1 & 1 & 0 & 1 & 0 & 0 & 0 & 1 & 0 & 0 & 1 \\
Q_{95} & 0 & 0 & 0 & 1 & 0 & 0 & 0 & 0 & 0 & 0 & 0 & 0 & 1 & 0 & 1 & 0 & 1 & 1 & 0 & 0 & 0 & 1 & 0 & 1 & 0 & 1 & 1 & 0 & 0 & 0 & 0 & 0 & 0 & 1 & 0 \\
\hline
P_{16} & 0 & 0 & 0 & 0 & 1 & 0 & 0 & 0 & 0 & 0 & 1 & 1 & 1 & 1 & 1 & 1 & 0 & 0 & 0 & 0 & 0 & 0 & 0 & 0 & 1 & 1 & 0 & 0 & 0 & 0 & 0 & 0 & 0 & 0 & 1 \\
P_{35} & 0 & 0 & 0 & 0 & 1 & 0 & 1 & 0 & 0 & 1 & 0 & 0 & 1 & 1 & 1 & 1 & 0 & 0 & 0 & 0 & 0 & 0 & 0 & 1 & 0 & 1 & 0 & 0 & 0 & 0 & 0 & 0 & 0 & 1 & 0 \\
P_{48} & 0 & 0 & 0 & 0 & 1 & 1 & 0 & 0 & 1 & 1 & 0 & 0 & 0 & 0 & 0 & 1 & 0 & 0 & 0 & 0 & 0 & 0 & 1 & 0 & 0 & 0 & 0 & 0 & 0 & 0 & 0 & 0 & 1 & 0 & 0 \\
P_{72} & 0 & 0 & 0 & 0 & 1 & 1 & 0 & 1 & 0 & 0 & 0 & 1 & 0 & 0 & 1 & 0 & 0 & 0 & 0 & 0 & 0 & 1 & 0 & 0 & 0 & 0 & 0 & 0 & 0 & 0 & 0 & 1 & 0 & 0 & 0 \\
\end{array}
\right)
$}
}~.~
\nn\\
\eea
The associated $U(1)$-charges coming from the $J$- and $E$-terms are summarized in
\beal{es0802}
{\footnotesize
 Q_{JE}=
 \resizebox{0.85\textwidth}{!}{$
\left(
\begin{array}{ccccc|ccccccccccc|cccccccccc|ccccccccc}
p_1 & p_2 & p_3 & p_4 & p_5&
u_1 & u_2 & u_3 & u_4 & u_5 & u_6 & u_7 & u_8 & u_9 & u_{10} &
u_{11} &
v_1 & v_2 & v_3 & v_4 & v_5 & v_6 & v_7 & v_8 & v_9 & v_{10} &
o_1 & o_2 & o_3 & o_4 & o_5 & o_6 & o_7 & o_8 & o_9
\\
\hline
 3 & 3 & 3 & 1 & 0 & 0 & 0 & 0 & 0 & 0 & 0 & 0 & 0 & 0 & 0 & 0 & 0 & 0 & 0 & 0 & 0 & 0 & 0 & 0 & 0 & 0 & -1 & -1 & -1 & -1 & -1 & 0 & 0 & 0 & 0 \\
 3 & 3 & 3 & 0 & 0 & 0 & -1 & 0 & 0 & 0 & -1 & 0 & 0 & 0 & 0 & 0 & 0 & 0 & 0 & 0 & 0 & 0 & 0 & 0 & 0 & 1 & -1 & -1 & 0 & -1 & -1 & 0 & 0 & 0 & 0 \\
 2 & 2 & 2 & 0 & 0 & 0 & 0 & 0 & 0 & 0 & -1 & 0 & 0 & 0 & 0 & 0 & 0 & 0 & 0 & 0 & 0 & 0 & 0 & 0 & 1 & 0 & 0 & -1 & 0 & -1 & -1 & 0 & 0 & 0 & 0 \\
 2 & 2 & 2 & 0 & 0 & 0 & 0 & 0 & -1 & 0 & 0 & 0 & 0 & 0 & 0 & 0 & 0 & 0 & 0 & 0 & 0 & 0 & 1 & 0 & 0 & 0 & 0 & 0 & -1 & -1 & -1 & 0 & 0 & 0 & 0 \\
 2 & 2 & 2 & 0 & 0 & 0 & 0 & -1 & 0 & 0 & 0 & 0 & 0 & 0 & 0 & 0 & 0 & 0 & 0 & 0 & 0 & 1 & 0 & 0 & 0 & 0 & -1 & -1 & -1 & 0 & 0 & 0 & 0 & 0 & 0 \\
 2 & 2 & 2 & 0 & 0 & 0 & -1 & 0 & 0 & 0 & 0 & 0 & 0 & 0 & 0 & 0 & 0 & 0 & 0 & 0 & 0 & 0 & 0 & 1 & 0 & 0 & -1 & -1 & 0 & -1 & 0 & 0 & 0 & 0 & 0 \\
 2 & 2 & 2 & 0 & 0 & 0 & -1 & 0 & -1 & 0 & -1 & 0 & 0 & 0 & 0 & 1 & 0 & 0 & 0 & 0 & 0 & 0 & 0 & 0 & 0 & 0 & 0 & 0 & 0 & -1 & -1 & 0 & 0 & 0 & 0 \\
 2 & 2 & 2 & 0 & 0 & 0 & -1 & -1 & 0 & 0 & -1 & 0 & 0 & 0 & 1 & 0 & 0 & 0 & 0 & 0 & 0 & 0 & 0 & 0 & 0 & 0 & -1 & -1 & 0 & 0 & 0 & 0 & 0 & 0 & 0 \\
 1 & 1 & 1 & 0 & 1 & 0 & -1 & -1 & -1 & 0 & -1 & 0 & 0 & 0 & 0 & 0 & 0 & 0 & 0 & 0 & 0 & 0 & 0 & 0 & 0 & 0 & 0 & 0 & 0 & 0 & 0 & 0 & 0 & 0 & 0 \\
 1 & 1 & 1 & 0 & 0 & 1 & 0 & -1 & -1 & 0 & 0 & 0 & 0 & 0 & 0 & 0 & 0 & 0 & 0 & 0 & 0 & 0 & 0 & 0 & 0 & 0 & 0 & 0 & -1 & 0 & 0 & 0 & 0 & 0 & 0 \\
 1 & 1 & 1 & 0 & 0 & 0 & 0 & 0 & 0 & 0 & 0 & 0 & 0 & 0 & 0 & 0 & 1 & 0 & 0 & 0 & 0 & 0 & 0 & 0 & 0 & 0 & -1 & -1 & 0 & 0 & 0 & 0 & 0 & 0 & 0 \\
 1 & 1 & 1 & 0 & 0 & 0 & 0 & 0 & 0 & 0 & 0 & 0 & 0 & 0 & 0 & 0 & 0 & 1 & 0 & 0 & 0 & 0 & 0 & 0 & 0 & 0 & -1 & 0 & -1 & 0 & 0 & 0 & 0 & 0 & 0 \\
 1 & 1 & 1 & 0 & 0 & 0 & 0 & 0 & 0 & 0 & 0 & 0 & 0 & 0 & 0 & 0 & 0 & 0 & 1 & 0 & 0 & 0 & 0 & 0 & 0 & 0 & 0 & -1 & 0 & -1 & 0 & 0 & 0 & 0 & 0 \\
 1 & 1 & 1 & 0 & 0 & 0 & 0 & 0 & 0 & 0 & 0 & 0 & 0 & 0 & 0 & 0 & 0 & 0 & 0 & 1 & 0 & 0 & 0 & 0 & 0 & 0 & 0 & 0 & -1 & 0 & -1 & 0 & 0 & 0 & 0 \\
 1 & 1 & 1 & 0 & 0 & 0 & 0 & 0 & 0 & 0 & 0 & 0 & 0 & 0 & 0 & 0 & 0 & 0 & 0 & 0 & 1 & 0 & 0 & 0 & 0 & 0 & 0 & 0 & 0 & -1 & -1 & 0 & 0 & 0 & 0 \\
 1 & 1 & 1 & 0 & 0 & 0 & 0 & 0 & 0 & 0 & -1 & 0 & 0 & 0 & 0 & 0 & 0 & 0 & 0 & 0 & 0 & 0 & 0 & 0 & 0 & 0 & 0 & -1 & 0 & 0 & -1 & 0 & 0 & 0 & 1 \\
 1 & 1 & 1 & 0 & 0 & 0 & 0 & 0 & -1 & 0 & 0 & 0 & 0 & 0 & 0 & 0 & 0 & 0 & 0 & 0 & 0 & 0 & 0 & 0 & 0 & 0 & 0 & 0 & -1 & -1 & 0 & 0 & 1 & 0 & 0 \\
 1 & 1 & 1 & 0 & 0 & 0 & 0 & -1 & 0 & 0 & 0 & 0 & 0 & 0 & 0 & 0 & 0 & 0 & 0 & 0 & 0 & 0 & 0 & 0 & 0 & 0 & 0 & -1 & -1 & 0 & 0 & 1 & 0 & 0 & 0 \\
 1 & 1 & 1 & 0 & 0 & 0 & 0 & -1 & 0 & 0 & -1 & 1 & 0 & 0 & 0 & 0 & 0 & 0 & 0 & 0 & 0 & 0 & 0 & 0 & 0 & 0 & 0 & -1 & 0 & 0 & 0 & 0 & 0 & 0 & 0 \\
 1 & 1 & 1 & 0 & 0 & 0 & -1 & 0 & 0 & 0 & 0 & 0 & 0 & 0 & 0 & 0 & 0 & 0 & 0 & 0 & 0 & 0 & 0 & 0 & 0 & 0 & -1 & 0 & 0 & -1 & 0 & 0 & 0 & 1 & 0 \\
 1 & 1 & 1 & 0 & 0 & 0 & -1 & 0 & 0 & 0 & -1 & 0 & 1 & 0 & 0 & 0 & 0 & 0 & 0 & 0 & 0 & 0 & 0 & 0 & 0 & 0 & -1 & 0 & 0 & 0 & 0 & 0 & 0 & 0 & 0 \\
 1 & 1 & 1 & 0 & 0 & 0 & -1 & 0 & 0 & 0 & -1 & 0 & 0 & 1 & 0 & 0 & 0 & 0 & 0 & 0 & 0 & 0 & 0 & 0 & 0 & 0 & 0 & 0 & 0 & 0 & -1 & 0 & 0 & 0 & 0 \\
 1 & 1 & 1 & 0 & 0 & 0 & -1 & 0 & -1 & 1 & 0 & 0 & 0 & 0 & 0 & 0 & 0 & 0 & 0 & 0 & 0 & 0 & 0 & 0 & 0 & 0 & 0 & 0 & 0 & -1 & 0 & 0 & 0 & 0 & 0 \\
\end{array}
\right)
$}
}~.~
\nn\\
\eea
The $U(1)$-charges coming from the $D$-terms are summarized in 
\beal{es0803}
{\footnotesize
 Q_{D}=
 \resizebox{0.85\textwidth}{!}{$
\left(
\begin{array}{ccccc|ccccccccccc|cccccccccc|ccccccccc}
p_1 & p_2 & p_3 & p_4 & p_5&
u_1 & u_2 & u_3 & u_4 & u_5 & u_6 & u_7 & u_8 & u_9 & u_{10} &
u_{11} &
v_1 & v_2 & v_3 & v_4 & v_5 & v_6 & v_7 & v_8 & v_9 & v_{10} &
o_1 & o_2 & o_3 & o_4 & o_5 & o_6 & o_7 & o_8 & o_9
\\
\hline
 1 & 1 & 1 & 0 & 0 & 0 & 0 & 0 & 0 & 0 & -1 & 0 & 0 & 0 & 0 & 0 & 0 & 0 & 0 & 0 & 0 & 0 & 0 & 0 & 0 & 0 & 0 & -1 & 0 & 0 & 0 & 0 & 0 & 0 & 0 \\
 1 & 1 & 1 & 0 & 0 & 0 & 0 & 0 & -1 & 0 & 0 & 0 & 0 & 0 & 0 & 0 & 0 & 0 & 0 & 0 & 0 & 0 & 0 & 0 & 0 & 0 & 0 & 0 & 0 & -1 & 0 & 0 & 0 & 0 & 0 \\
 1 & 1 & 1 & 0 & 0 & 0 & 0 & -1 & 0 & 0 & 0 & 0 & 0 & 0 & 0 & 0 & 0 & 0 & 0 & 0 & 0 & 0 & 0 & 0 & 0 & 0 & 0 & 0 & -1 & 0 & 0 & 0 & 0 & 0 & 0 \\
 1 & 1 & 1 & 0 & 0 & 0 & -1 & 0 & 0 & 0 & 0 & 0 & 0 & 0 & 0 & 0 & 0 & 0 & 0 & 0 & 0 & 0 & 0 & 0 & 0 & 0 & -1 & 0 & 0 & 0 & 0 & 0 & 0 & 0 & 0 \\
 0 & 0 & 0 & 0 & 0 & 0 & 0 & 0 & 0 & 0 & 0 & 0 & 0 & 0 & 0 & 0 & 0 & 0 & 0 & 0 & 0 & 0 & 0 & 0 & 0 & 0 & 1 & 0 & 0 & 0 & -1 & 0 & 0 & 0 & 0 \\
 -1 & -1 & -1 & 0 & 0 & 0 & 1 & 0 & 0 & 0 & 0 & 0 & 0 & 0 & 0 & 0 & 0 & 0 & 0 & 0 & 0 & 0 & 0 & 0 & 0 & 0 & 0 & 0 & 0 & 1 & 0 & 0 & 0 & 0 & 0 \\
 -1 & -1 & -1 & 0 & 0 & 0 & 0 & 1 & 0 & 0 & 0 & 0 & 0 & 0 & 0 & 0 & 0 & 0 & 0 & 0 & 0 & 0 & 0 & 0 & 0 & 0 & 0 & 1 & 0 & 0 & 0 & 0 & 0 & 0 & 0 \\
 -1 & -1 & -1 & 0 & 0 & 0 & 0 & 0 & 1 & 0 & 0 & 0 & 0 & 0 & 0 & 0 & 0 & 0 & 0 & 0 & 0 & 0 & 0 & 0 & 0 & 0 & 0 & 0 & 1 & 0 & 0 & 0 & 0 & 0 & 0 \\
\end{array}
\right)
$}
}~.~
\nn\\
\eea
We note that the charge matrices remain invariant under permutations of the extremal GLSM fields $(p_1,p_2,p_3)$ indicating that the global symmetry is enhanced to $SU(3)\times U(1)^2$.
This is expected by the construction of the general $J$- and $E$-terms in section \sref{sec:cp2intro} and the isometry group of $Y^{3,4}(\mathbb{CP}^2)$.

The toric data for the $2d$ $(0,2)$ theory is given by
 \beal{es0805}
{\footnotesize
 G_{t}=
 \resizebox{0.85\textwidth}{!}{$
\left(
\begin{array}{ccccc|ccccccccccc|cccccccccc|ccccccccc}
p_1 & p_2 & p_3 & p_4 & p_5&
u_1 & u_2 & u_3 & u_4 & u_5 & u_6 & u_7 & u_8 & u_9 & u_{10} &
u_{11} &
v_1 & v_2 & v_3 & v_4 & v_5 & v_6 & v_7 & v_8 & v_9 & v_{10} &
o_1 & o_2 & o_3 & o_4 & o_5 & o_6 & o_7 & o_8 & o_9
\\
\hline
 1 & 1 & 1 & 1 & 1 & 1 & 1 & 1 & 1 & 1 & 1 & 1 & 1 & 1 & 1 & 1 & 1 & 1 & 1 & 1 & 1 & 1 & 1 & 1 & 1 & 1 & 2 & 2 & 2 & 2 & 2 & 2 & 2 & 2 & 2 \\
 1 & 0 & -1 & 0 & 0 & 0 & 0 & 0 & 0 & 0 & 0 & 0 & 0 & 0 & 0 & 0 & 0 & 0 & 0 & 0 & 0 & 0 & 0 & 0 & 0 & 0 & 0 & 0 & 0 & 0 & 0 & 0 & 0 & 0 & 0 \\
 0 & 1 & -1 & 0 & 0 & 0 & 0 & 0 & 0 & 0 & 0 & 0 & 0 & 0 & 0 & 0 & 0 & 0 & 0 & 0 & 0 & 0 & 0 & 0 & 0 & 0 & 0 & 0 & 0 & 0 & 0 & 0 & 0 & 0 & 0 \\
 0 & 0 & 4 & 3 & 0 & 1 & 1 & 1 & 1 & 1 & 1 & 1 & 1 & 1 & 1 & 1 & 2 & 2 & 2 & 2 & 2 & 2 & 2 & 2 & 2 & 2 & 3 & 3 & 3 & 3 & 3 & 3 & 3 & 3 & 3 \\
\end{array}
\right)
$}
}~,~
\nn\\
\eea
where $o_k$ correspond to extra GLSM fields \cite{Franco:2015tna}.
The corresponding toric diagram is shown in \fref{f_toric_cp2_34}. 

\begin{figure}[H]
\begin{center}
\resizebox{0.35\hsize}{!}{
\includegraphics[height=6cm]{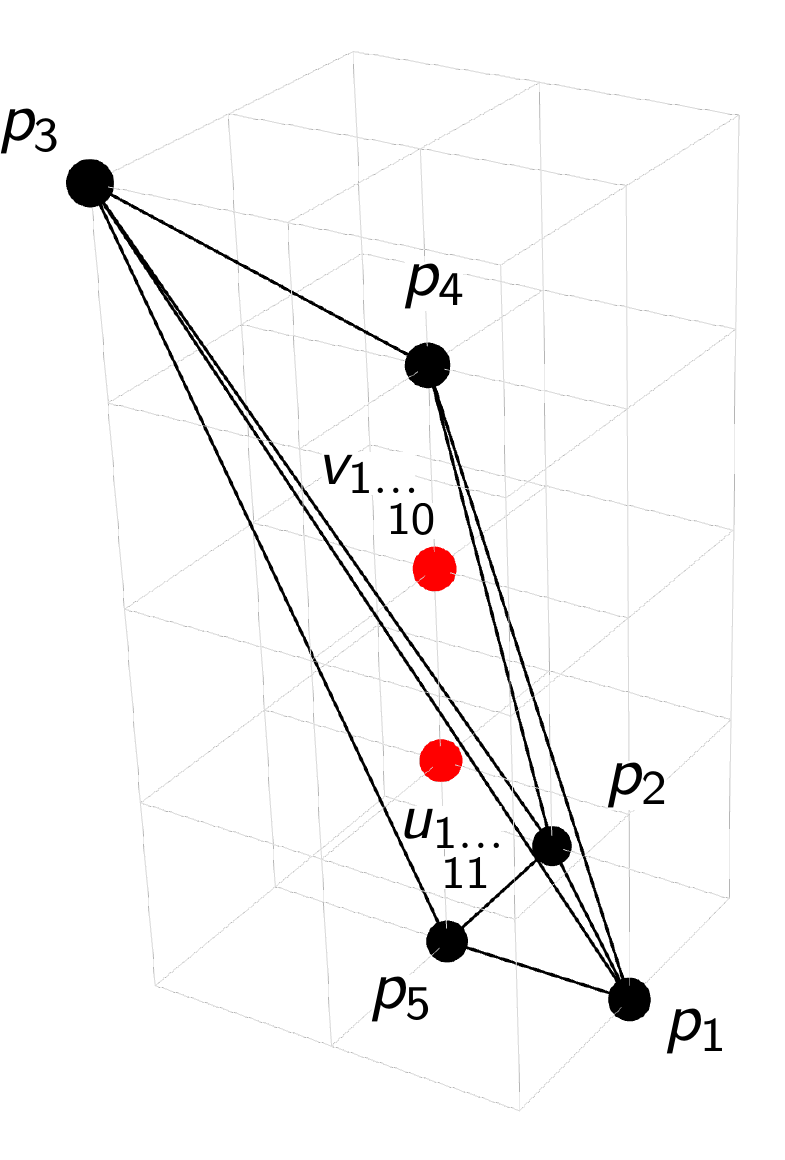} 
}
\caption{
Toric diagram of $Y^{3,4}(\mathbb{CP}^2)$.
\label{f_toric_cp2_34}}
 \end{center}
 \end{figure}

\section*{Acknowledgements}

S.F. was supported by the U.S. National Science Foundation grants PHY-1820721, PHY-2112729 and DMS-1854179.
D.G. was supported by the Basic Science Research Program of the National Research Foundation of Korea (NRF) under the Ministry of Education in Korea (NRF-2022R1A6A3A03068148), and by JST PRESTO Grant Number JPMJPR2117. 
D.G. would like to thank UNIST where this project was initiated.
R.K.-S. was supported by a Basic Research Project grant of the National Research Foundation of Korea (NRF-2022R1F1A1073128), and
a Start-up Research Grant for new faculty at UNIST. He is also partly supported by the BK21 Program (``Next Generation Education Program for Mathematical Sciences'', 4299990414089) funded by the Ministry of Education in Korea and the National Research Foundation of Korea (NRF).

\appendix

\section{Additional Examples}

\subsection{$J$- and $E$-terms for $Y^{1,k}(\mathbb{CP}^1 \times \mathbb{CP}^1)$ \label{sec:sap1}}

\paragraph{$Y^{1,0}(\mathbb{CP}^1 \times \mathbb{CP}^1)$.}

  \beq
\begin{array}{rccrcl}
 &   J &   & & E &  \\
 \Lambda_{12}^{i}  : &\epsilon_{ij}\epsilon_{kl} X^{k}_{23}   X^{j}_{34}   X^{l}_{41 } &~&    X^{i}_{12}   Q_{22 } &-&  Q_{11}   X^{i}_{12}\\    
 \Lambda_{23}^{2i}  : &\epsilon_{ij}\epsilon_{kl} X^{k}_{34}   X^{j}_{41}   X^{l}_{12 } &~&    X^{i}_{23}   Q_{33 } &-&  Q_{22}   X^{i}_{23}\\    
 \Lambda_{34}^{2i}  : &\epsilon_{ij}\epsilon_{kl} X^{k}_{41}   X^{j}_{12}   X^{l}_{23 } &~&    X^{i}_{34}   Q_{44 } &-&  Q_{33}   X^{i}_{34}\\   
 \Lambda_{41}^{2i}  : &\epsilon_{ij}\epsilon_{kl} X^{k}_{12}   X^{j}_{23}   X^{l}_{34 } &~&    X^{i}_{41}   Q_{11 } &-&  Q_{44}   X^{i}_{41}\\ 
 \end{array} 
\label{es2000}
 \eeq
 
 \paragraph{$Y^{1,1}(\mathbb{CP}^1 \times \mathbb{CP}^1)$.}

  \beq
\begin{array}{rccrcl}
 &   J &   & & E &  \\
 \Lambda_{12}^{i}  : & \epsilon_{ij}\epsilon_{kl} X^{k}_{23}   X^{j}_{34}   X^{l}_{41 }&~& P_{13}   X^{i}_{34}   Q_{42 } &-&   Q_{13}   X^{i}_{34}   P_{42}\\ 
 \Lambda_{43}^{1i}  : & \epsilon_{ij}\epsilon_{kl} X^{k}_{34}   X^{j}_{41}   Q_{13}   X^{l}_{34 } &~& X^{i}_{41}   P_{13 } &-&  P_{42}   X^{i}_{23 }\\ 
 \Lambda_{43}^{2i}  : & \epsilon_{ij}\epsilon_{kl} X^{k}_{34}   X^{j}_{41}   P_{13}   X^{l}_{34 } &~& X^{i}_{41}   Q_{13 } &-&  Q_{42}   X^{i}_{23 }\\ 
 \end{array} 
\label{es2001}
 \eeq

\subsection{$J$- and $E$-terms for $Y^{1,k}(\mathbb{CP}^2)$ \label{sec:sap2}}

\paragraph{$Y^{1,0}(\mathbb{CP}^2)$.}

  \beq
\begin{array}{rccrcl}
 &   J &   & & E &  \\
 \Lambda_{12}^{i}  : &   \epsilon_{ijk} X^{j}_{23 }   X^{k}_{31 } &~&   X^{i}_{12 }   Q_{22 } &-&  Q_{11 }   X^{i}_{12}\\  
 \Lambda_{23}^{2i}  : & \epsilon_{ijk}   X^{j}_{31 }   X^{k}_{12 } &~&   X^{i}_{23 }   Q_{33 } &-&  Q_{22 }   X^{i}_{23 }\\  
 \Lambda_{31}^{2i}  : & \epsilon_{ijk}   X^{j}_{12 }   X^{k}_{23 } &~&   X^{i}_{31 }   Q_{11 } &-&  Q_{33 }   X^{i}_{31 }\\  
 \end{array} 
\label{es2002}
 \eeq
 
 \paragraph{$Y^{1,1}(\mathbb{CP}^2)$.}

  \beq
\begin{array}{rccrcl}
 &   J &   & & E &  \\
 \Lambda_{12}^{i}  : &   \epsilon_{ijk} X^{j}_{23}   X^{k}_{31 } &~&      P_{12}   X^{i}_{23}   Q_{32 } &-&   Q_{13}   X^{i}_{31}   P_{12}\\      
 \Lambda_{33}^{2i}  : &   \epsilon_{ijk} X^{j}_{31}   P_{12}   X^{k}_{23 } &~&      X^{i}_{31}   Q_{13 } &-&  Q_{32}   X^{i}_{23 }\\      
 \end{array} 
\label{es2003}
 \eeq
 \\

\subsection{$J$- and $E$-terms for $Y^{2,k}(\mathbb{CP}^2)$ \label{sec:sap3}}

\paragraph{$Y^{2,0}(\mathbb{CP}^2)$.}

  \beq
\begin{array}{rccrcl}
 &   J &   & & E &  \\
 \Lambda_{12}^{i}  : &  \epsilon_{ijk}  X^{j}_{23 }  X^{k}_{31 } &~&     X^{i}_{15 }  Q_{52 } &-&   Q_{14 }  X^{i}_{42 }\\ 
 \Lambda_{26}^{2i}  : & \epsilon_{ijk}   X^{j}_{64 }  X^{k}_{42 } &~&     X^{i}_{23 }  Q_{36 } &-&   Q_{25 }  X^{i}_{56 }\\ 
 \Lambda_{34}^{2i}  : & \epsilon_{ijk}   X^{j}_{42 }  X^{k}_{23 } &~&     X^{i}_{31 }  Q_{14 } &-&   Q_{36 }  X^{i}_{64 }\\ 
 \Lambda_{45}^{i}  : &  \epsilon_{ijk}  X^{j}_{56 }  X^{k}_{64 } &~&     X^{i}_{42 }  Q_{25 } &-&   Q_{41 }  X^{i}_{15 }\\ 
 \Lambda_{53}^{2i}  : & \epsilon_{ijk}   X^{j}_{31 }  X^{k}_{15 } &~&     X^{i}_{56 }  Q_{63 } &-&   Q_{52 }  X^{i}_{23 }\\ 
 \Lambda_{61}^{2i}  : & \epsilon_{ijk}   X^{j}_{15 }  X^{k}_{56 } &~&     X^{i}_{64 }  Q_{41 } &-&   Q_{63 }  X^{i}_{31 }\\ 
 \end{array} 
\label{es2004}
 \eeq

 \paragraph{$Y^{2,1}(\mathbb{CP}^2)$.}

  \beq
\begin{array}{rccrcl}
 &   J &   & & E &  \\
 \Lambda_{12}^{i}  : &   \epsilon_{ijk}    X^{j}_{23}  X^{k}_{31 } &~&     P_{15}  X^{i}_{56}  Q_{62 } &-&  Q_{14}  X^{i}_{42 }\\     
 \Lambda_{26}^{2i}  : &   \epsilon_{ijk}    X^{j}_{64}  X^{k}_{42 }  &~&     X^{i}_{23}  Q_{36 } &-&  Q_{25}  X^{i}_{56 }\\     
 \Lambda_{34}^{2i}  : &   \epsilon_{ijk}    X^{j}_{42}  X^{k}_{23 } &~&     X^{i}_{31}  Q_{14 } &-&  Q_{36}  X^{i}_{64 }\\     
 \Lambda_{45}^{i}  : &   \epsilon_{ijk}    X^{j}_{56}  X^{k}_{64 }  &~&     X^{i}_{42}  Q_{25 } &-&   Q_{43}  X^{i}_{31}  P_{15}\\     
 \Lambda_{63}^{2i}  : &  \epsilon_{ijk}     X^{j}_{31}  P_{15}  X^{k}_{56} &~&     X^{i}_{64}  Q_{43 } &-&  Q_{62}  X^{i}_{23 }\\    
 \end{array} 
\label{es2005}
 \eeq
 
  \paragraph{$Y^{2,2}(\mathbb{CP}^2)$.}

  \beq
\begin{array}{rccrcl}
 &   J &   & & E &  \\
 \Lambda_{12}^{i}  : & \epsilon_{ijk}     X^{j}_{23 }  X^{k}_{31 }  &~&    P_{15 }  X^{i}_{56 }  Q_{62 } &-&   Q_{16 }  X^{i}_{64 }  P_{42}\\    
 \Lambda_{36}^{2i}  : &  \epsilon_{ijk}    X^{j}_{64 }  P_{42 }  X^{k}_{23 }  &~&    X^{i}_{31 }  Q_{16 } &-&  Q_{35 }  X^{i}_{56 }\\    
 \Lambda_{45}^{i}  : &\epsilon_{ijk}      X^{j}_{56 }  X^{k}_{64 }  &~&    P_{42 }  X^{i}_{23 }  Q_{35 } &-&   Q_{43 }  X^{i}_{31 }  P_{15}\\    
 \Lambda_{63}^{2i}  : & \epsilon_{ijk}     X^{j}_{31 }  P_{15 }  X^{k}_{56 }   &~&    X^{i}_{64 }  Q_{43 } &-&  Q_{62 }  X^{i}_{23 }\\   
 \end{array} 
\label{es2006}
 \eeq
 
   \paragraph{$Y^{2,3}(\mathbb{CP}^2)$.}

  \beq
\begin{array}{rccrcl}
 &   J &   & & E &  \\
 \Lambda_{12}^{i}  : &  \epsilon_{ijk}    X^{j}_{23}  X^{k}_{31 } &~&    P_{16}  X^{i}_{64}  Q_{42 } &-&   Q_{16}  X^{i}_{64}  P_{42}\\    
 \Lambda_{36}^{1i}  : &  \epsilon_{ijk}    X^{j}_{64}  Q_{42}  X^{k}_{23 } &~&    X^{i}_{31}  P_{16 } &-&  P_{35}  X^{i}_{56 }\\    
 \Lambda_{36}^{2i}  : &   \epsilon_{ijk}   X^{j}_{64}  P_{42}  X^{k}_{23 } &~&    X^{i}_{31}  Q_{16 } &-&  Q_{35}  X^{i}_{56 }\\    
 \Lambda_{45}^{i}  : &   \epsilon_{ijk}   X^{j}_{56}  X^{k}_{64 } &~&    P_{42}  X^{i}_{23}  Q_{35 } &-&   Q_{42}  X^{i}_{23}  P_{35}\\    
 \end{array} 
\label{es2007}
 \eeq

\bibliographystyle{JHEP}
\bibliography{mybib}

\end{document}